\documentclass[12pt]{article}
\usepackage{jheppub}
\usepackage{pstricks}
\usepackage{amsmath,amssymb, pxfonts}
\usepackage{graphicx}
\usepackage{ytableau}

\newcommand{\vev}[1]{\langle #1\rangle}
\newcommand{\ket}[1]{| #1\rangle}
\newcommand{\bra}[1]{\langle #1|}
\newcommand{\braket}[2]{\langle #1| #2\rangle}
\newcommand{\BR}{\mathbb{R}}

\def\Tr{\textrm{Tr}}

\newcommand{\1}{{\it id}}

\newcommand{\beq}{\begin{equation}}
\newcommand{\beqs}{\begin{equation*}}
\newcommand{\eeq}{\end{equation}}
\newcommand{\eeqs}{\end{equation*}}

\allowdisplaybreaks

\author{David Berenstein} \author{and Alexandra Miller}\affiliation{ Department of Physics, University of California, Santa Barbara, CA 93106}
\abstract{
We show that superpositions of classical states in quantum gravity with fixed topology can lead to new classical states with a different topology. We study this phenomenon in a particular limit of the LLM geometries. In this limit, the UV complete minisuperspace of allowed quantum states is exactly given by  the Hilbert space of a free chiral boson in two dimensions. We construct this chiral boson purely in terms of combinatorial objects associated with the permutation group. As a byproduct of this analysis, we re-derive the Murnaghan-Nakayama rule for characters of the permutation group. We are able to express this rule   in terms of operator relations for raising and lowering operators on the Hilbert space of states in a free fermion basis.
 Our construction provides a preferred notion of bulk locality by studying an appropriate notion of  D-brane state generating functions.
  We  describe how multi-droplet LLM geometries with different topologies give new classical limits of the free chiral boson, even though they can be written as superpositions of coherent states with trivial topology.  As a consequence, topology cannot be accessed by a single operator measurement in this quantum system. We study other non-linear measurements in the quantum wave-function, based on  uncertainty and entanglement between modes  of the chiral boson, that can be used as order parameters to measure the topology of such states.}

\preprint{}

\begin{document}


\title{Superposition induced topology changes in quantum gravity}

\maketitle

\section{Introduction}

The AdS/CFT correspondence \cite{Maldacena:1997re} has provided a detailed model of quantum gravity in terms of a dual quantum field theory. In particular, it has shown that the black hole information paradox \cite{Hawking:1974sw} is soluble within a quantum mechanical framework (at least in principle). 

It is natural to ask what else the AdS/CFT correspondence tells us about quantum gravity. The careful reframing of the Hawking paradox in terms of information theory, resulting in the AMPS paradox \cite{Almheiri:2012rt}, suggests that our ideas about quantum gravity might need rather large modifications in the presence of black hole horizons. 

Having a consistent theory of gravity in terms of a dual field theory turns things around. We can ask if basic assumptions about the nature of gravity are actually valid or not. 
We will focus in particular on the problem of measuring the topology of a spacetime. We can ask in the AdS/CFT context if this arises from an operator measurement or instead from some other procedure. 

For example, classical general relativity does not allow for topology changes and different topologies can many times be characterized by different topological invariants (which can be thought of as numbers) that can usually be computed in terms of the metric.
Our current semiclassical understanding of quantum gravity  suggests that we need to sum over different topologies of spacetime to obtain the correct 
answer for physical processes. 
This should lead to a picture of spacetime that microscopically has the attributes of a spacetime foam \cite{Wheeler,Hawking:1979zw}, but this picture based on the semiclassical quantization of gravity still suggests that the metric and the topology are observables.

 The simplest example of topology change in the AdS/CFT correspondence arises from the Hawking-Page phase transition for black holes in AdS \cite{Hawking:1982dh}. This transition can be understood as a confinement/deconfinement phase transition in gauge theory \cite{Witten:1998zw}.
This leads to a new topology of spacetime, certainly in the Euclidean field theory. However, one can argue that this change in topology ends up being hidden behind the horizon if one considers the real time dynamics 
of a black hole. The Einstein-Rosen bridge of the maximal extension of the $AdS$ black hole geometry connects to another region of spacetime that is completely hidden behind the horizon of the black hole and is not accessible to an observer on the boundary of AdS space in finite time measurements. 
More precisely, it is expected that the eternal black hole in AdS is equivalent to the field theory double of the thermal state \cite{Maldacena:2001kr}, so the other asymptotic boundary of the spacetime corresponds to a copy of the degrees of freedom of the first field theory that is completely independent of the original theory, except for the fact that the state in the double field theory is entangling the two copies of the field theory. In this sense, the topology change requires the addition of a hidden sector to the original field theory.

On the other hand, a true topology change that is not hidden behind a horizon can be obtained when studying bubbling solutions in AdS \cite{Lin:2004nb}. In this case, it is possible to show that various spacetime topologies can be supported by the same boundary field theory. Indeed, it is even possible to argue that the bubbles can be tiny and almost indistinguishable from perturbative excitations of the geometry.  The multi-bubbling solutions can sometimes be interpreted as spacetime foam.
A natural question to ask is if we can measure the geometry (and even more coarsely, the macroscopic topology) uniquely, for sufficiently smooth configurations and their superpositions, or if the notion of geometry and topology is highly dependent on the state that we are studying. That is, is the topology of spacetime a quantum observable in this minisuperspace model?

Since the full theory is a complete quantum mechanical system, we can consider arbitrary superpositions of geometries (with the same or different topology). Under the naive rules of semiclassical quantum mechanics,
one could argue that there should be a ``topology measuring" operator that distinguishes the different topologies, so that one could in principle divide the Hilbert space into  superselection sectors that are eigenstates of the topology operator.
However, Van Raamsdonk has suggested that topology and geometry arise from quantum entanglement \cite{VanRaamsdonk:2010pw}. In that case, it would be impossible to produce such a topology measuring operator. For example, entanglement of a factorization of the Hilbert space  can be obtained from superpositions of states each of which has no such entanglement. These could all have fixed topology, while the new superposed state could be a state with a different topology. In the case studied by Van Raamsdonk,
this was associated with the double field theory interpretation of the black hole geometry \cite{Maldacena:2001kr}, but  one might imagine that this could be applicable more generally \cite{Maldacena:2013xja}. This program for understanding geometry goes under the ``ER=EPR'' banner. 
One can also ask if these topology changes due to entanglement always require black hole horizons.

The main problem in being  precise about this idea is that models of holographic quantum field theories  where 
this could be analyzed tend to be too complicated to understand how to evaluate entanglement properties of most states, 
except perhaps for very special states like the field theory double. This is also beyond the Ryu-Takayanagi setup \cite{Ryu:2006bv}, as that setup assumes a single background geometry: it is not known how to make it compatible with superpositions of (macroscopically distinct) states.

In this paper, we argue that the topology of spacetime in quantum gravity cannot be measured by an operator.
 We do this in a situation that is free from black hole horizons and in a setup where a complete description of the quantum states that contribute to the phenomenon can be understood in excruciating detail: the corresponding states have been completely classified in the dual field theory and a complete basis for such states is known. 
The main argument has been presented by us already in \cite{Berenstein:2016pcx}. This argument depends on certain assumptions about the topology of  coherent states being trivial.
 In that work, it was also explained  how the topology of simple classes of states ends up encoded in the uncertainty and entanglement of various variables, echoing the 
observation of Van Raamsdonk. These are not operator measurements, but we argued that they are computable for a large class of interesting states which might not be fully classical, but that they are consistent with an interpretation of having a few excited quanta on top of a classical background. 
 
 More important than the statement that topology is not an operator,   is the fact that it can still be computed from the wavefunction of the system for states that are sufficiently classical.
 One can check that effective field theory makes sense in the vicinity of such states, although the effective cutoff required to make sense of effective field theory was out of reach to the techniques used in that paper.
 This paper provides the technical details that are required to explore these ideas in a controlled setting without making any approximations: we can analyze a complete description of the states that were studied  in \cite{Berenstein:2016pcx} as well as other states that were not covered there.  With this information we can explore the physics of the cutoff directly. This way we will find that the computed value of the topology varies as we vary the cutoff. In this sense, the topology that we will assign to spacetime depends on choices that we make. These choices are physical: a cutoff is usually  related to the limits of an experimental setup. In the right double scaling limit, one should see that all cutoffs can be pushed to infinity and that the usual classical picture of gravity is correct as a statement about these double scaling limits.

The main goal of the paper is to  setup a situation where the topology  of spacetime can be well understood in what turns out to be a {\em free field theory model of quantum gravity}.
That is, we will show that the topology of spacetime can already be argued to change in a regime where physics should be perturbative.  An advantage to having a setup in free field theory is that there will be a mode expansion. That mode expansion will  provide us with  various ways to factorize the Hilbert space in a canonical way. Those factorizations make it possible to compute entanglement entropies that have a physical meaning.

 The field theory  in question is the theory of a chiral free boson in two dimensions: this is the (UV complete) theory of gravity in the bulk. This quantum theory arises as a limit of the minisuperspace of half BPS states in ${\cal N}=4 $ SYM, but it can arise in many other holographic duals in a similar limit fashion. The supergravity dual configurations that correspond to smooth horizonless geometries with various topologies are well known and have been classified completely by Lin, Lunin and Maldacena \cite{Lin:2004nb}. They can be described as a black and white coloring of the plane, with some restrictions on the area of the colored regions that enforce the Dirac quantization condition.  We will call these the LLM solutions. If one quantizes the solutions around the ground state droplet (which is a circular disk), one gets chiral edge excitations of the droplet \cite{Grant:2005qc}. We want to analyze this system in the free field limit for the edge excitations.

This free field theory is interpreted as an effective closed string theory that is UV complete from the point of view of quantum mechanics: we do not need any information from outside the free chiral boson system to compute quantum mechanical amplitudes and probabilities. The main reason for requiring the gravity theory to be free is that it provides us with a canonical factorization of the full Hilbert space in terms of the mode decomposition of the free field. Again, this makes it possible to compute the entanglement and uncertainty properties of the different modes and to compare the results obtained this way with the topology of the corresponding LLM solution. 
 
To address various technical issues that are required to make the arguments more forceful and precise, the theory of the chiral boson is constructed in a novel way by considering  just the combinatorics of the symmetric group. This is an abstraction of the description of the dynamics of the states in the ${\cal N}=4 $ SYM dual that is necessary in order to be able to take the free field limit. This construction is independent of the gravity realization, but we can ask to what extent the gravity picture is forced on us and in particular, if the combinatorial construction suggests to us a particular notion of locality. 

We start with the idea that there are two natural basis for states. The first basis is what we would call the multi-trace basis in a matrix model. The second basis is the set of characters of a matrix in representations of the group. The first one is approximately orthogonal from large $N$ arguments \cite{Witten:1998qj}. The second one is exactly orthogonal by direct computation \cite{Corley:2001zk} in ${\cal N}=4 $ SYM, and also when used in the study of 2D Yang Mills theory \cite{Gross:1993hu}. 
 We will call the first the "string basis" and the second one the ``D-brane" basis. These can be described entirely in terms of group theory without any reference to matrices. 
 This property later lets us take a limit that can be interpreted as the exact $N=\infty$ limit of a matrix model and we can show directly that it corresponds to a free field theory.

 We construct a Hilbert space whose basis elements are either conjugacy classes or 
irreducible representations of the symmetric group (these are in 1-1 correspondence with each other).  The two preferred sets of states built from this data, the set of irreducible representations (we will call this the D-brane basis) and the set of conjugacy classes (the closed string basis), are related by a Fourier transform involving the characters of the group. We can show that this way we obtain 
a Hilbert space that has the same counting of states and energies as the Fock space of a chiral boson.
When we consider a particular generating series of D-brane states and their precise description in the string basis, an effective notion of locality emerges naturally, giving rise to a local field theory on a circle that is the chiral boson quantum field theory. That is, the circle on which the chiral boson theory lives on is deduced from the combinatorics of the symmetric group. 

It turns out that the simplest such multi-D-brane generating series can be shown to be exactly given by a coherent state of the chiral boson.  In this sense, the natural D-brane states are completely classical solutions of the free field theory and end up being associated with the same topology as the vacuum. We will then show that there
are other {\em bubbling solution states} that can be interpreted geometrically as states with a different topology, which is macroscopically very different from each of the coherent states that give rise to the state. 
The point is that coherent states are over-complete, so we can get any other state by superposing these classical states. Because these are superpositions of states with a fixed topology,  there cannot be  any quantum observable (in the sense of projectors in a Hilbert space) that 
can distinguish the two collections \cite{Berenstein:2016pcx}.

What we show instead is that the new bubbling states have an effective dynamics for nearby states whose collective dynamics is given by {\em multiple chiral bosons}, so that the states correspond to a new classical limit of the field theory. We will call these collective modes the IR (infrared) fields, while the original chiral field will be called the UV (ultraviolet) field theory. A similar statement about multiple chiral bosons is found in the work \cite{Koch:2008ah} as a suggestion for the dynamics around special sets of configurations, although one can already intuitively understand this from the original work \cite{Lin:2004nb}. Here we explain how this works in detail in a way that lets us go beyond simple Young diagram reference states. The work we do is aimed at being able to do computations. Our work makes it possible to
understand how an effective cutoff in these multiple chiral bosons is generated dynamically by the state. We also develop the tools that allow us to explore what happens when we move beyond the cutoff.

 The effective dynamics of these collective fields is subject to a stringy exclusion principle, and the UV field modes are in the vacuum well beyond this exclusion regime.
 The notion of new topology and geometry only makes sense for these states (and coherent excitations of their collective dynamics) when we study their physical properties at low and intermediate scales. They can be regarded as semiclassical states (superpositions of quantum excitations around a vacuum) for high energy observables. This is similar to the study of multi-throat configurations. We are also able to study in detail the amount of entanglement of the UV theory modes, essentially mode per mode, and to use that to characterize the topology of the new classical limits
 we need to compute other quantities in the quantum theory that end up being non-linear in the wave-function of the configuration. These are related to measuring how classical the state is (uncertainties), and how entangled the state is from a canonical preferred factorization of the full Hilbert space of states.

The paper is organized as follows.  In section two, we review LLM geometries, which are dual to half BPS states in ${\cal N}= 4 $ SYM.  These are the geometries we will study throughout the paper.  In section three, we build up the technology used to perform most of our calculations.  This comes from analyzing the Hilbert space formed by considering conjugacy classes and irreducible representations of the symmetric group.  This Hilbert space describes the free chiral boson in one dimension, which is also known to be equivalent to the space of half BPS states when $N\rightarrow \infty$.  In section four, we show how to build D-branes using both the conjugacy class basis and the irreducible representation basis.  To go between these basis, one needs the characters of the group. So, in section five, we introduce the Murnaghan-Nakayama rule, which provides a useful way to compute these characters.  Further, we explain how our Hilbert space can also be used to describe free fermions and show how the Murnaghan-Nakayama rule encodes the fermi statistics.

In section six, we get to the real meat of the paper.  Here, we study various states whose duals form classical geometries.  We show that one set of these come from coherent states of the raising and lowering operators in our oscillator bases.  However, we show that these are not the only states that give rise to classical geometries and further, the new states have a different topology than the coherent states.  We argue by contradiction that this leads to the fact that there cannot be a topology measuring operator in this set-up.  In section seven, we show how uncertainty measurements can provide a method to compute topology in not one, but several measurements.  We then discuss states that should not be thought of as having a classical dual.  Further, we analyze how to make progress studying more complicated geometries with folds.  Finally, in section eight, we discuss a second method for determining topology, this time from entanglement entropy calculations.

\section{Droplet geometries and the limits of semiclassical reasoning}

Half BPS states in ${\cal N}= 4 $ SYM for $U(N)$ gauge group (at weak coupling) are described exactly by the Hilbert space of $N$ free fermions in a harmonic oscillator potential \cite{Corley:2001zk,Berenstein:2004kk}. The dynamics is fully solvable 
and all the states can be counted. In the study of the AdS/CFT correspondence, the dual geometries have also been completely classified \cite{Lin:2004nb}: they are described in terms of incompressible droplets on a two plane. This is a semiclassical description of the phase space dynamics of the free fermions, which can also be associated with the study of the integer quantum hall effect on a plane \cite{Berenstein:2004kk} (indeed, the description that bosonizes small edge excitations of the droplets has been known previously \cite{Stone:1990ir}). One can work backwards from supergravity solutions to argue for the free fermion description \cite{Mandal:2005wv,Grant:2005qc}, but the analysis becomes convoluted when the droplets reach the minimal quantum size of $\hbar$. Indeed, there is more than one path to obtain certain solutions, especially the ones that have different topologies and it is not clear how the system takes care of over-counting in the supergravity regime (see for example \cite{Suryanarayana:2004ig}). This is taken care of by the ``matrix model" realization of the dynamics.
That is, the dual holographic CFT explains how to take care of the overcounting. 

What is clear is that the over-counting is in some sense handled by a stringy exclusion principle \cite{Maldacena:1998bw}. This idea was used to argue that BPS states bubble into giant gravitons \cite{McGreevy:2000cw}. In the case of giant gravitons, the stringy exclusion principle is stated by saying that trace modes become dependent for finite size matrices. 
In a sense, this gives a hard bound on the number of modes that are available to us and turns the effective dynamics of the traces  into an interacting theory. This is mainly the statement that there are constraints related to over-counting, so that there is no unconstrained free field theory description of the system. The  map of the states between the AdS and the CFT took much longer to sort out, especially since the discovery of the dual giant gravitons \cite{Grisaru:2000zn,Hashimoto:2000zp}. This was sorted out eventually in \cite{Balasubramanian:2001nh, Corley:2001zk}, and the geometric interpretation came out later \cite{Berenstein:2004kk,Lin:2004nb}. 

There is a set of natural questions that can be asked.

\begin{enumerate}
\item What is the set of states that can be accurately described by (semiclassical) droplet geometries?
\item How is the stringy exclusion principle implemented in detail for these different droplet geometries?
\item Are there topology measuring operators? 
\item Is there a limit of the system where the full dynamics of all the states is an unconstrained free field theory?
\item Are interactions necessary for understanding the topology changes? 
\end{enumerate}

The last question only makes sense if the answer to the fourth question is affirmative. Indeed, all questions become more interesting if the answer to the fourth question is affirmative. This is because in that case one would have a UV complete free theory in the  gravity variables, rather than the $N$ free fermion description.  This would mean that there is no intrinsic stringy exclusion principle in  the dynamics, and such a stringy exclusion principle would only appear effectively on particular solutions of the theory. Moreover, it is not clear that one can obtain a topology change in such a free theory. One is used to thinking about free theories as having a unique classical limit. A topology change would indicate that there is more than one possible classical limit.  The main goal of the paper is to argue that the answer to question 4 is YES, that the answer to questions 3, 5 is NO, and to produce detailed partial answers for questions 1 and 2. In particular, we will construct the multi-droplet dynamics from first principles around a preferred collection of states, in which the details of the stringy exclusion principle can be understood.

\subsection{Review of LLM geometries}

We will start the analysis with a review of the salient features of the LLM solutions in supergravity \cite{Lin:2004nb} that we need. The solutions that preserve half the supersymmetries in ${\cal N}=4 $ SYM preserve an $SO(4)\times SO(4) \times \BR $ bosonic symmetry of $SO(4,2)\times SO(6)$. The extra $\BR$ symmetry is split evenly between the $SO(2,4)$ and the $SO(6)$. A geometry with those symmetries  has the form
\begin{equation}
ds^2 = -\frac{y}{\sqrt{\frac 14- z^2}}(dt+V_i dx^i)^2+\frac{\sqrt{\frac 14- z^2}}{y}(dy^2+dx^i dx^i) +y\left(\sqrt{\frac{\frac 12 -z}{{\frac 12 }+z}}\right) d \Omega_3^2+y\left(\sqrt{\frac{\frac 12 +z}{{\frac 12 }-z}}\right) d\tilde \Omega_3^2
\end{equation}
 where $i=1,2$. The two copies of $d \Omega_3^2$ and $d\tilde \Omega_3^2$ are three-spheres that realize the $SO(4)\times SO(4)$ symmetry.
The metric is completely characterized by $z(y, x_i)$: the vector  $V$ satisfies a differential equation that ties it to $z$.  The function $z$ obeys a linear elliptic sourceless PDE:
\begin{equation}
\partial_i \partial_i z + y\partial_y \left(\frac{\partial_y  z}{y}\right)=0
\end{equation}
and requires a boundary condition at $y=0$. This locus is called the LLM plane. Non-singularity of the ten dimensional metric requires $z= \pm \frac 12$ at this locus (this forces only one of the two spheres to shrink to zero size, while the other stays finite). From here, one can compute
\begin{equation}
z(x_1,x_2, y) = \frac {y^2}{\pi} \int   \frac{z(w_1, w_2,0) \, dw_1 \, dw_2}{[(x_{1}-w_{1})^2+(x_{2}-w_{2})^2+y^2]^2}\label{eq:z}
\end{equation}
Notice that the integral is always convergent if $z(w_1, w_2,0) $ is bounded. This is guaranteed by the non-singularity condition. We can represent the areas of $\pm 1/2$ as a two coloring of the LLM plane. The area of each one of the two colored regions is quantized in fundamental units \cite{Lin:2004nb}.

This is a complete description of all the LLM solutions. We are interested in taking a limit where the areas of the regions with $z=\pm 1/2$ are both infinite. Moreover, we want the edge between the two areas to be compact.  There are two ways to do so. Each of them has their own advantages. 

The first way to do so is to consider a half filled plane, with two regions. Naively, this is the plane wave geometry with an infinite edge. To obtain the compact edge we perform a periodic identification with a translation along the edge. This is done without distorting the  geometry above. The disadvantage is that the asymptotic behavior of the geometry is not the one of $AdS_5\times S^5$ geometry any longer.

The second way to do so is to consider the strict $N\to \infty$ limit of excitations of $AdS_5\times S^5$ with finite energy. Since the edge of the droplet grows in size like $\sqrt N$ in fundamental units of the LLM area quanta, all the features of the solutions get compressed in the radial direction.  To see the topological features, we need to rescale the  coordinates to keep the coordinates of objects on the edge finite, even if the distance is effectively shrinking. This does not affect the topology of the configuration, but it distorts the geometry. We will study both of these. 
The reason is that both of them give rise to the same Hilbert space of states, even if they correspond to very different sets of geometries in ten dimensions. 

\subsection{Periodic LLM solutions}
 
First, we will just study the LLM solutions that are periodic, with an infinite black and white area after the periodic identification.
 
We will be particularly interested in solutions that are independent of one of the LLM plane variables. These are stationary and represent states that do not evolve under Hamiltonian evolution. 
In the semiclassical limit, these are eigenstates of the Hamiltonian.
 
In the geometry represented by \eqref{eq:z}, we can choose that variable  to be $w_2$. Then we have that 
\begin{equation}
z(x_1,x_2, y) = \frac {y^2}2\int  \frac{z(w_1, 0,0) dw_1 }{[(x_1-w_1)^2+y^2]^{3/2}}\label{eq:solpde}
\end{equation}
where $z(w_1,0,0)$ will alternate between values of $-1/2, 1/2$ in regions, and we will also assume that $x_2$ is a periodic coordinate. The master integral we need is
\begin{equation}
\int_a^b\frac {d w_1}{[(x_1-w_1)^2+y^2]^{3/2}}= 
\frac 1{y^2}  \left[\frac{(b-x_1)}{\sqrt{(b-x_1)^2+y^2}}-\frac{(a-x_1)}{\sqrt{(a-x_1)^2+y^2}}\right]
\end{equation}
This will give us the contribution for a region $a$ to $b$ where $z=+\frac 12$ (a $z=-\frac 12$ region will simply have an overall sign difference).  Notice the $1/y^2$ that appears here will cancel the numerator in \eqref{eq:solpde}.
Our vacuum will be the solution with 
\begin{equation}
z_0(w_1, w_2, 0)=\theta(w_1) -\frac 12
\end{equation}
for which
\begin{equation}
z(x_1, x_2, y) = \frac {x_1}{2\sqrt{x_1^2+y^2}}
\end{equation}
The excited solutions will be determined by requiring that the domain where
\begin{equation}
\Delta z(w_1,w_2,0)= z_0(w_1, w_2, 0)- z(w_1,w_2,0)\neq 0
\end{equation}
has compact support (remember we have imposed that $w_2$ is periodic), and that moreover 
\begin{equation}
\int [z_0(w_1,w_2,0)-z(w_1, w_2, 0)]\,dw_1 \, dw_2=0 \label{eq:conserve}
\end{equation}
We can understand the first condition by realizing that 
\begin{equation}
z(x_1,x_2, y) = z_0(x_1,x_2,y) + \sum_{\rm{images}}\frac {y^2}{\pi} \int _{\cal D}  \frac{-\Delta z(w_1, w_2,0) \, dw_1 \, dw_2}{[(x_{1}-w_{1})^2+(x_{2}-w_{2})^2+y^2]^2}\label{eq:solpde2}
\end{equation}
where $\cal D$ is the finite support where the two can differ. To implement the periodicity of $w_2$, we need to sum over images under discrete translation in $x_2$ (as indicated in the sum). The infinite sum goes as $1/x_1^3$ asymptotically at large $x_1$, rather than the naive $1/x_1^4$. This is due to the sum over images, which is clear for translation invariant solutions in equation \eqref{eq:solpde}. Let us explain this fact.

The sum can be performed explicitly, by writing the sum over images in detail
\begin{equation}
z(x_1,x_2, y) = z_0(x_1,x_2,y) + \sum_{n=-\infty}^{\infty}\frac {y^2}{\pi} \int _{\cal D}  \frac{-\Delta z(w_1, w_2,0) \, dw_1 \, dw_2}{[(x_{1}-w_{1})^2+(x_{2}-w_{2}-2\pi n)^2+y^2]^2}
\end{equation}
and where we have taken the period to be $2\pi$ (this is a convenient choice).   
\begin{eqnarray}
&& \sum_{n=-\infty}^{\infty} \frac{y^2}{\pi} \frac{1}{[(x_{1}-w_{1})^2+(x_{2}-w_{2}-2\pi n)^2+y^2]^2}\nonumber\\
& =&i \frac{y^2 \left[ \cot( \varphi) + \cot(-\bar \varphi) \right]}{4(2\pi)((w_{1}-x_{1})^{2}+y^{2})^{3/2}}-\frac{y^2}{16\pi}\frac{\left[  \csc^2(\varphi) + \csc^2(\bar \varphi)\right]}{((x_1-w_1)^2+y^2)}
\end{eqnarray}
where 
\begin{equation}
\varphi=\frac{(w_{2}-x_{2})+i \sqrt{(w_{1}-x_{1})^{2}+y^2}}2
\end{equation}
We would like to see how this expression behaves asymptotically, at large $x_{1}$.  Let's look at the expression one term at a time.  The terms of the form $\csc(a+ib)$ will decay exponentially fast, beating out the polynomial in $x_1$, which multiplies them.  So, these terms can be dropped.  The cotangent pieces, however, can be re-expressed using
\begin{equation}
\cot(a+ib)=\frac{1-i\tan a \tanh b}{\tan a +i\tanh b}
\end{equation}
and again, as $b\to \infty$, we find that $\tanh(b)\to 1$ up to exponentially suppressed terms. We then get that
 \begin{equation}
z(x_1,x_2, y) \to  z_0(x_1,x_2,y) + \int _{\cal D}   \frac{-\Delta z(w_1, w_2,0) y^2 \left( 2 \right)}{4(2\pi)((w_{1}-x_{1})^{2}+y^{2})^{3/2}}dw_1 dw_2
\end{equation}
Notice that this matches \eqref{eq:solpde} when we make $\Delta z(w_1, w_2,0)$ independent of the second variable, and integrate over the range ($2\pi$).

This implies that $z(x_1,x_2, y)$ and $z_0(x_1,x_2,y) $ have the same asymptotics at finite $y$ and $x_1\to \pm \infty$ up to order $1/x_1^3$, while the second condition improves the match to a higher order. Expanding the expression asymptotically in powers of $x_1^{-1}$ we find that
 \begin{equation}
\Delta z(x_1,x_2, y) = -y^2 \int _{\cal D} {\Delta z(w_1, w_2,0) \, dw_1 \, dw_2}\left(\frac {1}{4 \pi x_1^3}+ \frac {3 w_1}{4 \pi x_1^4}+O(x_1^{-5})  \right)\label{eq:multip}
\end{equation}

This shows how the first term vanishes with the condition \eqref{eq:conserve}. This is a coordinate choice: the coordinate $x_1$ is only well defined up to translation invariance, as moving the origin of $x_1$ changes the
domain where $\Delta z \neq 0$.

The next term is the first subleading term of the gravity solution and its coefficient can be interpreted as the energy (see also the similar analysis in \cite{Mosaffa:2006qk})
\begin{equation}
E\propto  \int _{\cal D} {\Delta z(w_1, w_2,0) \,  w_1 \, dw_1 \, dw_2}\label{eq:dropenergy}
\end{equation}
This is positive, because $\Delta z \geq 0$ for $w_1>0$ and $\Delta z \leq 0$ for $w_1<0$. We see that compactness of $\cal D$ gives rise to a finite energy. The area quantization condition forbids us from trying to place smaller droplets very far away without incurring a large energy cost.
In principle, it is possible to produce finite energy configurations if we allow a droplet to have  a thinning finger whose width decreases as it tries  to reach infinity: these will inevitably become smaller than Planck size features in the metric and we should be worried about using those solutions in the semiclassical regime.

This set of solutions is interpreted as a set of droplets for an incompressible fluid on the $x_1,x_2$ plane (we can choose $z(x_1, x_2, 0)=1/2$ as the liquid, and the other region as an absence of liquid). 
The first equation tells us that we have a finite energy solution, and the second one says that
the number of liquid particles is conserved. 
The semiclassical Dirac quantization condition requires that the  areas of the individually connected (compact) regions with $z(w_1,w_2,0)>0$ and $z(w_1,w_2,0)<0$ both be 
quantized. Each different droplet topology corresponds to a different spacetime topology. Our set of semi-classical coherent states will be droplet geometries  that satisfy all of these properties.
In the particle/hole language, we have both an infinite number of particles and vacancies. This tells us that there are no upper bounds on the energy of either a single particle being moved into the vacancy region, or for a vacancy being 
moved into a particle region. These are ``giant gravitons" and ``dual-giant gravitons." Saying that there is no upper bound on any of their energies is roughly stating  that there is no stringy exclusion principle.

Once we have these semiclassical states, we can build a Hilbert space by taking superpositions of these geometries. Notice that coherent states are usually over-complete, so the set of solutions by itself does not tell us how this completion is supposed to work in practice. This is what the dual gauge theory actually accomplishes. Once we resolve this problem of what the correct theory is, we can reconstruct the geometric states and ask questions about quantum gravity. 

\begin{figure}[ht]
\begin{center}
\includegraphics[width=4cm]{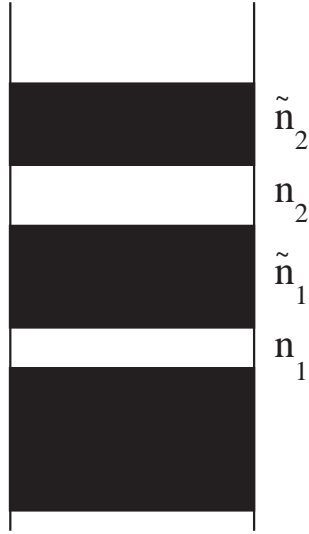}\caption{Periodic LLM solutions are characterized by strips. The quantities $n_1, \tilde n_1, n_2, \tilde n_2\dots$ are quantized and can be taken to be integers.}\label{fig:strips}
\end{center}
\end{figure}

As discussed previously, we are also particularly interested in geometries that have an extra translation symmetry. The translation symmetry comes from the lack of dependence on the periodic variable $x_2$. The black and white pattern is therefore represented by a black and white pattern on a cylinder that is rotationally invariant.
The area of a (finite) droplet is then the size of the periodic variable $2\pi$ times the height of the region. This is quantized. Therefore any such configuration is described by a set of integers: the ordered heights of the regions. These alternate between the two colors. Since the basic configuration has the bottom half of the cylinder filled and the top half empty, the configurations must be asymptotically colored in the same way. This means that there are as many finite white regions as there are finite black regions. This can be represented as in the figure \ref{fig:strips}.

It is worth noticing that no additional information than the integers $n_i, \tilde n_i$ is required. This is because condition \eqref{eq:conserve} gives an equation for the ground state level 
relative to the strip configuration.  This level is obtained by  requiring that the area in black above the `zero level' is equal to the area in white under the 'zero level'. There is only one such level.
The important point for us is that the (translationally invariant) geometry is determined by the collection of integers $n_1,\tilde n_1, \dots$.  These are all unconstrained integers. In that sense, we should think of the geometry as being free from a 'stringy exclusion principle' that would limit the integers somehow.

\subsection{Zooming onto the edge}

\begin{figure}[ht]
\begin{center}
\includegraphics[height= 3.5 cm]{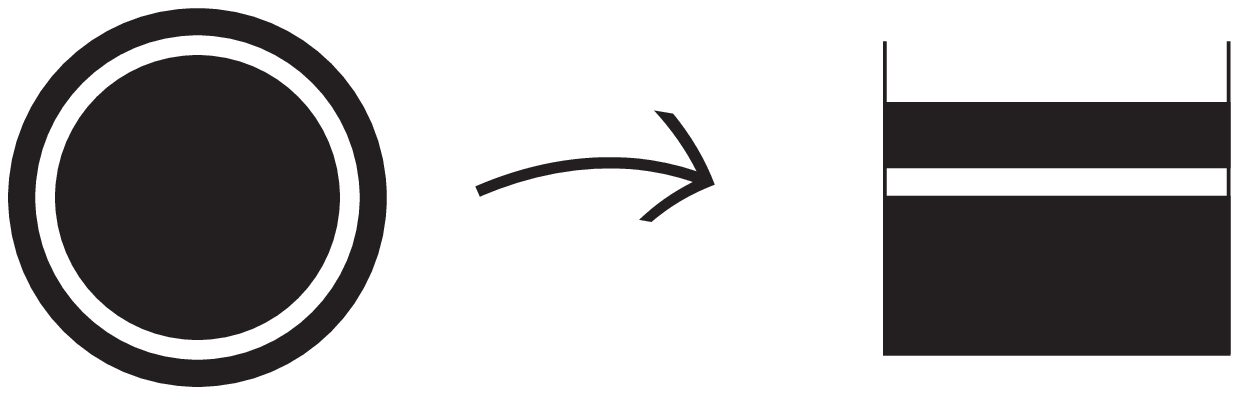}
\end{center}\caption{ Pictorial description of the change of coordinates $(w, \bar w) \to (y,\theta)$. The change of variables is area preserving. }\label{fig:coordch}
\end{figure}

Now, consider the case where we take $N\to \infty$, while keeping the energy fixed, of an LLM geometry that asymptotes to $AdS_5\times S^5$ (in the dual field theory we are considering a state with finite scaling dimension, but $N\to \infty$). We will also require that the geometry has an extra symmetry, this time of rotations around a center of the geometry in polar coordinates. The corresponding solutions are given by concentric rings in black and white. Since the area of the droplet is $N$, when we scale $N\to \infty$ the area becomes infinite. It is also important to understand that the energy of the solution is given by \cite{Lin:2004nb}
\begin{equation}
E= \frac 12 \int_{\cal D} z(w, \bar w,0) \, w \, \bar w \, d^2 \, w-\frac 12\int_{{\cal D}_0} z_0(w,\bar w,0) \, w \, \bar w \, d^2 w
\end{equation}
 where the domain ${\cal D}$ is the region of the droplet covered in black. We would like to measure the energy relative to that of the ground state, which is why we subtract off the integral over the ground state black droplets (we label this ${\cal D}_0$). 
We can check that the full integral is over the region where $\Delta z\neq 0$, like in the previous subsection. The region ${\cal D}_0$ is a circular droplet centered around the origin with radius 
$r\simeq \sqrt N$ in fundamental units. Because the radius of the droplet scales with $N$, the values of $w, \bar w$ where $z$ differs from $z_0$ also scale in the same way. 
We also want to keep the areas of small subdroplets fixed and of order one. To understand these configurations, it becomes convenient to change coordinates to a variation of angular polar coordinates $y= r^2/2-r_0^2/2,\theta$ that are centered near the edge of the droplet, so that $dy\, d\theta \simeq r\, dr\, d\theta= d^2 w$ is the natural area element. This means that we do the change of coordinates in a way that preserves the area measure.  In this way, quantization of the circular area droplets is given immediately by requiring that the black and white rings have quantized $y$ sizes. This change of coordinates was sketched in \cite{Berenstein:2016pcx}, but with most of the details on scaling of coordinates ommitted. 
This is depicted pictorially in the figure \ref{fig:coordch}.

 Because the area of the new region is required to have the same area as the original droplet, one can check readily that 
$\int_{\cal D} z(w, \bar w,0) \, r_0^2 \, d^2 w-\int_{{\cal D}_0} z_0(w,\bar w,0) \, r_0^2 \, d^2 w=0$.

This way we find that the energy is given also by a simple expression
\begin{equation}
E\propto \int_{\cal \tilde D} \Delta z(y, \theta) \, y \, dy \, d\theta \label{eq:height}
\end{equation}
This is very similar to equation \eqref{eq:dropenergy} if we identify the two sets variables $w_1\to y$, $w_2\to \theta$.  The advantage is that now we can take $N\to \infty$, keeping $y$ fixed and the areas fixed. Indeed, we find that the description of the droplets that survive the limit become identical to the pictorial representation in figure \ref{fig:strips}, even though the ten dimensional geometries are very different.

What needs to be understood is that the variable $y$ being of order one implies that $2y \simeq r^2 -r_0^2 = (r_0+\delta r)^2- r_0^2\simeq 1$. This means that $\delta r \simeq 1/\sqrt N$. 
In practice, this means that in order to take the limit we have zoomed onto the edge of the droplet. In the $y$ coordinates taking $N\to \infty$ is trivial, but in the regular polar coordinates $r$, the  shift in the radial coordinate scales as $1/\sqrt N$ relative to the edge. From the point of view of the topology of the configurations, it should not matter how we scale the coordinates. Topology is after all a very coarse 
feature.

 However, from the point of view of geometry (which includes distances), the features that we are trying to zoom in to are typically smaller than the Planck scale. The main reason for taking the limit is that at finite $N$, the negative $y$ coordinate 
has a finite depth of order $N$. When we take $N\to \infty$ we remove this constraint and can work on the cylinder. This means that the numbers $n_i, \tilde n_i$ are unconstrained and there is no stringy exclusion principle. What is important for us is that in this limit the corresponding modes of the supergravity theory become free \cite{Witten:1998zw}. This is because the energy is of order one rather than $N^2$.

It is important to understand that taking this limit at the level of geometries for different topologies is suspect. In a free limit of a quantum field theory we would expect that the classical solutions that survive would be related to coherent states of a field. These would be solutions where the edge of the droplet is deformed, but there is no topology change. It is natural to ask if both of these types of solutions can be thought of as allowed classical limits simultaneously in the supergravity theory or not. This is not resolved directly within the supergravity theory.

 The claim that topology cannot be measured by an operator in 
\cite{Berenstein:2016pcx} depends on this assumption being true. If both types of "classical limits" survive in the quantum theory with different topology, then the fact that coherent states are complete means that states with the new topology (different than the ground state) can be made by superposing coherent state geometries. The coherent states are expected to all have trivial topology.
This means that one can induce a topology change (of different classical limits)  by superposition of solutions in one (trivial) topological class.
The main goal of the paper is to explain how this works in detail in the Hilbert space of states of the free theory that arises from the $N\to \infty$ limit of half BPS states of the ${\cal N}=4$ SYM 
theory.

A particularly important claim of the LLM setup \cite{Lin:2004nb} is that these concentric ring solutions are a complete basis of states for the Hilbert space of BPS states in ${\cal N}=4 $ SYM. That is because these solutions can be put into 1-1 correspondence with Young tableaux. The Young tableaux states are a complete basis of states of the dual ${\cal N}=4$ SYM theory \cite{Corley:2001zk}. 
Such classification of states has been exploited in many setups (see for example \cite{Balasubramanian:2005mg, Balasubramanian:2008da}). It is important to notice that most of these will have topological features on scales that are much shorter than the $AdS$ planck scale. The typical such solution should not be treated as classical objects in these extreme situations.

\section{A Hilbert space from group theory}\label{sec:gt}

As is well understood in the theory of orbifolds, twisted sector states for closed strings are assigned conjugacy classes of the group \cite{Dixon:1985jw,Dixon:1986jc}. Similarly, D-brane charges are associated with representations 
of the group \cite{Douglas:1996sw}. 

In the study of finite groups, one can use character tables to go from conjugacy classes to projectors on representations \cite{FultonHarris}. These two sets are in one to one correspondence with each other, and there is a linear map between the vector spaces that they generate at the group algebra level.
This linear map can be promoted into a change of basis in a Hilbert space, so that one furnishes a model where D-branes can be understood as linear combinations of strings. This is a model where a D-brane is thought of as a soliton for a (closed) string theory: the D-brane state  can be written directly in terms of string states.
In this toy model, we start from nothing but the group, so we might want to think of this procedure as orbifolding the theory of a target space which is a point.
Our goal is to eventually apply this procedure to the symmetric group, where additional structures are present and to connect this information with the study of matrix models.

 The idea is rather simple. Consider a finite group $\Gamma$, with elements $\sigma$, and the usual transform
in the group algebra 
\begin{equation}
P_R = \frac 1{|\Gamma|}\sum_{\sigma\in \Gamma} \chi_R(\sigma) \, \sigma
\end{equation}
where $R$ labels the irreducible representations of $\Gamma$. The object $P_R$ is a projector in the group algebra that projects into the irreducible representation $R$ and $\chi_R(\sigma)$ is the character of $\sigma$ in the irreducible representation $R$. 
Because the characters only depend on the conjugacy class of $\sigma$, we can convert this sum into an equation relating the conjugacy classes and the projectors themselves
\begin{equation}
P_R=\frac 1{|\Gamma|}\sum_{[\sigma]\in \rm{Conj}[\Gamma]} \chi_R([\sigma]) \, d_{\sigma} \, [\sigma]
\end{equation}
where $d_{\sigma}$ is the degeneracy of the class (number of elements in the class), and this defines $[\sigma]$, which is simply the average over the elements of the class. 

Now, we can promote this equation to an equation in a (finite) Hilbert space of states, where we have two basis: one furnished by strings (conjugacy classes), and another one labeled by irreducible representations $R$, which we call the D-branes. Again, these are labels that we add to the two basis to reiterate the classification properties of the corresponding objects in orbifold theories.
We thus write
\begin{equation}
\ket{[R]}=\frac 1{|\Gamma|}\sum_{[\sigma]\in \rm{Conj}[\Gamma]} \chi_R([\sigma]) \, d_{\sigma} \, \ket{[\sigma]}
\end{equation}
and treat both the $[R]$ and the $[\sigma]$ as non-zero elements in a  Hilbert space. One can expect that strings associated with different conjugacy classes are orthogonal to each other (they are different twisted sectors), and that different D-branes are also orthogonal to each other (they have different D-brane charges). Therefore, we require that 
\begin{eqnarray}
\braket{[\sigma]}{[\sigma']}= \delta_{[\sigma],[\sigma']} \, f_{[\sigma]}\\
\braket{[R]}{[R']}= \delta_{[R],[R']} \, g_{[R]}
\end{eqnarray}
 This is a non-trivial set of orthogonality conditions: it is not guaranteed that they can be made compatible. The inverse transformation, relating the conjugacy classes and the representations is 
 \begin{equation}
 |[\sigma]\rangle = \sum_{[R]} \chi_R([\sigma^{-1}]) \, | [R] \rangle \label{eq:inverse}
\end{equation} 
where we use the characters of the conjugacy class of the inverse elements.  Remember that $\chi_R([\sigma^{-1}]) =\chi_R([\sigma])^*$.
Using the conjugacy class of the identity (which has the identity as its only element), we find 
\begin{equation}
\braket{[\1]}{[R]}= \frac{\chi_{R}([\1])}{|\Gamma|} f_{[\1]}
\end{equation}
Now, expressing the vector $\ket{[\1]}$ in terms of the inverse function \eqref{eq:inverse}, we find that 
\begin{equation}
\sum_{[R^{\prime}]} \chi_{R^{\prime}}([\1]) \braket{[R^{\prime}] }{[R]} = \chi_R[\1] g_{[R]}
\end{equation}
Comparing the two expressions, we find that
\begin{equation}
g_{[R]} = \frac 1 {|\Gamma|} f_{[\1]}
\end{equation}
so the norm of the kets associated with the representations is independent of the representation.

Similarly, we can now apply this information to the trivial representation $\bigtriangleup$, which is such that $\chi_{\bigtriangleup}([\sigma])=1$ for all $[\sigma]$. We find that
\begin{equation}
\braket{\bigtriangleup}{[\sigma]}= \sum_{[R]} \chi_R([\sigma])  \braket{\bigtriangleup}{[R] }= \chi_{\bigtriangleup}([\sigma]) \braket{\bigtriangleup}{\bigtriangleup}= g_{\bigtriangleup}
\end{equation} 
Writing $\ket{\bigtriangleup}$ in terms of the class vectors, we find
\begin{equation}
\braket{\bigtriangleup}{[\sigma]}= \frac {d_\sigma}{|\Gamma|}\chi_{\bigtriangleup}{[\sigma]}^* f_{[\sigma]} = \frac{d_{\sigma}}{|\Gamma |} f_{[\sigma]}
\end{equation}
Putting these two together we get that
\begin{equation}
f_{[\sigma]} =g_{\bigtriangleup} \frac {|\Gamma |}{d_\sigma}\label{eq:trnorm}
\end{equation}
So, we find that the representations are all normalized to the same value (which we can choose to be equal to one) and the class function kets are proportional to the inverse of the degeneracy of the class.

Since we only used the kets $\ket{\bigtriangleup}$ and $\ket{\1}$ to normalize everything, we need to check consistency. This is straightforward. The orthogonality of representation kets becomes equivalent to the orthogonality 
of the rows of the character table. The orthogonality of the conjugacy classes ends up being related to the orthogonality of the columns of the character table (see \cite{FultonHarris}, pp 17, exercise 2.21).

\subsection{Fock space from the symmetric group}

At this point, this can be thought of as a curiosity. The reason is that as written, this exercise should be thought of as a first quantized setup. Single string states can be reshuffled via a ``Fourier transform" into  single D-brane states. We have made no mention of multi-string states.

This all changes when we consider the symmetric group $S_n$. As argued in \cite{Dijkgraaf:1996xw,Dijkgraaf:1997vv}, the conjugacy classes of the symmetric group should be associated with multi-strings, rather than single strings. Similarly, one can imagine that the representation theory left hand side should be generically associated with multiple D-branes, rather than a single D-brane. In essence, when we consider the symmetric group, the theory should be for all practical purposes second-quantized. This becomes precise when we consider the more general object generated by the symmetric group $S_n$ for all values of $n$
\begin{equation}
{\cal H} =\oplus_{n\geq 0} {\cal H}_{S_n}
\end{equation}
with the Hilbert space constructed as above. We can now also add (multi-) strings to a configuration. Consider a conjugacy class in $S_n$ and a group element $g_1\in S_n$ that represents it. Similarly, consider another conjugacy class and a group element  $g_2\in S_m$ that represents it. There is a natural embedding of the product of symmetric groups into a larger symmetric group
$\mu: S_n \times S_m \to S_{n+m}$. Here the $S_{n+m}$ acts as permutations of a set of $n+m$ elements and the $S_n$ acts only on the first $n$ elements, while the $S_m$ acts only on the last $m$ elements. This gives a natural 
element of $S_{n+m}$ for the two group elements, namely $\mu(g_1\times g_2)\simeq \mu(g_1\times 1) \circ\mu(1\times g_2)$. Now, it is important to notice that this defines a unique conjugacy class in $S_{n+m}$ which is irrespective 
of the representatives $g_1, g_2$ that were chosen. This is done as follows.

A group element of $S_n$ can be written in a cycle presentation of permutations of the set $\{ 1, \dots n \}$.  As is well known, the conjugacy classes are in one to one correspondence with the lengths of the  cycle decomposition. A conjugacy class thus gives a partition of $n = \sum_s w_s s$ where we have $w_s$ cycles of length $s$. We do this for $g_1$, so that
$[g_1]\simeq \prod_s [s]^{w_s}$, where we pick the cycles of length $[s]$ as generators, and similarly for $[g_2]\simeq \prod_s [s]^{w'_s}$. With this we find that 
\begin{equation}
[g_1]\otimes [g_2] \simeq \prod_s [s]^{w_s+w'_s} \simeq [g_1\otimes g_2] \simeq [g_2]\otimes[g_1]
\end{equation}
which shows that the cycle decomposition is irrespective of the elements of the class that we pick. 
We also find that the product of conjugacy classes is commutative, and the set is generated by the primitive cycles of length $[s]$. We will call $[s]:= t_s$ so that the set of states can be thought of as (a particular completion of) the  set of polynomials in an infinite set of variables $\{t_s\}$. The decomposition of the Hilbert space into the different ${\cal H}_{S_n}$ can be thought of as  being graded by $n$, and the grading is additive on the product we defined. The product operation we defined is just the product of polynomials when extended linearly. The degree of $[s]$ is $s$. We will call this function for a monomial the energy of the state.
A conjugacy class associated to the monomial $t_1^{w_1} \dots t_k^{w_k}$ will be said to have $w_1$ particles of energy $1$, $w_2$ particles of energy $2$, etc.

The most important question for us is to address is how we are going to relate the inner products of the different ${\cal H}_{S_n}$ to each other. 
There are two routes we can take. The first route is to declare that the irreducible representations of all the symmetric groups have norm equal to one. This can be twisted by the energy, so an irreducible of energy $n$ has norm $|T|^n\simeq \exp( E \log(|T|))$. A second route  is to assume that the norm of a two particle state, where the two particles have different energy, is the product of the norms of the corresponding single particle states. This is a factorization condition. This will be shown to be equivalent to the first route after a computation.

To do a computation, we need to find both the dimension of the symmetric group $|S_n|=n!$ and also the number of elements in a conjugacy class $d_\sigma$ (remember we have  labeled these as $\prod_s [s]^{w_s}\simeq t_1^{w_1} \dots t_k^{w_k}$ with $\sum_k w_k k = n$). The number of elements of a conjugacy class of the symmetric group is known to be given by
\begin{equation}
d_\sigma= \frac{n!}{ \prod_k w_k ! k^ {w_k}}=\frac{|\Gamma|}{ \prod_k w_k ! k^ {w_k}} \label{eq:degen}
\end{equation} 
from equation \eqref{eq:trnorm}, we find that 
\begin{equation}
||t_1^{w_1} \dots t_k^{w_k}||^2=\,_n\braket{[\sigma]}{[\sigma]}_n =\,_n\braket{\bigtriangleup}{\bigtriangleup}_n \frac {|\Gamma|}{d_\sigma}= \,_n \braket{\bigtriangleup}{\bigtriangleup}_n \prod_k w_k ! k^ {w_k}
\end{equation}
So, if we choose $\,_n \braket{\bigtriangleup}{\bigtriangleup}_n= |T|^n$, the right hand side becomes equivalent to a norm on a bosonic Fock space where to each $t_k$ we assign a raising operator of norm squared $k |T|^k $. That is, we associate a raising and a lowering operator pair $(a_k^\dagger, a_k)$, with commutation relations
\begin{equation}
[a_k, a^\dagger_k] = k |T| ^k \label{eq:oscnorm}
\end{equation}
with all others vanishing.  Obviously, the simplest choice is to take $|T|=1$. The raising operator acts by multiplication by $t_k$, namely $a^\dagger_k\rightarrow t_k$, and the adjoint acts by $k\partial_{t_k}$. The inner product can be computed using
\begin{equation}
\braket{g}{f}= \int \prod_k dt_k d\bar t_k \exp\left(-\sum_k \frac{t_k \bar t_k}{k}\right) \bar g(\bar t) f(t)\label{eq:gaussmeasure}
\end{equation}
where the normalization factor of the measure is such that 
\begin{equation}
\braket  1 1= 1
\end{equation}

If we take the factorization condition instead, we find that 
\begin{equation}
|t_a t_b| =  a b \,_a \braket{\bigtriangleup}{\bigtriangleup}_a \,_b \braket{\bigtriangleup}{\bigtriangleup}_b = a b \,_{a+b} \braket{\bigtriangleup}{\bigtriangleup}_{a+b}
\end{equation}
for all $a,b$ with $a+b$ constant. This would suggest that 
\begin{equation}
\,_a \braket{\bigtriangleup}{\bigtriangleup}_a \,_b \braket{\bigtriangleup}{\bigtriangleup}_b= \,_{a+b} \braket{\bigtriangleup}{\bigtriangleup}_{a+b}
\end{equation}
but this only seems to work if $a\neq b$ and  $a,b\neq 0$. 

With this, one can show that it works for all $a,b$. Consider $\,_2 \braket{\bigtriangleup}{\bigtriangleup}_2$.  By the naive factorization condition, it is independent of $\,_1 \braket{\bigtriangleup}{\bigtriangleup}_1$.
From here, we can form $\,_3 \braket{\bigtriangleup}{\bigtriangleup}_3= \,_2 \braket{\bigtriangleup}{\bigtriangleup}_2 \,_1 \braket{\bigtriangleup}{\bigtriangleup}_1$ uniquely, and similarly $\,_4 \braket{\bigtriangleup}{\bigtriangleup}_4= \,_3 \braket{\bigtriangleup}{\bigtriangleup}_3 \,_1 \braket{\bigtriangleup}{\bigtriangleup}_1=  \,_2 \braket{\bigtriangleup}{\bigtriangleup}_2 \,_1 \braket{\bigtriangleup}{\bigtriangleup}_1^2$.
When we get to $\,_5 \braket{\bigtriangleup}{\bigtriangleup}_5$, there is a consistency condition
\begin{eqnarray}
\,_5 \braket{\bigtriangleup}{\bigtriangleup}_5&=& \,_3 \braket{\bigtriangleup}{\bigtriangleup}_3 \,_2 \braket{\bigtriangleup}{\bigtriangleup}_2=  \,_4 \braket{\bigtriangleup}{\bigtriangleup}_4 \,_1 \braket{\bigtriangleup}{\bigtriangleup}_1\\
&=& \,_2 \braket{\bigtriangleup}{\bigtriangleup}_2^2  \,_1 \braket{\bigtriangleup}{\bigtriangleup}_1= \,_2 \braket{\bigtriangleup}{\bigtriangleup}_2 \,_1 \braket{\bigtriangleup}{\bigtriangleup}_1^3
\end{eqnarray}
Clearing out the common terms, we find that $ \,_2 \braket{\bigtriangleup}{\bigtriangleup}_2=  \,_1 \braket{\bigtriangleup}{\bigtriangleup}_1^2$, and one can then easily show that  we get $\,_n \braket{\bigtriangleup}{\bigtriangleup}_n = \,_1 \braket{\bigtriangleup}{\bigtriangleup}_1^n$ for all other $n$.

\subsection{Physical Interpretations}

The Hilbert space we have constructed is the Hilbert space of a chiral boson in one dimension. We have a single oscillator of energy $k$ and left moving momentum $k$ for each $k$. This is natural considering that in 
the computation of elliptic genera one builds an extra chiral circle \cite{Dijkgraaf:1996xw}. We have the momentum modes of the chiral boson field theory, but we still need to argue that there is preferred notion of a local field that also arises from this construction. This will be taken up in the next section.

One should also remember that there is a straightforward relation between a chiral boson field theory and edge states in a quantum hall droplet
\cite{Stone:1990ir}. The relation uses the theory of symmetric polynomials in many variables to go from traces to Schur polynomials. The Schur functions are associated with representations of the symmetric group $S_k$ (described by Young tableaux) , and one can directly write these in terms of free fermions. The same combinatorics appears in the study of half BPS states in ${\cal N}=4 $ SYM \cite{Corley:2001zk} and the representation of the physics in terms of droplets and fermions was explored in detail in \cite{Berenstein:2004kk}. Surprisingly, the description of half BPS geometries in the gravity dual of ${\cal N}=4$ SYM is also in terms of droplets: a droplet configuration with specified shape corresponds to a particular solution of the supergravity equations of motion \cite{Lin:2004nb}, as we described in the pervious section. It has also been argued recently that similar physics might control a wide class of AdS/CFT dual configurations when one studies elements of the chiral ring that are extremal \cite{Berenstein:2015ooa}. It was this particular observation that led us to try to understand the problem using group theory of the symmetric group without any particular matrix model in mind.
The map that realizes this correspondence assigns
\begin{equation}
\Tr(\tilde Z^\ell) \simeq t_\ell
\end{equation}
where $\tilde Z$ can be either an elementary  matrix valued field (like in ${\cal N}=4$ SYM), or a composite matrix field (this would be common in toric field theories or  some simple orbifolds). The main reason that this works is that to any conjugacy class of the symmetric group one naturally associates a multi-trace object constructed from the permutation itself. This was critical in the computations of \cite{Corley:2001zk} that show that Schur functions associated with different Young tableaux are orthogonal. Here we have reversed the arguments: assuming orthogonality of Young tableaux and conjugacy class states leads to a unique consistency condition that gives precisely a free chiral boson.
Thus, in this setup, we do not have any non-trivial three point functions: we are strictly in the $N=\infty$ limit. The free fermions that realize the correspondence for the free field that we have here are associated with a system of free fermions for a Cuntz oscillator. The Cuntz oscillator algebra is defined by
\begin{equation}
\beta^\dagger \ket n = \ket {n+1}
\end{equation}
usually with $\beta\ket 0=0$. For us, we have an infinite sea of fermions, so all we require is that the set of $\ket n$ is labeled by integers (both positive and negative). We set by convention the Fermi-sea energy at $n=0$.
The Cuntz oscillators appear repeatedly also in the study of open spin chain dual states to open strings in the AdS/CFT correspondence \cite{Berenstein:2005fa} and their coherent states are especially important to describe the ground state energies of the corresponding open spin chains 
\cite{Berenstein:2013eya,Berenstein:2014isa}.
What we have described here  corresponds to the strict $N=\infty$ limit of the corresponding matrix models.

In a certain sense, the picture we have been advocating above is building  a bridge between matrix string theory and an abstract version of a (holomorphic) matrix model, in a vein similar to \cite{Verlinde:2002ig}, but with very few states to consider. The picture we have is greatly inspired by the observation that requiring orthogonality of the Young tableaux states and a large $N$ counting for correlators, at leading order in $N$ produced a result where the norm of Young tableaux states is independent of the shape of the tableaux and only depends on the number of boxes \cite{Berenstein:2015ooa}. Here we have even removed the large $N$ counting hypothesis and replaced it by the weaker orthogonality of multi-string states. The fact that the harmonic  oscillators that are constructed in this fashion have the correct statistics to correctly describe quantum fields in one dimension is derived from the compatibility of the two basis of states and the Fourier transform that relates them to each other.

\section{D-brane creation operators and constructing local fields }\label{sec:LF}

\subsection{The D-brane}

It is instructive now to consider the simplest (trivial) representation of the symmetric group, which we have labeled ${[\bigtriangleup]_n}$ in the previous section. This is associated with a Young tableaux with $n$ boxes and only one row. Using our polynomial formulation, we have that
\begin{equation}
\ket{\bigtriangleup}_n= \sum_{ \overrightarrow{\omega} \, \in \, p(n)} \prod_k \frac 1 {k^{w_k} w_k!} (t_k)^{w_k}   \label{eq:trivialrep}
\end{equation}
where $p(n) := \{ \, \overrightarrow{\omega} \, | \sum_k k \omega_k =n\}$ are the partitions of $n$ and we have used the fact that all characters are one for the trivial representation.  We also have included the explicit value of the degeneracy of corresponding conjugacy classes in the sum, extracted from equation \eqref{eq:degen}.  Notice that this is an equation relating states.  In the raising/lowering operator notation, this state would have been created by
\begin{equation}
\ket{\bigtriangleup}_n= \sum_{ \overrightarrow{\omega} \, \in \, p(n)} \prod_k \frac 1 {k^{w_k} w_k!} (a^{\dagger}_k)^{w_k} \ket 0
\end{equation}
which would be equivalent.  From here we find that
\begin{equation}
\Lambda^n \ket{\bigtriangleup}_n= \sum_{ \overrightarrow{\omega} \, \in \, p(n)}  \prod_k\left (\frac {(\Lambda^{k}) }{k} t_k \right)^{w_k} \frac 1{ w_k!} 
\end{equation}
We notice that apart from the constraint $\sum_k k w_k = n$, the right hand side represents a series expression for an  exponential function. It is convenient to sum over $n$ and consider a generating series for these representations, so that we can write
\begin{equation}
\ket{\bigtriangleup; \Lambda}=\sum_{n=0}^{\infty} \Lambda^n \ket{\bigtriangleup}_n= \sum_{n=0}^{\infty} \sum_{ \overrightarrow{\omega} \, \in \, p(n)} \prod_k\left (\frac {(\Lambda^{k}) }{k} t_k \right)^{w_k} \frac 1{ w_k!}  =  \exp\left(\sum_k \Lambda^k\frac {t_k}{k}\right)   \label{eq:defexp}
\end{equation}
The trivial representations of each $S_n$ correspond to the totally symmetric representations of the group $U(N)$ with $n$ boxes (as shown below), and have an interpretation as a single dual giant graviton \cite{Corley:2001zk}.
That is, it has an interpretation as a single D-brane in an $AdS_5\times S^5$ geometry.

\begin{equation}
\ket{\Delta}_n=\begin{ytableau}
\ & &\none[\dots]&n
\end{ytableau}
\end{equation}

Here, we find that when we move away from thinking of $\Lambda$ as  formal parameter, and rather think of it as an actual c-number, the right hand side can be interpreted as a coherent state of the harmonic oscillators represented by $t_k$. We will now push the idea  that a special generating series of interesting objects should be more than a formal expression and actually have physical meaning. The reason this is a coherent state is that it is an exponential of a linear combination of raising operators. We still need to find the range of $\Lambda$ that is appropriate for this expression.

We can consider the norm of the state
\begin{equation}
\braket{\bigtriangleup; \Lambda}{\bigtriangleup; \Lambda}= \sum_n (\bar \Lambda \Lambda)^n = \frac 1{1-\bar \Lambda \Lambda} \label{eq:norm1}
\end{equation}
and we see that it is convergent for $|\Lambda|<1$. This can be similarly obtained from the exponential and the gaussian measure \eqref{eq:gaussmeasure}. The proof is instructive. Consider 
\begin{eqnarray}
\braket{\bigtriangleup; \Lambda}{\bigtriangleup; \Lambda}&=&\int \prod_k dt_k \bar dt_k\exp\left(-\sum_k \frac{t_k \bar t_k}{k}\right) \exp\left(\sum_k \bar \Lambda^k\frac {\bar t_k}{k}\right)\exp\left(\sum_k \Lambda^k\frac {t_k}{k}\right)\\
&=& \int \prod_k dt_k \bar dt_k\exp\left(-\sum_k \frac{(t_k-\bar \Lambda^k) (\bar t_k-\Lambda^k)}{k}\right) \exp\left( \sum_k  \frac{ \Lambda^k\bar \Lambda^k}{k}\right)
\end{eqnarray}
where to arrive at the second line we completed the square. Shifting the integration variables, we find that
\begin{equation}
\braket{\bigtriangleup; \Lambda}{\bigtriangleup; \Lambda}=\exp\left( \sum_k  \frac{ \Lambda^k\bar \Lambda^k}{k}\right)= \exp( -\log(1-\Lambda \bar \Lambda))= \frac 1{1-\Lambda \bar \Lambda} \label{eq:norm2}
\end{equation}
where we have  recognized the Taylor series for $\log(1-x)$ in the exponential. If we didn't already know that the $\ket{\bigtriangleup}_n$ were orthonormal, the coefficients in the Taylor series that appear in \eqref{eq:norm1} after expanding in \eqref{eq:norm2} would have shown that.

The fact that this generating series produces a coherent state for the oscillators is important in more than one way. First, it shows that the D-brane can be thought of as a ``soliton" of the free field theory: a non-dissipating solution of the classical equations of motion. This is how, in the weak string limit, a D-brane can be thought of as a localized classical source for string fields, where away from the D-brane location one has a solution of the classical equations of motion. This is usually encoded in how different boundary states overlap by considering how closed strings propagate from one boundary state to the other \cite{Polchinski:1995mt}. Another example where D-branes decay into string fields can be found in \cite{Sen:2002in}.

Since this D-brane state is a coherent state, it is an eigenstate of the lowering operators represented by $(t_k)^\dagger\simeq k\partial_{t_k}$. We find that for these states
\begin{equation}
\vev{a_k}_{\Delta; \Lambda} = \vev{(t_k)^\dagger}_{\Delta; \Lambda}=\Lambda^{k},\quad \vev{a^{\dagger}_k}_{\Delta; \Lambda} = \vev{t_k}_{\Delta; \Lambda}=\bar\Lambda^{k} \label{eq:important_vev}
\end{equation}
This behavior is expected from the collective coordinate treatment in setups at finite $N$ \cite{Berenstein:2013md,Berenstein:2015ooa}. In these other approaches, the coherent states in question are described in terms of Slater determinants of coherent states for generalized oscillator algebras.

The states we have found are also of minimal uncertainty for all the oscillators. This will become important when we try to describe other classical limits later in the paper.

The next thing that is interesting to compute is the average energy per oscillator in each one of these states. This is captured by
\begin{equation}
\vev{E_k}_{\Delta; \Lambda}= \vev{k t_k\partial_{t_k} }_{\Delta; \Lambda} = (\Lambda \bar \Lambda)^k 
\end{equation} 
so that the expectation value of the energy is
\begin{equation}
\vev{E}_{\Delta; \Lambda} = \sum_{k>0} (\Lambda \bar \Lambda)^k = \frac {\Lambda\bar \Lambda} {1-\Lambda \bar \Lambda}
\end{equation}
the energy carried by the state is large in the limit where $|\Lambda|\to 1$, but in this limit the state becomes non-normalizable. It makes sense to consider states near this limit, where $\Lambda\simeq (1-\epsilon)^{1/2} \exp({i \gamma})$ with $\epsilon$ infinitesimally small and acting as a cutoff. In that case, the energy per oscillator degree of freedom goes to one, but  this means that each oscillator has on average low occupation number 
\begin{equation}
\vev{ \hat N_k}_{\Delta; \Lambda \rightarrow \exp({i \gamma})}= \vev{E_k}_{\Delta; \Lambda \rightarrow \exp({i \gamma})}/k = 1 /k. \label{eq:avoccnumb}\end{equation} The excitations are then still a coherent state for all oscillators, and if we cut off the degrees of freedom in the UV, we find a finite energy lump determined by the cutoff.  The failure of the state to be normalizable is due to the infinitude of modes that can be excited, not to any one oscillator mode going bad on its own. Also, the amplitudes for the different modes are phase correlated. This is important. The reason is that we want to build a field out of the oscillators $a_k, a^\dagger_k$.
The geometry where the fields live should be on a circle (we have argued that we have the degrees of freedom of a chiral boson in the previous section).

\subsection{Field of the brane}

The finiteness properties and phase correlations we have found suggest that we can think of the field generated by the D-brane state as being a classical profile everywhere except at the position of the brane itself. We will use this intuition to argue that there is a preferred linear combination of the oscillators that gives nice properties for the field profile in the limit we want. We will posit that the field operator take the following hermitian combination as an ansatz
\begin{equation}
\hat\phi(\theta) = \sum_{k > 0}  f_k[ a_k \exp(- i k \theta) + a_k ^\dagger \exp(i k\theta) ]\label{eq:fieldcoeff}
\end{equation}
where $f_k$ is a set of positive numbers. Although we could have added phases to $f_k$, the phase correlation of the modes suggest that we set all the phases equal to each other.
Scale invariance of the chiral boson suggests that we take $f_k \simeq 1/|k|^\alpha$ for some exponent $\alpha$ which is yet to be determined. Replacing the expectation values from
\eqref{eq:important_vev} and taking the limit, we find that the profile associated with the D-brane is given by
\begin{equation}
\phi(\theta) =\vev{\hat\phi(\theta)}_{\Delta; \Lambda = \exp (i \gamma)} = \sum_{k> 0} f_k [\exp( i k (\gamma-\theta))+\exp( -i k (\gamma-\theta)) ]
\end{equation}
The phases give (maximal) constructive interference at $\theta= \gamma$, which we will call the position of the D-brane. This is why it is important to choose all phases as we did: it produces the maximum possible 
constructive interference of the profile at a  point. This is tantamount to saying that we have localized the peak as much as possible.
 Everywhere else, the phase sums can cancel enough that the result is finite. Taking  $ f_k = 1/|k|^\alpha$, the result can be written in terms of 
Polylog functions.
We will now argue that if we instead choose $f_k= 1$, the answer becomes extremely simple, and we should therefore choose this value. This choice will have preferable geometric consequences. We find
 \begin{equation}
\phi(\theta) =\vev{\hat\phi(\theta)}_{\Delta; \Lambda = \exp (i \gamma)} = \sum_{k\neq 0} \exp(i k (\gamma-\theta))= 2\pi \delta(\gamma-\theta)-1 \label{eq:geomecond}
\end{equation}
That is,  the field away from the position of the brane becomes constant and has an exactly flat shape in the tail of the D-brane. We will promote this property to the reason why we make this choice for the field. 
This is a geometric condition.
Because the field does not have a mode at $k=0$, it is required that
\begin{equation}
\int_0^{2\pi} d\theta \, \phi(\theta)=0
\end{equation}
which is verified readily. This is actually true for any choice of the $f_k$, subject to convergence at  $|\Lambda|<1$. 
For the particular choice we made, we have a quantization condition on the area under the $\delta$-function distribution, which is $2\pi$. 
The energy can also be written simply in terms of $\phi(\theta)$. More precisely
\begin{equation}
E[\phi] = \left\langle \sum_k  a_k^\dagger a_k \right\rangle = \left\langle \frac 1{2\pi} \int d \theta \left(  \frac 12 : \hat \phi(\theta)^2 : \right) \right\rangle = \frac 1{2\pi} \int d \theta  \left(\frac 12 \phi(\theta)^2\right)\label{eq:locality1}
\end{equation}
where we use the normalization of the modes in \eqref{eq:oscnorm} with $|T|=1$.
This shows that the choice for the field coefficients in equation \eqref{eq:fieldcoeff} is also determined by being able to write a local expression for the energy: a single integral of the field and a finite number of its derivatives. The rightmost term in \eqref{eq:locality1} is the classical contribution to the energy for a smooth $\phi(\theta)$.

There is a second natural choice for a field. This is the field that is obtained by considering $|\Lambda|=1$ and looking for the combination of modes that appears in the exponent of \eqref{eq:defexp}. The idea is that when we declared the generating function \eqref{eq:defexp} to be singled out by our representation basis and the convergence properties, the limit defined a preferred combination of modes.

 It is convenient to call this field $\chi$ and define it to be
\begin{equation}
\hat{\chi}(\theta) = \sum_{k>0} \frac{1}{ i k} \exp(i k \theta) \, a_k^\dagger + c.c
\end{equation}
The factor of $i$ in the denominator is a convention we choose, which gives
\begin{equation}
\ket{\bigtriangleup; \Lambda\to \exp(i \gamma)} = \, :\exp(i \chi(\gamma) ): \ket 0
\end{equation}
One easily finds that the field $\chi$ and the field $\phi$ are related by
\begin{equation}
\partial_\theta \chi(\theta)= \phi(\theta)
\end{equation}
so that locality in the sense of $\chi$ (in terms of the smooth variable $\theta$) ends up being equivalent to locality in the sense of $\phi$. Indeed, the field $\chi$ is what we would usually call the free boson and the field $\phi$ is the associated current. The local energy is the standard stress tensor for the chiral boson. This matches the derivations in \cite{Stone:1990ir} very well.

The field $:\exp(i \chi(\theta)) :  $ is usually thought of as a fermionic field written in the bosonized language (see \cite{Polchinski:1998rr}, pp 11, eq. 10.3.10).

\subsection{The anti-brane and its field}

The other natural representation of the symmetric group is the alternating  representation. This representation is also one dimensional, and the character counts how many transpositions (modulo 2) are in a group element.
For a cycle $[s]$, the sign assigned to it is
\begin{equation}
sign[s] = (-1)^{s-1}
\end{equation}
This is multiplicative, meaning that $sign[a \circ b]=sign[a] sign[b]$ . When we consider the equivalent of equation \eqref{eq:trivialrep}, we get that the sum over characters is 
\begin{equation}
\ket{\bigtriangledown}_n= \frac{1}{|\Gamma|} \sum_{[\sigma] \in S_n} \chi_{\bigtriangledown}([\sigma]) d_{\sigma} | [\sigma]\rangle = \left| \sum_{\overrightarrow{\omega} \, \in \, p(n)} \prod_k \frac 1 {k^{w_k} w_k!}( (-1)^{k-1} t_k)^{w_k} \right\rangle \label{eq:trivialrep}
\end{equation}
where we used the multiplicative rule on the right hand side. We have labeled the states with $\bigtriangledown$ instead of $\bigtriangleup$. In the language of Young diagrams, this representation corresponds to a single column with $n$ boxes. 
\begin{equation}
\ket{\bigtriangledown}_n=\begin{ytableau}
*(white)\\
*(white)\\
\vdots \\
*(white)
\end{ytableau}
\end{equation}
This also corresponds to a totally antisymmetric representation of $U(N)$ with $n$ indices. These states correspond to giant graviton states \cite{Balasubramanian:2001nh} in the AdS/CFT correspondence.  We now want to do a similar generating function to the one in \eqref{eq:defexp} for these states. Consider the following
\begin{eqnarray}
\ket{\bigtriangledown, -\Omega}&=& \sum_n \Omega^n \ket{\bigtriangledown}_n  \\
&=&  \sum_n  \sum_{\overrightarrow{\omega} \, \in \, p(n)} \Omega^n \prod_k \frac 1 {k^{w_k} w_k!}( (-1)^{k-1} t_k)^{w_k}  \\
&=&  \sum_n  \sum_{\overrightarrow{\omega} \, \in \, p(n)}  \prod_k \frac 1 { w_k!}\left(\frac{\Omega^k }{k} (-1)^{k-1} t_k\right)^{w_k}  \\
&=&  \sum_n  \sum_{\overrightarrow{\omega} \, \in \, p(n)}  \prod_k  \frac 1 { w_k!}\left(- \frac{(-\Omega)^k }{k}  t_k\right)^{w_k}  \\
&=&   \exp\left( -\sum_k \frac{(-\Omega)^k}{k} t_k\right) 
\end{eqnarray}
where we have made a sign convention choice in how we wrote $\Omega$ in the generating series, versus how we wrote it in the previous state. We find that with this convention
\begin{equation}
\ket{\bigtriangledown, \Lambda}=  \exp\left(-\sum_k \Lambda^k\frac {t_k}{k}\right) 
\end{equation}
so that in the same limit as before we have that 
\begin{equation}
\ket{\bigtriangledown; \Lambda\to \exp(i \gamma)} = :\exp(-i \chi(\gamma) ): \ket 0
\end{equation}
which is the other fermion field. The natural notion of locality derived from the states $\ket{\bigtriangleup; \Lambda}$ and $\ket{\bigtriangledown; \Lambda}$ are the same. 
 It is now trivial to show that for these new solutions
\begin{equation}
\vev{\hat\phi(\theta)}_{\bigtriangledown; \Lambda=\exp (i \gamma)}= -\sum_{k\neq 0} \exp(i k (\gamma-\theta))= -2\pi \delta(\gamma-\theta)+1 \label{eq:delta2}
\end{equation}
There is also a symmetry that sends $\bigtriangleup\leftrightarrow \bigtriangledown$. This is `particle-hole' duality and is implemented by $\chi(\theta)\to -\chi(\theta)$, and $\phi(\theta)\to -\phi(\theta)$. In some finite matrix models built from microscopic fermionic degrees of freedom, this can be implemented exactly \cite{Berenstein:2004hw}. 

We will call the two families of operators \begin{equation}
B_{\pm, \Lambda} = \exp\left(\pm\sum_k \Lambda^k\frac {t_k}{k}\right)
\end{equation}  acting on any state  the D-brane creation operators. That is, up to including ${:\exp(\pm i \chi(\gamma)):}$, which is non-normalizable.

\subsection{Multiple brane states}

Consider acting with a D-brane creation operator on a state that already has one D-brane. At the level of the oscillator representation of states, this is straightforward. We can write
\begin{equation}
B_{+,\Lambda_1} B_{+ ,\Lambda_2} \ket 0=  \exp\left(\sum_k \left(\Lambda_1^k+\Lambda_2^k\right) \frac {t_k}{k}\right) 
\end{equation}
We easily find that this is also a classical state, which results from the superposition of the two profiles of the individual D-branes. The classical field is characterized by
\begin{equation}
 \vev{(t_k^\dagger)}_{\Delta,\Lambda_1;\Delta,\Lambda_2}= \Lambda_1^k+\Lambda_2^k\quad \vev{t_k }_{\Delta,\Lambda_1;\Delta,\Lambda_2}= \bar\Lambda_1^k+\bar\Lambda_2^k
\end{equation}
and
\begin{equation}
\vev{\hat\phi(\theta)}_{\Delta,\Lambda_1;\Delta,\Lambda_2}=2 \Re e \left( \frac {\Lambda_1e^{-i \theta}}{1-\Lambda_1e^{-i \theta}}+ \frac {\Lambda_2e^{-i \theta}}{1-\Lambda_2e^{-i \theta}}\right)
\end{equation}
Similarly, we can write
\begin{equation}
B_{+,\Lambda_1} B_{-,\Lambda_2} \ket 0= \exp\left(\sum_k \left(\Lambda_1^k-\Lambda_2^k\right) \frac {t_k}{k}\right)
\end{equation}
\begin{equation}
\vev{\hat\phi(\theta)}_{\Delta,\Lambda_1;\bigtriangledown,\Lambda_2}=2 \Re e \left( \frac {\Lambda_1e^{-i \theta}}{1-\Lambda_1e^{-i \theta}}- \frac {\Lambda_2e^{-i \theta}}{1-\Lambda_2e^{-i \theta}}\right)
\end{equation}
Notice that if we take $\Lambda_1=\Lambda_2$ in the second profile, the fields cancel. This tells us that the two types of D-brane states annihilate one another into the vacuum \footnote{The more precise version of this cancellation is that the OPE expansion of the two fermion fields contains the identity}. 

We will now use the D-brane basis (the basis using irreducible representations of $S_n$), to better understand how these generating functions behave.  We will use Young tableaux to compute the multi-brane states.  The correct multiplication table is governed by the Littlewood-Richardson coefficients \cite{Corley:2001zk} (see also the discussion in \cite{Berenstein:2015ooa}). Remember that $\ket{\Delta}_n$ is associated with a tableaux with $n$ boxes in a row.   These are the objects that appear in $\ket{\bigtriangleup,\Lambda}$.  We want to compute
\begin{equation}
B_{+,\Lambda_1} B_{+,\Lambda_2} \ket 0 = \sum_{n,m} \Lambda_1^n \Lambda_2^m  \,\, 
\begin{ytableau}
\ & &\none[\dots]&n
\end{ytableau}
 \times \begin{ytableau}
\ & &\none[\dots]&m
\end{ytableau}
\end{equation}
The multiplication of two of these objects is similar to addition of angular momentum in $U(2)$, so that 
\begin{eqnarray}
&\begin{ytableau}
\ & &\none[\dots]&n
\end{ytableau} \ \times \begin{ytableau}
\ & &\none[\dots]&m
\end{ytableau}&\\
= &\underbrace{\begin{ytableau}
\ & &\none[\dots]&
\end{ytableau}}_{n+m}+
\underbrace{\begin{ytableau}
\ & &\none[\dots]&\\
\ 
\end{ytableau}}_{n+m-1}+\dots+\begin{ytableau}
\ & &\none[\dots]&\none[\dots]&\none[\dots]&n\\
\ & &\none[\dots]&m
\end{ytableau}&
\end{eqnarray}
where we assume $n\geq m$ in the second line, but we can also have the opposite ordering in which case we exchange $n,m$. We see that all states have only two rows in their Young diagrams. We will group together all diagrams with two rows of lengths $r$ and $s$ with $r\geq s$. To find the coefficient of these, we need to consider the decomposition $r+s= n+m$, such that $|n-m|\leq s$ and to sum over these possibilities. When we perform this sum we obtain
\begin{equation}
\Lambda_1^r \Lambda_2^s+ \Lambda_1^{r-1} \Lambda_2^{s+1} +\dots + \Lambda_1^s \Lambda_2^r= \frac{\Lambda_1^{r+1}\Lambda_2^s-\Lambda_1^s \Lambda_2^{r+1} }{\Lambda_1-\Lambda_2}
\end{equation}
That is, we can write identically that 
\begin{equation}
(\Lambda_1-\Lambda_2) B_{+,\Lambda_1} B_{+,\Lambda_2} \ket 0= \sum_{r\geq s}   \begin{vmatrix} \Lambda_1^{r+1}& \Lambda_1^s\\
\Lambda_2^{r+1}&\Lambda_2^s\end{vmatrix} \begin{ytableau}
\ & &\none[\dots]&\none[\dots]&\none[\dots]&r\\
\ & &\none[\dots]&s
\end{ytableau}\label{eq:genff}
\end{equation}
where the coefficients are very easily written in terms of determinants. This suggests that the full set of states with two classical D-branes acting on the vacuum can be interpreted as a set of fermionic wave functions (Slater determinants), as long as we multiply them on the left by a Vandermonde determinant made from the $\Lambda$ parameters. The point is that the complexity of the coefficients of the tableaux in equation \eqref{eq:genff} is very small. Notice that even though $r\geq s$, this implies that $r+1>s$, so the Slater determinants never vanish, unless we have that $\Lambda_1=\Lambda_2$ (this is trivial in the left hand side, as the full ket is multiplied by $\Lambda_1-\Lambda_2$). Notice that if both D-brane states are normalizable, so is their product.
The same arguments work for the product of various D-brane states, but the combinatorics of multiplying the Young tableaux are more complicated. The simplest way to understand it is through the relation between free fermions and formal matrix models for a generalized oscillator (this is described in \cite{Berenstein:2015ooa}).

We can also compute the norm of the state to obtain
\begin{eqnarray}
||(\Lambda_1-\Lambda_2) B_{+,\Lambda_1} B_{+,\Lambda_2} \ket 0||^2=& \sum_{r\geq s} \begin{Vmatrix} \Lambda_1^{r+1}& \Lambda_1^s \nonumber\\
\Lambda_2^{r+1}&\Lambda_2^s\end{Vmatrix} ^2= \sum_{r\geq s} ||\Lambda_1^{r+1}\Lambda_2^s-\Lambda_1^s \Lambda_2^{r+1} ||^2\\
=& \sum_{r\geq s}||\Lambda_1^{r+1} \Lambda_2^s||^2+\sum_{s\geq r}||\Lambda_1^{r} \Lambda_2^{s+1}||^2\nonumber\\
&-\sum_{r\geq s} \Lambda_1^{r+1} \Lambda_2^s \bar \Lambda^s_1\bar \Lambda^{r+1}_2-\sum_{s\geq r}\Lambda_1^{r} \Lambda_2^{s+1} \bar \Lambda^{s+1}_1\bar \Lambda^{r}_2\nonumber\\
=&\sum_{r,s} ||\Lambda_1^r \Lambda_2^s||^2 - \Lambda^r_1\Lambda_2^s \bar \Lambda_1^s\bar \Lambda_2^r\nonumber\\
=& \frac 1{1-||\Lambda_1||^2 }\frac 1{1-||\Lambda_2||^2}- \frac 1{1-\Lambda_1\bar \Lambda_2}\frac 1{1-\Lambda_2\bar \Lambda_1}\nonumber
\end{eqnarray}
where we have relabeled $r,s$ in some of the sums in the second line, and next we add and subtract the $r+1=s$ and $s+1=r$ contribution to obtain unrestricted sums that can be evaluated explicitly on the last line.
This  expression is equal to 
\begin{equation}
||(\Lambda_1-\Lambda_2) B_{+,\Lambda_1} B_{+,\Lambda_2} \ket 0||^2= || \ket {\Delta,\Lambda_1}||^2 || \ket {\Delta,\Lambda_2}||^2- \braket{\Delta,\Lambda_2}{\Delta,\Lambda_1}\braket{\Delta,\Lambda_1}{\Delta,\Lambda_2}
\end{equation}
after we recognize  the result \eqref{eq:norm1} and we think of $\Lambda, \bar \Lambda$ as independent variables. 
This results in the typical norm for two-particle states  in fermion systems: the product of the norms minus the exchange contribution.

The multiplication rules for two $B_-$ operators will give a similar result, but with Young tableaux with two columns, rather than two rows. This follows from the $\phi\to -\phi$ symmetry, which flips tableaux along the
main diagonal. The other example with two branes is what happens when we multiply $B_{+,\Lambda_1}$ and $B_{-,\Lambda_2}$. This is the most interesting example because one can get a cancellation between the two. It is instructive to see how this comes about from multiplying the corresponding Young tableaux, as follows
\begin{equation}
\begin{ytableau}
\ & &\none[\dots]&n
\end{ytableau} \ \times \begin{ytableau}
\ \\\none[\vdots]
\\ m
\end{ytableau} =\begin{ytableau}
\ &\none[\dots]&n & \\
\none[\vdots]\\
m
\end{ytableau} +\begin{ytableau}
\ &\none[\dots]&n \\
\none[\vdots]\\
m\\ \
\end{ytableau} 
\end{equation}
and to each of these we associated the coefficient $\Lambda_{+,1}^n (-\Lambda_{-,2})^m$. Now, we only get two possible tableaux on the right hand side. If we fix the Young diagram to have $r$ boxes on the first row and $s$ boxes on the first column, we get by summing over possibilities $\Lambda_1^r(-\Lambda_2)^{s-1}+\Lambda_1^{r-1}(-\Lambda_2)^s= \Lambda_1^{r-1}(-\Lambda_2)^{s-1}(\Lambda_1-\Lambda_2)$, except in the case of no boxes, where we get $1$. That is, we find that 
\begin{equation}
 B_{+,\Lambda_1} B_{-,\Lambda_2} \ket 0 = 1 +\sum_{r\geq1, s\geq 1}  \Lambda_1^{r-1}(-\Lambda_2)^{s-1}(\Lambda_1-\Lambda_2)\ \begin{ytableau}
\ &\none[\dots]&r \\
\none[\vdots]\\
s
\end{ytableau}\label{eq:BpBm}
\end{equation}
and now the right hand side simplifies if we divide by $(\Lambda_1-\Lambda_2)$ (rather than when we multiply by it). If we compute the norm of the state (when dividing by $(\Lambda_1-\Lambda_2)$) we get that
\begin{equation}
\left |\left|\frac{ B_{+,\Lambda_1} B_{-,\Lambda_2} \ket 0}{\Lambda_1-\Lambda_2}\right|\right|^2= \frac 1{||\Lambda_1-\Lambda_2||^2}+|\ket{\bigtriangleup, \Lambda_1}|^2|\ket{\bigtriangledown,\Lambda_2}|^2
\end{equation}
and apart from the first term, it shows that the two types of fermionic ``particles" are distinguishable. In this setup, when $\Lambda_1\to \Lambda_2$, the first term develops a pole. That is, the $1$ dominates the norm of the state, 
but this can be subtracted if we are careful, and then we can get a smooth two particle state in the limit.

Formally, when we consider a state $\ket{\psi}$, which results from applying various $B_{\pm}$ operators, we can identify the expectation values as supertraces of a   complex supermatrix
\begin{equation}
 \bra{\psi}a_k\ket{\psi}= \sum \Lambda_{+i}^k-\sum_j\Lambda_{-j}^k= Str \begin{pmatrix}{\bf\Lambda_+}^k &\vdots&0\\
\hdotsfor{3}\\
0&\vdots&{\bf\Lambda_-}^k
\end{pmatrix} = Str[{\bf \Lambda}^k]
\end{equation}
where we identify the different values $\Lambda_{+,-}$ with the corresponding eigenvalues. The values of $\Lambda$ are then interpreted as collective coordinates for the D-brane states. 
The interesting prefactor of the wave functions where we multiply by 
\begin{equation}
\prod_{i< j} (\Lambda_{+,i}-\Lambda_{+,j})\prod_{i< j}(\Lambda_{-,i}-\Lambda_{-,j})\prod( \Lambda_{+,i}-\Lambda_{-,j})^{-1}
\end{equation}
is a super-Vandermonde determinant. 

\section{Free fermions and the Murnaghan-Nakayama rule}\label{sec:MN}

We see that it is rather helpful to be able to go from the conjugacy class basis to the representation basis efficiently. Therefore it makes sense to understand how to compute the characters $\chi_R(\sigma)$ more precisely in order to be able to make progress. The main tool to do so is the Murnaghan-Nakayama rule, as described in appendix \ref{app:MN}. This gives a recursive way to compute the characters of the symmetric group. We will now see that this rule is essentially 
encoding the fact that the tableaux states correspond to free fermions.

We will prove this fact now. To do so, let us analyze the main result of the appendix, where the Murnaghan-Nakayama rule can be rewritten as an operator equation in the Hilbert space of states
\begin{equation}
s\partial_{t_s} \ket R = \sum_{\hbox{hooks of length $s$}} (-1)^{f_{\rm{hook}}} \ket{\tilde R_{\rm{hook}}} \label{eq:MNruleop}
\end{equation}
where $f_{\rm{hook}}$ is the number of rows spanned by the hook, minus one, and $\tilde R_{hook}$ is the Young tableaux $R$ with the skew hook corresponding to the hook that has been singled out removed.
A skew hook is a set of boxes at the edge of the tableaux whose removal produces an allowed tableaux, and they are in one to one correspondence with regular hooks
(see appendix \ref{app:MN}). 

 Remember that $s\partial_{t_s} \simeq a_s$ is the lowering operator of the mode $s$ in the Fock space. This equation above can also be read as follows
\begin{equation} \label{eq:MNruledotprod}
\bra{\tilde R} a_s \ket R= (-1)^{f_{\rm{hook}}}
\end{equation} 
or taking adjoints
\begin{equation}
\bra R a_s^\dagger \ket {\tilde R} = (-1)^{f_{\rm{hook}}}
\end{equation}
where $\ket {\tilde R}$ is a particular diagram appearing in the sum with one hook removed.  For instance, consider the state corresponding to the representation given by

\begin{equation}
\ket R =
 \ytableausetup{boxsize=1em, aligntableaux = center}
\begin{ytableau}
*(white) & *(white) & *(white) & *(white)  \\
*(white) & *(white) & *(white) & *(white) 
\end{ytableau}
\end{equation}

We could apply the lowering operator $a_3$.  Equation \ref{eq:MNruleop} gives

\begin{equation}
a_{3}\;
 \ytableausetup{boxsize=1em, aligntableaux = center}
\begin{ytableau}
*(white) & *(white) & *(white) & *(white)  \\
*(white) & *(white) & *(white) & *(white) 
\end{ytableau}
= 
\begin{ytableau}
*(white) & *(white) & *(white) & *(white)  \\
*(white) 
\end{ytableau}
-
\begin{ytableau}
*(white) & *(white) & *(white)   \\
*(white) & *(white) 
\end{ytableau}
\end{equation}

We could then dot this with the state given by

\begin{equation}
\ket {\tilde{R}} =
 \ytableausetup{boxsize=1em, aligntableaux = center}
\begin{ytableau}
*(white) & *(white) & *(white)   \\
*(white) & *(white)  
\end{ytableau}
\end{equation}

And we find

\begin{equation}
\bra {\tilde{R}} a_3 \ket {R} =-1= (-1)^{1}
\end{equation}
 
 as expected.

We will now show how this arises using free fermion intuition.  Proving equation \eqref{eq:MNruledotprod} is equivalent to proving equation \eqref{eq:MNruleop}, which in turn gives a proof of the Murnaghan-Nakayama rule. 

We need to think of what $a_s^\dagger$ is doing in terms of the eigenvalue representation (the fermions in Matrix models). The idea is that to each eigenvalue of an infinite matrix we associate 
a Cuntz oscillator pair $\beta_\ell \beta_\ell^\dagger = 1 $ (which commute with each other for different $\ell$) and we will treat these eigenvalues as Fermions (see \cite{Berenstein:2015ooa} for details on how to build fermionic systems from general oscillators). The operator
\begin{equation}
a_s^\dagger = \sum_{\ell=0}^\infty (\beta_\ell^\dagger)^s= Tr((\beta^\dagger)^s)
\end{equation}
is a trace of the powers of the raising operator $\beta^\dagger$ thought of as a matrix. 
The ground state of the multiple particle system $\bullet$ is defined by the Slater determinant
\begin{equation}
\bullet \simeq \lim_{N\to \infty} \ket{-1/2} \ket{-3/2} \dots \ket{-(2N-1)/2}_{\rm{antisymm}} 
\end{equation}
where we have set the Fermi sea at energy zero, and all the (infinite tower of) negative energy states are occupied. If all the $\ket j$ are orthonormal, the procedure of antisymmetrization 
gives a normalization factor in front of the state with an ${\cal N}=((N+1)!)^{-1/2}$ to obtain a normalized state. This is common to all states in what follows.  Pictorally, we will represent our ground state as

\begin{equation}
\begin{pspicture}(2,4)
\psdots[dotstyle=o, dotsize=6pt](1,2.2)(1,2.6)(1,3.4)(1,3.8)(1,3)
\psdots[dotsize=4pt](1,1.4)(1,1)(1,0.6)(1,0.2)(1,1.8)
\psline(1,0)(1,4)
\psline(0.5,2)(1.5,2)
\end{pspicture}
\end{equation}

Notice we have chosen to label the energy of each particle at half integers, rather than filling the Fermi sea to zero and having the particles occupy integer energies.  We did this because it makes the particle/hole duality more explicit and symmetric. This is simply a convention and everything would follow in the same way if all particle energies were shifted by $1/2$, as it is always energy differences that are measured.

A complete basis of states is given by
\begin{equation}
\ket{\{n \}}\simeq \lim_{N\to \infty} \ket {n_1} \ket{n_2} \dots \ket{n_{N}}_{\rm{antisymm}} 
\end{equation}
with $n_1>n_2>n_3 >\dots>n_N$, half integers, and for all sufficiently large $j$ we require that $n_j=-\frac{2j-1}{2}$.

We will now show how to go directly from one of these states to a Young tableaux representation, which we will associate with it.  Consider the numbers given by $r_j= n_{j}-n_j^0=n_{j}-(\frac 12 - j)$, e.g. $r_1= n_{1}-(- \frac 12 )$, $r_2= n_{2}-(-\frac 32 )$.  These numbers will give the differences between the particles excited positions and their ground state position.  We can check easily that $r_i-r_{i+1}= n_{i}-n_{i+1}-1\geq 0$, since the $n_i$ are strictly decreasing integers and moreover that $r_i=0$ for sufficiently large $i$. We assign to this set a Young diagram with rows of length $r_1, \dots r_s$ for all the $r_k$ that are different from zero. This is an allowed tableaux because the integers are non-increasing. Details of the pictorial representation of this assignment can be found in appendix \ref{app:fermions}. We can clearly invert this map, because knowing $r_i$ is equivalent to knowing the $n_i$. Now, instead of 
$\ket{\{n\}}$ we use the tableaux $\ket {\tilde R}$.

Now we act with $a_s^\dagger= Tr((\beta^\dagger)^s) $ on $\ket{\tilde R}$ and get that 
\begin{equation}
a_s^\dagger\ket{\tilde R}\propto \lim_{N\rightarrow\infty} \sum_{\ell=1}^N  \ket {n_1} \ket{n_2} \dots\ket {n_{\ell-1}} \ket{n_\ell+s} \ket{n_{\ell +1}}\dots\ket{n_{N}}_{\rm{antisymm}} 
\end{equation}
If the state has two of the $n_j$ equal to each other, the antisymmetrization procedure will remove the term from the sum. 
Now, we need to check if the new state has the $n_{s}$ in decreasing order or not. If it does not, we need to reshuffle the $n_s$, by moving the $n_{\ell+s}$ to the left until we get a proper order. We do this by transposition of nearest neighbors, moving $n_\ell+s$ as we go along. Each such transposition is an exchange of two fermions, so it costs a factor of $(-1)$. The sign we get is $(-1)^{\# \rm{transpositions}}$.

As an example of how this works, we would have

\begin{equation}
\begin{split}
a_2^\dagger \left( \ket {5/2} \ket{3/2} \ket{-1/2} \ket{-7/2} \ket{-9/2} \dots\ket{n_{N}}_{\rm{antisymm}} \right) \\
 = \ket {9/2} \ket{3/2} \ket{-1/2} \ket{-7/2} \ket{-9/2} \dots\ket{n_{N}}_{\rm{antisymm}} \\
   - \ket {7/2} \ket{5/2} \ket{-1/2} \ket{-7/2} \ket{-9/2} \dots\ket{n_{N}}_{\rm{antisymm}} \\
 + \ket {5/2} \ket{3/2} \ket{-1/2} \ket{-3/2} \ket{-9/2} \dots\ket{n_{N}}_{\rm{antisymm}} \\
 - \ket {5/2} \ket{3/2} \ket{-1/2} \ket{-5/2} \ket{-7/2} \dots\ket{n_{N}}_{\rm{antisymm}}
 \end{split}
\end{equation}

Notice the second and fourth terms have negative signs because we had to perform one transposition on each to find a state with the proper ordering.  Notice also that we dropped all terms that would have had two $n_i$'s that are equal to each other and so would go away upon antisymmeterization.  Now we want to think about how the transpositions affected the Young diagrams these states correspond to.

Because various of the $n_i$ have been moved to the right, we find that for these that have moved we get that $n_{i+1}^{\rm{new}}=n_i^{\rm{old}}$, so $r_{i+1}^{\rm{new}} =n_{-1+i}^{\rm{old}}-(-i)= r_i^{\rm{old}} +1$. That is, in the Young diagram we have moved the row $i$ to the $i+1$ row and added one extra box to the right (this is equivalent to moving the corner of the row one to the right and one down). There is still the one that got moved to the left. This one row was moved upward by $\#\rm{transpositions}$, and each such transposition makes the corresponding  row we are tracking shorter by one (this is the opposite of the $+1$ to the right that we have found for the others). The net effect is that we have added just $s$ boxes to the Young diagram and gotten a normalized state for each $\ell$ that is allowed.
That is, we get that 
\begin{equation}
a_s^\dagger \ket{\tilde R}= \sum_{\rm{hooks}} (-1)^{\# \rm{transpositions}} \ket {R_{\rm{allowed}}}
\end{equation}
The motion to the right and down produces a skew hook of length $s$ added to the original tableaux (we just color in the new boxes in a different color than those of $\tilde R$). The sign we find is  the same sign that  is assigned by the Murnaghan-Nakayama rule. The number of transpositions is the number of rows that have changed minus one!

As a simple example consider
$ a_{5}^{\dagger}\;
 \ytableausetup{boxsize=1em, aligntableaux = center}
\begin{ytableau}
*(white) & *(white) & *(white) 
\end{ytableau}
$. 

What does this correspond to in the Fermi sea picture of appendix \ref{app:fermions}?  We have 5 units of energy we are adding and there are several options we have for where to put them.  The state we are starting with is

\begin{equation}
\begin{pspicture}(2,4)
\psdots[dotstyle=o, dotsize=6pt](1,2.2)(1,2.6)(1,3.4)(1,1.8)(1,3.8)
\psdots[dotsize=4pt](1,1.4)(1,1)(1,0.6)(1,0.2)(1,3)
\psline(1,0)(1,4)
\psline(0.5,2)(1.5,2)
\end{pspicture}
\end{equation}
We could imagine giving all the energy to the already excited particle.  This gives

\begin{equation}
\begin{pspicture}(2,5.5)
\psdots[dotstyle=o, dotsize=6pt](1,2.2)(1,2.6)(1,3.4)(1,1.8)(1,3.8)(1,4.2)(1,4.6)(1,5.4)(1,3)
\psdots[dotsize=4pt](1,1.4)(1,1)(1,0.6)(1,0.2)(1,5)
\psline(1,0)(1,5.5)
\psline(0.5,2)(1.5,2)
\pscurve[linewidth=2pt]{->}(0.9,3.1)(0.5,4)(0.9,4.9)
\end{pspicture}
\end{equation}
which we know corresponds to a totally symmetric diagram with 8 boxes.  

However, we could have made a different choice and excited the top particle out of the fermi sea.  This would give
\begin{equation}
\begin{pspicture}(2,4)
\psdots[dotstyle=o, dotsize=6pt](1,2.2)(1,2.6)(1,1.8)(1,3.8)(1,1.4)
\psdots[dotsize=4pt](1,1)(1,0.6)(1,0.2)(1,3)(1,3.4)
\psline(1,0)(1,4)
\psline(0.5,2)(1.5,2)
\pscurve[linewidth=2pt]{->}(0.9,1.5)(0.5,2.3)(0.9,3.3)
\end{pspicture}
\end{equation}
Notice that to get to this position, we had to pass the already excited particle (we had to perform a transposition), and therefore had to pick up a minus sign.  We find, then, that the negative sign, which previously was simply a part of the MN rule, actually encodes the exchange statistics of fermions.

Carrying on, we we could imagine exciting the second particle from the fermi sea, but we know this should not be allowed by the Pauli exclusion principle:
\begin{equation}
\begin{pspicture}(2,4)
\psdots[dotstyle=o, dotsize=6pt](1,2.2)(1,2.6)(1,1.8)(1,3.8)(1,3.4)
\psdots[dotsize=4pt](1,1)(1,0.6)(1,0.2)(1,3)(1,1.4)
\psline(1,0)(1,4)
\psline(0.5,2)(1.5,2)
\pscurve[linewidth=2pt]{->}(0.9,1.1)(0.5,2)(0.9,2.9)
\psline[linecolor=red](1.5,2.5)(0.5,3.5)
\psline[linecolor=red](0.5,2.5)(1.5,3.5)
\end{pspicture}
\end{equation}
If we think about this in terms of Young tableau, we see that this would have corresponded to a diagram that is not allowed.  Specifically, it would have corresponded to something of the form

\begin{equation}
\begin{ytableau}
*(white) & *(white) & *(white) \\
*(white) & *(white) & *(white) &*(white) \\
*(white)
\end{ytableau}
\end{equation}
We know that this is not an allowed diagram and that fact encodes the exclusion principle.

The end result is that

\begin{equation}\label{a5onsym3}
a_{5}^{\dagger}\;
 \ytableausetup{boxsize=1em, aligntableaux = center}
\begin{ytableau}
*(white) & *(white) & *(white) 
\end{ytableau}
=
\begin{ytableau}
*(white) & *(white) & *(white) & *(blue) & *(blue) & *(blue) & *(blue) & *(blue) 
\end{ytableau}
 - 
\begin{ytableau}
*(white) & *(white) & *(white) & *(blue)  \\
*(blue) & *(blue) & *(blue) & *(blue) 
\end{ytableau}
+
\begin{ytableau}
*(white) & *(white) & *(white) \\
*(blue) & *(blue) & *(blue) \\
*(blue) \\
*(blue)
\end{ytableau}
-
\begin{ytableau}
*(white) & *(white) & *(white) \\
*(blue) & *(blue)  \\
*(blue) \\
*(blue) \\
*(blue)
\end{ytableau}
+
\begin{ytableau}
*(white) & *(white) & *(white) \\
*(blue) \\
*(blue) \\
*(blue) \\
*(blue) \\
*(blue)
\end{ytableau}
\end{equation}

The outcome of this computation is that 
\begin{equation}
\bra R a_s^\dagger \ket{\tilde R} = (-1)^{\# \rm{transpositions}}= (-1)^{f_{\rm{hook}}}
\end{equation}
as we wanted to prove. Moreover, the sign of the Murnaghan-Nakayama rule is nothing other than the Fermi statistics. The boxes colored in blue are exactly the skew hooks that can be added to the original tableaux. These are always at the border of the tableaux such that adding them produces an allowed tableaux.

The upshot is that rather than trying to compute the $\chi_R(\sigma)$ directly, we compute the action of the $t_s$ and its adjoints on the basis of Young tableaux (the D-brane basis).
This algebraic action is simple and will let us establish a lot of facts in the next sections.

\section{Multi-edge geometries: new classical limits with different topology}\label{sec:multiedge}

In this section we will think of the classical field $\phi(\theta)$ as a displacement of the geometric interface between two fluids, made of particles and holes. This is the main viewpoint in treating the system as a set of free fermions as exemplified in the description of the quantum hall effect \cite{Stone:1990ir}. This is a geometric interpretation that also appears naturally in studying bubbling solutions \cite{Lin:2004nb}, where the two fluids in question arise as the two possible values of a function on  a plane that give rise to a regular  BPS geometry in ten dimensions. Some of the treatment here follows the 
previous work by the authors \cite{Berenstein:2016pcx}. 

Let us start with a simple identity for the classical energy of a configuration, where we use
\begin{equation}
E[\phi] = \frac 1{2\pi}\int d\theta \, \frac 12 \, \phi(\theta)^2= \frac 1{2\pi}\int d\theta \left[\int_0^{\phi(\theta)\geq 0} h(\theta) \, dh- \int_0^{-\phi(\theta)>0} (-h(\theta)) \, dh\right] 
\end{equation}
That is, we have divided the coordinate $\theta$  into the regions $\phi(\theta)\geq 0$ and the regions $\phi(\theta)<0$. This makes sense in the classical theory for smooth functions, but not quite in the quantum theory.
The first region is associated with moving particles into spaces that had been occupied by holes, and the second one is associated with moving holes where there were particles. Both particles and holes are conserved because $\int \phi(\theta) \, d\theta=0$. We want to associate a function with this change of occupation, that takes the value $+1$ when particles occupy a hole state, and takes the value $-1$ when holes occupy previously occupied particle states. The function should otherwise vanish. This function is the relative density of particles with respect to the ground state 
$\tilde \rho(\theta, h)= \rho_+(\theta,h) - \rho^0_{+}(\theta,h)$. It can also be symmetrically constructed from the hole density with few modifications. The energy can then be expressed as 
\begin{equation}
E[\phi]= \frac 1{2\pi}\iint  \tilde \rho(\theta, h) \, h \, dh \, d\theta \label{eq:densitydef}
\end{equation}
Notice that this formula is very similar to \eqref{eq:dropenergy}. The main difference is that in \eqref{eq:dropenergy} one is allowed to have arbitrary regions with $\rho\neq 0$, while in
\eqref{eq:densitydef} we have a description not only with fixed topology, but also  with a unique height for the boundary between holes and particles for each $\theta$.
 The field $\phi$ is given by
\begin{equation}
\phi(\theta) = - \int dh\,h\,\partial_h \tilde \rho(\theta,h)    \label{eq:phiedge}
\end{equation}
This uses the fact that $\rho$ can be written as Heaviside step functions whose derivative is a delta function, precisely at the height of the droplet. The contribution at $h=0$ vanishes. Alternatively, we can integrate by parts to find that
 \begin{equation}
\phi(\theta) = \int \tilde \rho(\theta,h) \, dh  \label{eq:naivephi}
\end{equation}
and the conservation of particles and holes can be expressed as
\begin{equation}
\iint \tilde \rho(\theta, h) \, d\theta \, d h  =0 
\end{equation}
Basically, we are expressing the states as pictures on a two dimensional cylinder.
We will use this identity together with previous observations (particularly equation \eqref{eq:geomecond}) to understand how new classical limits can appear from different quantum states that are not classical in the oscillator representation.

\subsection{Fixed energy single D-brane state}

Let us start with a single D-brane state, but instead of considering the  coherent states \eqref{eq:defexp}, we want to consider a state of the form $\ket{\bigtriangleup}_n$ for a fixed $n$: a single D-brane with fixed energy, and we want to think of it formally as a superposition of non-normalizable states $\ket{\bigtriangleup, \exp(i\gamma)}$ by doing a Fourier transform. The state $\ket{\bigtriangleup}_n$ is an eigenstate of the momentum operator, so it is translation invariant. Because the state is a superposition of classical states, it is not a classical state with respect to the usual variables $a_k^\dagger, a_k$ any longer. For example, it does not factorize into a product of coherent states because it has a fixed energy and it is not in the vacuum. Indeed, it is possible to show that 
in this state $\vev{a_k}= \vev{a^\dagger_k}=0$, yet the energy is not zero (as one would conclude for coherent states). The naive classical state we would associate with this profile is the ground state. Because in the end the corresponding state is not a coherent state, we will be interested eventually in characterizing to what extent it violates the properties of the coherent state. For example, if a state has minimal uncertainty then it is a coherent state. Conversely, a non-trivial uncertainty serves as a measure of how much the state differs from a coherent state. Similarly, a coherent state is a product state ( a pure state mode per mode). Entanglement between the modes measures to what extent the mode per mode quantum state is not pure. We will take the problem of measuring these later in the paper. 

What we want to do is we want to find an {\em alternative} classical description of this particular state so that when we insert the right value of $\tilde \rho(\theta, h) $ in equation \eqref{eq:densitydef}, we get the right energy. We want to  declare that this state can be classical as well, even if it is not a coherent state. Our goal is to come up with a prescription for how to do this consistently.

Moreover, we want the
 relative density to be translation invariant, so $\tilde \rho$ is independent of $\theta$. To be classical, it should take the same prescribed nominal values from before $\tilde \rho= +1, -1, 0$. In principle, a value in between can be obtained 
 from statistically averaging  states (a  density matrix state rater than a pure state). Those will not be treated as classical states but as statistical states. Here, we want the state to be pure, so no averaging should be performed. 

Now, let us look precisely at \eqref{eq:geomecond}. The idea behind building the new classical solution is that acting with a D-brane state lowers the level of $\phi$ by a prescribed amount: $-1$ in our conventions, and the area under the delta function is identified with the amount of area that a single D-brane (particle) occupies.  We need to move this occupied area somewhere else, but we want to leave the lowering of the level interface exactly as the $-1$ demands: this is after all the classical field everywhere else away from the $\delta$ function distribution.
Because of translation invariance, we should add horizontal strips with $\tilde \rho=1 $ to the picture in order to conserve area. Since we are acting with a single D-brane, this horizontal strip should be connected (a single object), and of width one because of area conservation. We are building this picture by hand, but we are inspired by the description of states in \cite{Lin:2004nb}. 
To match the energy of the state, there is only one place where we can put the strip: the topmost edge of the strip should be at height $n$.  

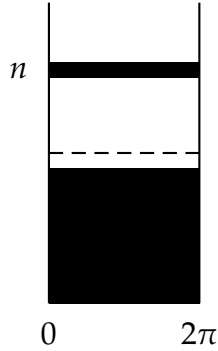
\begin{figure}[ht]
\begin{center}
\begin{pspicture}(4,4)
\pscustom{
\gsave
    \psline(1,0)(1,1.8)
    \psline(1,1.8)(3,1.8)
    \psline(3,1.8)(3,0)
    \psline(3,0)(1,0)
    \fill[fillstyle=solid,fillcolor=black]
\grestore
}
\pscustom{
\gsave
    \psline(1,3)(1,3.2)
    \psline(1,3.2)(3,3.2)
    \psline(3,3.2)(3,3)
    \psline(3,3)(1,3)
    \fill[fillstyle=solid,fillcolor=black]
\grestore
}
\psline(1,0)(1,4)
\psline(3,0)(3,4)
\psline[linestyle=dashed](1,2)(3,2)
\rput(0.6, 3.1){$ n$}
\rput(1, -0.4){$ 0$}
\rput(3, -0.4){$ 2\pi$}
\end{pspicture}
\end{center}
\caption{The black and white LLM plane drawing corresponding to a D-brane with fixed energy $n$.}
\end{figure}

If we use equation \eqref{eq:naivephi}, we find that the field $\phi(\theta)$ vanishes, just as expected from the expectation values of the quantum state. However, when we analyze it from the point of view of \eqref{eq:phiedge} we realize that the expectation value of the field vanishes by adding three contributions
\begin{equation}
\phi= \int dh \, h \, (\delta(h-n) - \delta(h-n+1) +\delta(h+1))= n - (n-1) - 1 \label{eq:multi-value}
\end{equation}
one from each of the edges of the $\rho$ density. That is, we should think of the field $\phi(\theta)$ essentially as becoming a multi-valued function of $\theta$, and the expectation value of the field is obtained by summing over these values with signs. We have gone from one well defined classical edge to three. This should be thought of as a topology transition. Notice that the superposition of states that gave rise to the state we want can be performed for finite values of $\Lambda$ as well. In this interpretation, superpositions of classical coherent states with one topology can give rise to a topology change.

Notice that this new state is macroscopically different from each of the states that is used to make it, even statistically. For example, all the coherent states have $\tilde \rho(\theta, h)\geq 0$  strictly non-increasing as a function of $h$ when $h\geq 0$. So when we take a classical statistical average of these states, this property should still hold. The new $\tilde \rho$ does not have this property.

 If for all original classical states we assign the same topology where the edge is a circle, then we find that there is no operator measurement 
in the Hilbert space that can distinguish the strip from the coherent states \cite{Berenstein:2016pcx}. The argument is by contradiction.  It proceeds as follows:

Suppose first that all coherent states are associated with  a fixed (trivial) topology.
Imagine now  that such an operator exists. The operator should be such that all states with a  trivial topology have the same eigenvalue.
 We can imagine building one such operator if we simply count the total number of edges to characterize the topology.   We then get the same number for all coherent states: 1 (because they are described by a single height function), and so we are describing the identity operator. The strip topology state is a superposition of these, so the operator should evaluate to $1$ as well, which clearly fails to count the number of edges appropriately. Therefore, there is no such topology measuring operator.

This means that topological type should be associated with details of a particular classical approximation of a state, and not with measurement of an operator. Notice that at this stage we have made no reference to entanglement as a source for topology changes \cite{VanRaamsdonk:2010pw, Maldacena:2013xja}. We are also stating that the set of classical states is even more overcomplete than regular coherent states. That there might be an overcompleteness that exceeds the standard overcompleteness of coherent states has been hinted at in \cite{Papadodimas:2015jra} for situations involving the interior of a black hole.
It was also argued there (following \cite{Motl}) that gravitational physics requires state dependence, which implies that gravitational physics can not be encoded in operators. These works, in some sense, already assume that the $ER=EPR$ conjecture is true \cite{Maldacena:2013xja}, so that non-trivial entanglement measures a topology change. As is well known, entanglement is not an operator measurement in Hilbert space, although it can be computed in simple enough setups.

Back to our main discussion, since 
the strip state is not a coherent state, one can presume it is a state with larger uncertainty on each oscillator. We can actually compute these uncertainties using the techniques developed in section \ref{sec:MN}. We will see later that the spirit of \eqref{eq:multi-value} can be made precise for a large number of quantum states in the field theory, and the microscopic quantum field $\phi$ can be written as a sum with signs of other effective fields that appear as quantizations of small deviations away from a particular multi-strip configuration that should be thought of as a `ground state' classical configuration with a coherent state excitation of its collective coordinates.

Let us finish analyzing this one configuration at a large but finite energy $n$. The Young diagram describing the state has $n$ boxes and is completely horizontal.
\begin{equation}
\ket{\Delta}_n=\begin{ytableau}
\ & &\none[\dots]&n
\end{ytableau}
\end{equation}
We can evaluate the expectation values of the mode number operators $E_s=a_s^\dagger a_s$, $\hat N_s= s^{-1} a_s^\dagger a_s$ by the Murnaghan-Nakayama rule (see appendix \ref{app:MN}) in a straightforward fashion
\begin{equation}
_n\bra \Delta a_s^\dagger a_s \ket{\Delta}_n = \;_{n-s}\langle\Delta | \Delta\rangle_{n-s} = \left\{\begin{array} {ll} {1}& \hbox{ if $s\leq n$}\\
0& \hbox{ Otherwise}\end{array}\right.
\end{equation}
so the average energy per mode is constant up to mode $n$ where it cuts off abruptly. The average occupation number per mode $s$ is $1/s$, exactly as in equation \eqref{eq:avoccnumb}. This shows why the state is a very close approximation to a regularized "fermion field state."

It also makes sense to ask which of the states $\ket{\Delta, \Lambda}$ is the best approximation to $\ket\Delta_n$. The normalized probability amplitude to go from one to the other is
\begin{equation}
_n\langle \Delta | \Delta; \Lambda\rangle= \Lambda^n \sqrt{1-\Lambda\bar \Lambda}
\end{equation}
So that 
\begin{equation}
|_n\langle \Delta | \Delta; \Lambda\rangle|^2= (\bar \Lambda \Lambda)^n(1-\Lambda\bar \Lambda)\label{eq:cohover}
\end{equation}
This is maximized for 
\begin{equation}
\Lambda\bar \Lambda= \frac{n}{n+1}
\end{equation}
For this value of $\Lambda\bar \Lambda$, we would find that the energy of the best coherent state approximation to the state to be exactly $n$, the energy of the state. For large $n$, the size of the overlap is of order $(en)^{-1}$ where $e$ is Euler's constant.
It is also easy to check the equations $a_s= a_1^s$ for these states, just as expected from the fact that all the coherent states that can be superposed to obtain the state satisfy them.

This means that the state $\ket{\Delta}_n$ should be thought of as an overlap of order $n$ approximately orthogonal coherent states. That is roughly the number of states that would be needed to get 
the probabilities $\sum_i |_n\langle \Delta | \Delta; \Lambda_i\rangle|$ to add up to order one: what we need in order to say that we closely approximate the state $\ket{\Delta}_n$.

One can use this result to state that a superposition of a large number of states in quantum gravity might have very different properties (even different topology) than any of the individual states that make up the system. This is a result that has been argued quite effectively in \cite{Almheiri:2016blp}, where the goal was to show that entanglement entropy can be thought of as an operator for a sufficiently small superposition of classical states, but not when we take the limit.

One can do a similar analysis with a single hole state of fixed energy $\ket{\bigtriangledown}_n$ and the results are very similar. 

Now, we can  also evaluate the quantities
\begin{equation}
\,_n\bra{\Delta} a_s a_s^\dagger \ket{\Delta}_n = \vev{ (a_s-\vev{a_s}) (a_s^\dagger -\vev{a_s^\dagger})} = s + \vev{a_s^\dagger a_s}= s+1 \label{eq:unc}
\end{equation}
for $s\leq n$. These can be thought of as the net quantum uncertainty of the solution (including the ground state uncertainty, which is $s$). This shows that the state has low uncertainty (mode per mode) and can therefore be 
interpreted classically.

\subsection{More general one stripe geometries}

Now we will turn our attention to other states that have a similar interpretation. These are "translation invariant condensates of branes". The idea is to look at states whose Young diagram looks as follows
\begin{equation}
\ket{\square_{L,M}} = \begin{ytableau}
\ & &\none[\dots]&L\\
\none[\vdots] & \none[\ddots]&\none[\ddots]& \none[\vdots]\\
M& &\none[\dots]& 
\end{ytableau}
\end{equation}
That is, a rectangular Young diagram with $L$ columns and $M$ rows.

We can visualize this state in the fermion picture (assuming $L>M$) as follows
\begin{equation}
\begin{pspicture}(2,4)
\psdots[dotstyle=o, dotsize=6pt](1,2.2)(1,2.6)(1,1.8)(1,3.8)(1,1.4)
\psdots[dotsize=4pt](1,1)(1,0.6)(1,0.2)(1,3)(1,3.4)
\psline(1,0)(1,4)
\psline(0.5,2)(1.5,2)
\rput(0, 3.2){$ M \ \Big{\{ }$}
\rput(0, 1.6){$ M \ \Big{\{ }$}
\rput(2, 2.4){$ \ \Big{\} } \;(L-M) $}
\end{pspicture}\label{eq:LMexcitation}
\end{equation}
where there is a gap of size $L$ between the filled sea and the excitation, and then the excitation has width $M$ (this corresponds to $M$ filled states).
This is natural from identifications in the figure 8 of \cite{Lin:2004nb} . These states are also translation invariant and share many of the properties of the previous state $\ket{\bigtriangleup_n}$, which would correspond to 
$\ket{\square_{n,1}}$. The natural identification with states in the droplet picture is to extend the representation \eqref{eq:LMexcitation} to a stripe configuration where the holes and filled states become extended on an interval
$(0,2\pi)$. The gap between the filled regions is of width $L$, and the filled top region is of width $M$.  The corresponding LLM droplet picture is shown below.

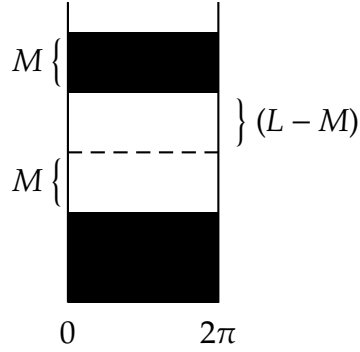
\begin{figure}[ht]
\begin{center}
\begin{pspicture}(4,4)
\pscustom{
\gsave
    \psline(1,0)(1,1.2)
    \psline(1,1.2)(3,1.2)
    \psline(3,1.2)(3,0)
    \psline(3,0)(1,0)
    \fill[fillstyle=solid,fillcolor=black]
\grestore
}
\pscustom{
\gsave
    \psline(1,2.8)(1,3.6)
    \psline(1,3.6)(3,3.6)
    \psline(3,3.6)(3,2.8)
    \psline(3,2.8)(1,2.8)
    \fill[fillstyle=solid,fillcolor=black]
\grestore
}
\psline(1,0)(1,4)
\psline(3,0)(3,4)
\psline[linestyle=dashed](1,2)(3,2)
\rput(0.6, 3.2){$ M \ \Big{\{ }$}
\rput(0.6, 1.6){$ M \ \Big{\{ }$}
\rput(4, 2.4){$ \ \Big{\} } \;(L-M) $}
\rput(1, -0.4){$ 0$}
\rput(3, -0.4){$ 2\pi$}
\end{pspicture}
\end{center}
\caption{The black and white LLM plane drawing corresponding to a rectangular Young tableaux with $L\times M$ boxes.}
\end{figure}
 
Since these states are of fixed energy and $a_s^\dagger, a_s$ change the energy, we also have that $\vev {a_s}_{LM}= \vev{a_s^\dagger}_{LM}=0$. Again, if this were a classical state in the usual classical limit of regular coherent states we would find that it should correspond to  the vacuum. We want to interpret this state also as a classical state, with a black and white pattern as described. The configuration now has three edges: one at the very topmost of the most excited state, one (anti-) edge where the gap ends, and one more edge where the infinitely deep sea ends. We use the labeling edge for an edge where the filled states are below and the empty states above. We will use the label anti-edge for the opposite set, where the empty sites are below and occupied sites are above.  The energy of this state is $LM$.

What we call the field $\phi$ again becomes multi-valued, with one contribution from each (anti-) edge. The vanishing of the (zero mode of the ) field is governed by adding three contributions
\begin{equation}
\phi= \int dh \, h \, (\delta(h-L) - \delta(h-(L-M)) +\delta(h+M))=L - (L-M) - M \label{eq:multi-value2}
\end{equation}
 very similar to \eqref{eq:multi-value}.  
 
 \subsection{Excitations of striped geometries}

 In general, we expect that we could start with one of these translationally invariant striped geometries and deform each edge independently, As shown schematically below.
 
 \begin{figure}[ht]
\begin{center}
\begin{pspicture}(4,4)
\pscustom{
\gsave
    \psline(1,0)(1,1.2)
    \psline(1,1.2)(1.6,1.2)
    \pscurve(1.6,1.2)(1.8,1.3)(2,1.1)(2.2,1.2)
    \psline(2.2,1.2)(3,1.2)
    \psline(3,1.2)(3,0)
    \psline(3,0)(1,0)
    \fill[fillstyle=solid,fillcolor=black]
\grestore
}
\pscustom{
\gsave
    \psline(1,2.8)(1,3.6)
    \psline(1,3.6)(1.8,3.6)
    \pscurve(1.8,3.6)(1.9,3.5)(2.1,3.6)(2.2,3.7)(2.4,3.6)
    \psline(2.4,3.6)(3,3.6)
    \psline(3,3.6)(3,2.8)
    \psline(3,2.8)(2.8,2.8)
    \pscurve(2.8,2.8)(2.7,2.9)(2.6,2.8)(2.5,2.7)(2.4,2.8)
    \psline(2.4,2.8)(1,2.8)
    \fill[fillstyle=solid,fillcolor=black]
\grestore
}
\psline(1,0)(1,4)
\psline(3,0)(3,4)
\psline[linestyle=dashed](1,2)(3,2)
\rput(1, -0.4){$ 0$}
\rput(3, -0.4){$ 2\pi$}
\end{pspicture}
\end{center}
\caption{A schematic depicting possible independent deformations of each edge of an LM state.}
\end{figure}
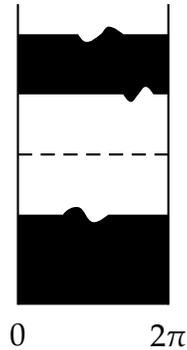

Recall that the energy is given by
\begin{equation}
E[\phi]= \frac 1{2\pi}\iint  \tilde \rho(\theta, h) \, h \, dh \, d\theta 
\end{equation}
Further, let us recall that $\phi\simeq h$ on the edge, while $\tilde \rho=1,0,-1$. We want to consider small deformations of each height, keeping the local density at nominal values, and the area of each region fixed. This is formalizing the idea that the field $\phi$ becomes multivalued, with $\phi(\theta)$ the height at each (anti) edge. 
The total field sums over these contributions $\phi(\theta) = \phi^1(\theta)+\phi^2(\theta) -\tilde \phi^1(\theta)$. Here, we indicate the fields in order from the top down, with $\tilde\phi$ representing the anti-edge fields. To keep the area of each region fixed, we require that for each field $\int d\theta\,\delta \phi^i= \int d\theta\,\delta \tilde \phi^i=0$, and we can substitute in the energy as an area integral to find that
\begin{equation}
E[\phi] = \frac{1}{2\pi} \sum_{i,j} \int d\theta \frac {(\phi^i+\tilde \phi^j)}2 (\phi^i-\tilde \phi^j) = \frac{1}{2\pi} \sum_{i,j} \int d\theta \frac 12 ((\delta \phi^i)^2-(\delta \tilde \phi^j)^2)+ E_{LM}
\end{equation}
We find that the energy splits into a (local) sum over fluctuations of each edge independently, but the ones associated with $\tilde \phi$ have a negative sign. This is an indication that the reference state 
$\ket{\square_{L,M}}$ is not the ground state. The fields $\tilde \phi$ can be thought of as ghosts (in the sense of negative energy states, not negative norm states). In the paper \cite{Koch:2008ah}, such states are called {\em counter-gravitons}. The fields $\phi^{(i)}$ and $\tilde \phi^{(i)}$ will be the collective excitations of the configuration.
 
What we have done is formalize the suggestion of \cite{Koch:2008ah} that makes $\phi$ decompose linearly into pieces that work on each (anti-)edge independently. Now, we can take the Fourier transform to obtain the mode decomposition, as in \cite{Berenstein:2016pcx}.

We find that
\begin{equation}
a_s^\dagger = b^{(1)\dagger}_s+b^{(2)\dagger}_s-  c^{(1)}_s\label{eq:Bog}
\end{equation} 
where the $b^{(i)}$ are excitations of the edges, and the $c^{( i)}$ modes are excitations of the anti-edges.
We need to explain this equation. The mode $a_s^\dagger$ increases the energy by $s$ units. This is the equation of motion of the mode $a_s^\dagger$. The modes $b^{(1)\dagger}_s,b^{(2)\dagger}_s$ also increase the energy by the same amount (which is equal to the momentum of the mode).
The lowering mode for excitations of the anti edge destroys $(-s)$ units of energy, so it acts in the same way to increase the energy.  This is the only way to make the equations consistent with energy conservation and with a linear expression for the modes $a_s^\dagger$ in terms of the collective excitations of the state $\ket{\square_{LM}}$. As long as we can show that the $b, c$ have the correct commutation relations, we find that the equation \eqref{eq:Bog} indicates that the UV fields (those that are canonical modes of the chiral free field) are related by a Bogolubov transformation to the collective fields (which we are calling the IR theory). It is not a complete Bogolubov transformation, because such transformations preserve the number of oscillators. The linear transformation can be completed to a full Bogolubov transformation if we add additional orthogonal fields to the $a_s$ that are made of linear combinations of the $b,c$ modes (we will expand on this idea in the next section). 

The important point is that equation \eqref{eq:Bog} is an approximation to the fluctuations about the state. It is clear that if the fields $\phi^{(i)}$ get too large that the edges will collide changing the topology. Our purpose right now is to actually derive this transformation from first principles by analyzing the physics of the state $\ket{\square_{LM}}$ independent of the geometric intuition. We want to show that the decomposition can actually be derived from the combinatorial structure of the full Hilbert space of states.

The reference vacuum state $\ket{\square_{LM}}$ is defined by $b \ket{\square_{LM}}= c\ket{\square_{LM}}=0$. We will call an excitation a {\em classical configuration} if it is a coherent state excitation of the low lying modes of $b^{(i)}$ and $c^{( i)}$, assuming that they have canonical commutation relations. This can be derived semiclassically by following \cite{Grant:2005qc}, but again, it does not explain what to do when the $\phi$ get large, so the commutation relations are approximately true on a sector of low amplitude. 

Our goal now is to construct the operators $b, c$ explicitly in the quantum theory and show that they have the correct (canonical) commutation relations when acting on a particular subspace of the Hilbert space of states, so that the equation \eqref{eq:Bog} is indeed a Bogolubov transformation. To do this,  the results of section \ref{sec:MN} and appendix \ref{app:MN} are essential. This is the point where we are able to improve the discussion in \cite{Koch:2008ah} substantially because we will have no guesswork. Moreover, we will be able to understand the cutoffs in field space implicitly and to study the cutoff dependence of various quantities.

The idea is to consider nearby excited states relative to $\ket{\square_{L,M}}$, where, by nearby, we mean young tableaux states that differ from $\ket{\square_{L,M}}$ by a few boxes. This is depicted in equation \eqref{eq:smallfluct}.

\begin{equation}
\ket{R} = \begin{ytableau}
\ & && & & &L&  *(blue)\\
\ & && & & & \\
\ & && & & & \\
\none[\vdots] & \none[\ddots]&\none[\ddots]&  \none[\ddots]&  \none[\ddots]&\none[\ddots]&\none[\vdots]\\
 & && & & & \\
M & && & & &*(red) \\
*(green) 
\end{ytableau}\label{eq:smallfluct}
\end{equation}
The states $\ket{R}$ are those that differ from the reference state $\ket{\square_{L,M}}$ by adding (few) boxes in the corners marked by blue and green, and removing (few) boxes in the corner marked by red.
Each of these can be done independently at large $L,M$. What this means is that the Hilbert space of nearby states factorizes into 
\begin{equation}
{\cal H}_{\rm{nearby}} \simeq {\cal H}_{\rm{blue}}\otimes {\cal H}_{\rm{green}} \otimes {\cal H}_{\rm{red}} \label{eq:nearby}
\end{equation}
Now, each of ${\cal H}_{\rm{blue}}, {\cal H}_{\rm{green}} ,{\cal H}_{\rm{red}} $ can be written as a small Young diagram. For ${\cal H}_{\rm{blue}}$ and ${\cal H}_{\rm{green}}$ this is pretty obvious. For ${\cal H}_{\rm{red}}$ all we need to do is rotate the empty corners marked in red by $(180)^o$ and we get a proper Young diagram. 

Relative to the reference state, the Young diagrams with blue and green boxes will have norm one, and so will the diagram with antiboxes. This is special to the original theory being described with Young diagrams states that all have the same norm \footnote{This is easy to modify for theories of generalized free fermions as in \cite{Berenstein:2015ooa} because the norms will factorize for each different color of Young diagram. See also \cite{Koch:2016jnm} and references therein for how to compute energies of excitations in the  special case of ${\cal N}=4$ SYM. This is not essential for this paper.}. Now, we can do the Fourier transform from Young diagrams to a Fock space description with canonical  raising and lowering operators. This is essentially due  to the uniqueness of the relations discussed in section \ref{sec:gt}. For each of these, we can apply the results of the previous section and the appendix \ref{app:MN}. We find that 
\begin{equation}
{\blue\bra {R} a^\dagger_s \ket {\tilde R}= (-1)^{f_{\rm{hook}}} }
\end{equation}
as long as the skew  hook is an allowed transition from $\tilde R$ to $R$ with $s$ boxes.
We have a similar expression for the other colored operators. The one difference for the red boxes is that as the red diagram is growing, the original Tableaux is being chipped away, but this is done by a  skew hook that is at the interface of the red tableaux and the reference state: it is also an allowed skew hook of length $s$ for the complement of the red tableaux. These are the operators that we have identified as $b_s^{(i)\dagger}$ and $b_s^{(i)}$, and for the antiboxes these are the $c^{( i)\dagger}, c^{( i)}$ operators. Because the Hilbert spaces factorize we obtain the following exact commutation relations
\begin{eqnarray}
\ [b_s^{(i)}, b_t^{(j)\dagger}] &=& \delta^{ij} s \delta_{st}\\
\ [c_s^{(i)}, c_t^{(j)\dagger}] &= &\delta^{ij} s \delta_{st}
\end{eqnarray}
and all other commutators vanish. These commutation relations are true only when evaluated  in states that are sufficiently close to the reference state, so that the boxes and anti-boxes do not interfere with each other.  This is implicit in the full discussion, but it is worth emphasizing that these equations are not true for arbitrary excitations of the original system, only so in the effective field theory of nearby configurations. The extreme value where they could be right would be  halfway along the reference state sides, so it is only for labels that are less than  $L/2,M/2$, and total differences of energies that are much less than $LM$. Beyond this scale we would claim that the modes become related to each other and there are fewer independent states. These give relations between the 
operators at high order.  We call such transitions, where the number of states is reduced relative to the very naive counting of independent free fields, 
the stringy exclusion principle for the collective dynamics in the same sense as 
\cite{Maldacena:1998bw,McGreevy:2000cw}.

We will label the  nearby states as follows 
\begin{equation}
\ket{\square_{LM} +{\blue R_1} - {\red  R_{\tilde 1}} +{\green R_2}}\simeq \ket{R_1}\otimes \ket{R_2} \otimes \ket{ R_{\tilde 1}}
\end{equation}
in a way that makes it clear that we are adding and substracting Young diagrams from each corner as is appropriate to the nature of the corner.

From here, one can easily check that the following is true
\begin{equation}
a_s \ket{\square_{LM} +{\blue R_1} - {\red  R_{\tilde 1}} +{\green R_2}} =\sum_{\rm{hooks}} (-1)^{f_{\rm{hook}}} \ket{\tilde R}
\end{equation}
where $\tilde R$ differs from the original state by removing a skew hook of length $s$. These can only be removed from either of the $R_i$, or added to the $ R
_{\tilde i}$. These are the only places where small hooks can be subtracted (or added). This means that 
\begin{eqnarray}
a_s \ket{\square_{LM} +{\blue R_1} - {\red  R_{\tilde 1}} +{\green R_2}}&= &\sum (-1)^{f_{\rm{hook}}} \ket{\tilde R_1}\otimes \ket{R_2} \otimes {\ket R_{\tilde 1}}
+\sum\ket{ R_1}\otimes ( -1)^{f_{\rm{hook}}} \ket{\tilde R_2} \otimes {\ket R_{\tilde 1}}
\nonumber \\&+& \sum\ket{ R_1}\otimes \ket{R_2} \otimes  ( -1)^{f_{\rm{hook}}} {\ket {\tilde R_{\tilde 1}}}
\end{eqnarray}
This translates to 
\begin{equation}
a_s= b_s^{(1)}+b_s^{(2)} + c_s^{\dagger (1)}
\end{equation}
so we get what we want in equation \eqref{eq:Bog} up to a sign for $c^\dagger_s$. Following the discussion in section \ref{sec:LF} this sign can be changed by a particle-hole transformation, which is a symmetry of the algebra of the raising and lowering operators that we have constructed. The choice we make is a convention that needs to be established, and this is used to match better our geometric intuition. There is no deeper meaning to it.

We have proven what we set out to: the UV modes can be written as a linear superposition of  the collective modes of the configuration (the infrared modes) when acting on nearby states. Moreover, the collective modes ave canonical commutation relations.
It is the presence of these collective modes that lets us know for sure  that we have changed the topology. Their number dictates the number of edges (anti-edges).

\section{Topology from uncertainty measurements}

Now, we can further evaluate how non-classical these states are from the point of view of the original $a_s^\dagger, a_s$ oscillators, by computing quantities that appear in  \eqref{eq:unc}.
That is, we want to compute
\begin{equation}
\langle a_s a_s^\dagger\rangle_{LM}= \langle a^\dagger_s a_s+s\rangle_{LM}
\end{equation}

Using the MN rule, the $ \langle a^\dagger_s a_s\rangle$ are evaluated by counting  all the skew hooks of length $s$ (this is the same as the number of hooks of length $s$, see appendix \ref{app:MN}). One can easily see that each hook has its corner on a diagonal band, as in equation \eqref{eq:diag}. 
\begin{equation}
\ket{\square_{LM}}\simeq \begin{ytableau}
\ & && & & &  \\
\ & && & & &*(green) \\
\ & && & &*(green) & \\
\ & && &*(green) & &
\end{ytableau}\label{eq:diag}
\end{equation}
so we get that for low $s$ (that is, $s\leq \min(L,M)$)
\begin{equation}
\langle a_s a_s^\dagger\rangle_{LM}= 2 s
\end{equation}
For simplicity we will assume that $L\geq M$, so that the first place where things change is at $s=M+1$. At this point the diagonal bands are of constant length $M$. Things change again when 
we want very large hooks at $s=L$, where the available diagonal bands start shrinking.
We get the answer
\begin{equation} 
\langle a_s a_s^\dagger\rangle_{LM} = \left\{\begin{array} {ll} { 2 s}& \hbox{ if $s\leq M$}\\
{M+s} & \hbox{ if $L\geq s\geq M$}\\
{M+L} & \hbox{ if $L+M \geq s\geq L$} \\
{s} & \hbox{ Otherwise}\end{array}\right.
\end{equation}
We can also write this equation equivalently in terms of the average occupation per mode $N_s = \frac{a^\dagger_s a_s}{s}$. It is convenient to do this in two different (equivalent)  ways so that 
\begin{equation} 
\tilde N_s = \langle N_s +1 \rangle_{LM} = \left\{\begin{array} {ll} { 2   }& \hbox{ if $s\leq M$}\\
{M/s+1}  & \hbox{ if $L\geq s\geq M$}\\
(M+L)/s   & \hbox{ if $L+M \geq s\geq L$} \\
{1}   & \hbox{ Otherwise}\end{array}\right. \label{eq:tildeNs}
\end{equation}
and 
\begin{equation} 
\langle N_s  \rangle_{LM} = \left\{\begin{array} {ll} { 1   }& \hbox{ if $s\leq M$}\\
{M/s}  & \hbox{ if $L\geq s\geq M$}\\
(M+L)/s  -1  & \hbox{ if $L+M \geq s\geq L$} \\
{0}   & \hbox{ Otherwise}\end{array}\right.
\end{equation}
the number $\tilde N_s$ can be computed for more general multi-strip geometries for low $s$ using the Bogolubov transformation. It is given by
\begin{equation}
\tilde N_s = \langle a_s a_s^\dagger\rangle_{\rm{multi-stripe}}/s \simeq \frac 1 s \sum_i \langle b_s^{(i)}b_s^{(i)\dagger}\rangle_{\rm{multi-stripe}}= n_{\rm{edges}}
\end{equation}
while
\begin{equation}
N_s = \langle a_s^\dagger a_s \rangle_{\rm{multi-stripe}}/s \simeq \frac 1 s \sum_i \langle c_s^{(i)}c_s^{(i)\dagger}\rangle_{\rm{multi-stripe}}= n_{\rm{anti-edges}}
\end{equation}
We can draw a few consequences from these equations. 

First, since the numbers $N_s$ are generally of order one, the states that we have considered so far have very low uncertainty in the fluctuations, comparable in size to the usual quantum uncertainty of the ground state. That uncertainty is multiplied by an integer for low $s$. In this sense, they should be regarded as being classical, because uncertainties are of typical quantum size. This integer that we get is exactly the number of edges. In this sense, the number of edges is measurable in the size of the quantum fluctuations of the UV modes of the theory (those that are given a priori without any reference to the particular state). It is important that there are a large number of modes for which this number is the same. This means that with a simultaneous measurement of various quantities that commute with each other we can do enough statistics to compute the topology (without destroying the full information of the state, but already knowing that the state is a rectangular tableau). We measure the topology by census-taking and finding consensus. 

 This is where the size of $L,M$ actually start mattering. At roughly the same place where the stringy exclusion principe becomes important (at modes of order $L,M$) the numbers that effectively measure the topology of the state start changing.

In the large $L,M$ limit, the number $N_s$ becomes a continuous function of the rescaled parameters $x=s/M$ (or $s/L$) and interpolates smoothly between $n_{\rm{edges}}$ for $x<<1$ and $1$ for large $x$.
This can be interpreted as the effective number of edges at the scale associated with $x$. In this sense, the measurement of topology is energy dependent. Since the energy goes like $E\simeq L^2$ at least for the square tableau, the stringy exclusion energy scales as $L\simeq \sqrt E_{LM}$ and can be effectively very high.

\subsection{Coherent states of edge oscillators}

We can now consider general classical coherent states of the $b,c$ oscillators and as we have said above, we will think of these as new classical configurations. 
The coherent state is defined by the equations
\begin{equation}
(b_s-\vev {b_s}) \ket{\rm{Coh}_{LM}}=0 
\end{equation} 
and similar for the $c$. These belong the small Hilbert space for sufficiently small shifts (the tails at high occupation number  will have negligible probability). The subscript $LM$ is to indicate that this is a coherent state relative to the $b,c$ oscillators of the $LM$ state.

When we set out to compute the numbers $N_s$ as above, they will depend greatly on the properties of the state we pick. The occupation number  itself is {\em not robust} against taking general coherent states. What we mean by this is that we can change it by a large fraction (even in a lot of modes). However, consider the following operators that are shifted by a c-number
\begin{equation}
a_s -\vev {a_s} = \sum_i b_s^{(i)}- \vev{b_s^{(i)}} - \sum_j (c_s^{(j)\dagger}
-\vev{c_s^{(j)\dagger}})  \label{eq:shiftedmodes}
\end{equation}
Because the $b,c$ operators shifted by a c-number still satisfy the canonical commutation relations (the shift  is an automorphism of the algebra of $b,c$ operators),  then we can repeat the calculations we did before including these shifts.

We find that for these states the same computation that we did before holds with the shifted oscillators. That is, we find that
\begin{equation}
\frac 1s \, \vev {|a_s -\vev {a_s} |^2}_{\rm{Coh}_{LM}} =n_{\rm{edges}}
\end{equation}
We can measure the number of edges if we measure the uncertainty, not the number operator itself. The expectation value of the number operator is the same as the uncertainty if the shift vanishes, but the uncertainty is not in general the number operator. 

This means that  to measure $n_{\rm{edges}}$, we end up evaluating a non-linear function of the wave function. This is because the shift $\vev {a_s}$ depends on the state! In a certain sense this should not be a surprise.  
We already argued that the topology cannot be measured by an operator, because all coherent states relative to the trivial vacuum of the chiral boson are of the same topology. But a non-linear function of the wave function is not an operator measurement. It is something that can be computed in quantum mechanics, and that moreover can be recovered with a set of observations 
on the system with different variables that do not commute with each other: a polynomial function involving the number operator and the expectation values of the raising and lowering operators. Once we measure the expectation values of the field $\vev{\phi(\theta)}$ we can recover the shifts we need. Given these shifts, we can evaluate an effective number operator for the shifted variable. This is a protocol for measuring the topology. It just cannot be done with one single observation.
  
Notice also that at least in this example, even though we can measure the topology with a  few observations for low energy modes (from the UV point of view), measuring the values of $L,M$ themselves requires getting to the scale of the stringy exclusion principle (although the product $LM$ is readily measured by the energy). This is not expected to persist in more general circularly symmetric LLM geometries, because the corresponding generalized effective Bogolubov transformation has coefficients that depend on $L,M, N$  \cite{Koch:2008ah} (the details of this operation are beyond the scope of the present paper). This issue of indistinguishability of states based on simple measurements of the energy has also been alluded to in \cite{Balasubramanian:2006jt,Skenderis:2007yb}, but it is also important to understand that the problem persists if we only have the expectation value of the UV fields and the energy.  To reconstruct the classical geometry we need the expectation values of all of the $b,c$ modes, not just the $a$ modes. The $a$ modes are  the naive classical data needed on the boundary of AdS to specify the classical fields associated to $\phi$ in the bulk. The collection of  $a$ modes is the list of `single particle' supergravity modes that 
can be excited. This description does not take into account that the field $\phi$ is effectively multi-valued.
This has been discussed recently in the work by the authors of the present paper in \cite{Berenstein:2016mxt}, where it is argued that it gives an example where bulk reconstruction from classical boundary data fails (in the sense of \cite{Hamilton:2005ju,Kabat:2011rz}). It is not clear at this stage if this is special to LLM states, or applicable in more general settings (formally, it works in all cases where the states are the large $N$ limit of the extremal chiral ring \cite{Berenstein:2015ooa}). This needs to be investigated further.

Indeed, following \cite{Skenderis:2007yb} we can consider more general operators of the low momentum modes (of the UV theory) and we find
\begin{equation}
\vev{ a_s^\dagger a_{s_1} \dots a_{s_m}}_{LM} =0
\end{equation}
with $m>1$.
This result is easily obtained by using the Bogolubov transformation and Wicks theorem with respect to the vacuum of the $b,c$ oscillators and it's true as long as $s<M$.
What this means is that there are no obvious correlations
between the low lying modes,  nor any quantum correlation that can be used to measure $L,M$ directly without going to high energy.

Another question that can be asked is how close this $\ket{\square_{LM}}$ state is to a regular coherent state of the free chiral boson. A rough estimate  of a similar overlap to \eqref{eq:cohover} suggests that 
\begin{equation}
|\braket{\square_{LM}} {\vec \Lambda}|^2 \simeq  \frac 1 {L^M} \simeq \exp(-M \log L)
\end{equation}
which is exponentially suppressed at large $M,L$. This means that the new classical state is very far from any one {\em traditional classical state}. Alternatively, we can say that the state $\ket{\square_{LM}}$ can only be approximated well by an exponentially large superposition of classical states.

On taking double scaling limits in $L,M$, the new classical reference state is essentially orthogonal to all other standard classical states of the chiral boson theory and should not be thought of any longer as a Schr\"odinger cat state, but a classical state in its own right. This classical state has different topology than the standard classical states and this automatically implies that it has a different geometry.
This is the sense in which these new states represent different classical limits of the chiral free boson theory. In particular, the existence of fluctuation fields $\delta \phi^{(i)}, \delta \tilde \phi^{(i)}$ with a well defined action on the small 
Hilbert space of the corners defines a semi-classical quantization on top of the classical state and can be used to argue that one can do effective and unitary quantum field theory in the background of the state $\ket{\square_{LM}}$. 

\subsection{Beyond classical states}

If we consider a semiclassical state (the classical coherent of the $b,c$ oscillators  state with a  few excitations of the $b,c$ quanta), our consensus measurement can still be used to get to the topology. 
The point is that the few extra quanta can only affect a few of the modes for these measurements. The majority will have the same value of the uncertainty as before, and the majority vote will win.

Consider now a very different type of state: the thermal state at temperature $T>>1$. This is a stand-in for the typical state of high energy. A lot of the physics associated with this state in the LLM setup has been addressed in \cite{Balasubramanian:2005mg}. For us, the energy per mode is equal to $T$ for low enough modes, up to the cutoff scale (of order $T$), but the state also satisfies $\vev {a_s}=0$. This means that the occupation number per mode is $N_s \simeq T/s$ and is a rapidly varying function with $s$. For us this means that the effective topology that counts the number of edges in the geometry is varying rapidly with scale: the state should not be thought of as a classical geometric state with a fixed topology (the topology can not be measured and have a meaningful answer in the consensus part of the test).

We can also consider the triangular diagram of equation 131 in \cite{Balasubramanian:2005mg}, given by tableaux of the typical  form
\begin{equation}
\ket {R_{2 \rm{step}}} = \begin{ytableau}
\ & && & & &  & && & & & && && && && && &\\
\ & && & & & & && & & & && && && && &\\
\ & && & & & & && & & & && && && &\\
\ & && && && && & & & && && &\\
\ & && & & &  & && & & & && &\\
\ & && & & & & && & & & &\\
\ & && & & & & && & & \\
\ & && && & & && \\
\ & && & & & &  \\
\ & && & &  \\
\ & &&  \\
\ & 
\end{ytableau}\label{eq:2-step}
\end{equation}
with $L$ rows.
It is easy to check that there are no hooks with a length that is a multiple $3$.  That is, there are no hooks of length $3k$ for every $k$. So, we find that $\tilde N_{3k}=0$ for all $k$. This would suggest that the topology is that of 
the vacuum. However, when we consider other modes we find that 
\begin{equation}
\langle a_s^\dagger a_s \rangle \simeq \langle a_1^\dagger a_1 \rangle= L
\end{equation}
where $L$ is the number of hooks of length $1$, so again the energy per mode that is not a multiple of three is roughly fixed, but the effective number of edges varies wildly and we fail to find consensus. Now the number of edges is not even a smooth function for the rescaling parameter $x\simeq s/L$.
 It is only states that have few corners that 
are deemed sufficiently geometrical, to the extent that one can fix their topology by checking that $N_s$ is independent of $s$ for all $s$ that are small (below a suitable stringy exclusion principle energy).

These states fail to be classical also in that any attempt to produce oscillators like the $b,c$ oscillators fails because the naive stringy exclusion principle is very small: the equivalent of the red corners can at most remove one or two boxes before interfering with the addition of boxes in the anti-corners. In a sense, this is seen in that the edge of the tableaux is rough (very jagged) rather than a straight line.

These other examples show that the geometric states $\ket {\square_{LM}}$ are essentially characterized by having  low, but on average essentially constant, occupation number per low momentum (energy) mode of the UV theory. Also, the different modes must be very correlated to each other in order to be able to find fluctuation fields like the $b,c$ systems above that implement the required partial Bogolubov transformation from the collective dynamics to the UV modes.

\subsection{Geometries with folds}

\begin{figure}[ht]
\begin{center}
\begin{pspicture}(4,4)
\pscustom{
\gsave
    \psline(1,0)(1,1.8)
    \pscurve(1,1.8)(1.6,2.0)(1.2,2.2)(2.0,2.3)(3,1.8)
    \psline(3,1.8)(3,0)
    \psline(3,0)(1,0)
    \fill[fillstyle=solid,fillcolor=black]
\grestore
}
\psline(1,0)(1,4)
\psline(3,0)(3,4)
\psline[linestyle=dashed](1,2)(3,2)
\rput(1, -0.4){$ 0$}
\rput(3, -0.4){$ 2\pi$}
\end{pspicture}\label{fig:fold}
\end{center}
\caption{An example of a state with a number of edges that is dependent on $\theta$.}
\end{figure}
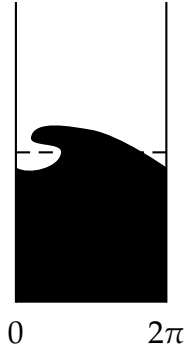

Thus far, we have primarily considered only a couple classes of states: coherent states and striped (i.e. LM) states.  These are nice because we know how to write them down in terms of oscillators and/or young tableaux, which we know how to deal with.  Of course, there are many other possible geometries in the full set of LLM states.  There is a general class that is particularly tricky, so we will touch on this class here.  These are the states that wrap in such a way that their number of edges that is a function of $\theta$.  For instance, the state drawn in figure \ref{fig:fold}.

If one were to do a local measurement of one of these states, then one would expect to find that the number of edges that varies with location.  Here, we would like to determine if there is a way to compute $n_{\rm{edges}}(\theta)$ from uncertainty computations. Notice that here we do not have an obvious candidate for the state in terms of Young tableaux. Instead we expect the state to be complicated in that basis.   The reason to say so is that such geometries also arise in the study of the $c=1$ matrix model \cite{Das:1995gd} and in that case they result from Hamiltonian evolution of more standard coherent states. But already coherent states superpose a lot of Young tableaux, and such evolution would scramble a lot of the phases between the different basis elements.
 The fact that such states exist point to the non-triviality of the classical limit in the $c=1$ matrix model.

The local commutation relation of our field operator can be computed using $\left[a_{k},a_{j}^{\dagger}\right]=k\delta_{k,j}$.  We find 
\begin{equation}
\left[\hat\phi\left(\theta\right),\hat\phi\left(\theta^{\prime}\right)\right]=\frac{i\left(2\pi\right)}{2}\left[\partial_{\theta}\delta\left(\theta^{\prime}-\theta\right)-\partial_{\theta^{\prime}}\delta\left(\theta^{\prime}-\theta\right)\right]
\end{equation}
As before with striped states, we will want to decompose our operators into pieces that act on each edge or anti-edge.  Here, we will do this as follows
\begin{equation}
\hat\phi\left(\theta\right)=\sum_{I=0}^{n_{\rm{edges}}(\theta)} \phi^{I}\left(\theta\right)-\sum_{J=0}^{n_{\rm{anti-edges}}(\theta)} \tilde{\phi}^{J}\left(\theta\right)
\end{equation}
where, as before, the tilde refers to operators that act on an anti-edge.  Also note the operator-denoting hats are dropped on the edge fields for notational convenience.  We see that unlike before, this decomposition is $\theta$-dependent.  That is, the range the indices run over depends on theta, just as the number of edges does.  

It is easy to see that these subfields should obey the same local commutation relations as the field $\hat \phi(\theta)$ with signs that depend on if the field is at a regular edge or an anti-edge
\begin{equation}
\left[\phi^{I}\left(\theta\right),\phi^{J}\left(\theta^{\prime}\right)\right]=\frac{i\left(2\pi\right)}{2}\left[\partial_{\theta}\delta\left(\theta^{\prime}-\theta\right)-\partial_{\theta^{\prime}}\delta\left(\theta^{\prime}-\theta\right)\right]\delta^{IJ}
\end{equation}
and
\begin{equation}
\left[\tilde{\phi}^{I}\left(\theta\right),\tilde{\phi}^{J}\left(\theta^{\prime}\right)\right]=-\frac{i\left(2\pi\right)}{2}\left[\partial_{\theta}\delta\left(\theta^{\prime}-\theta\right)-\partial_{\theta^{\prime}}\delta\left(\theta^{\prime}-\theta\right)\right]\delta^{IJ}
\end{equation}
These signs ensure that the commutation relations of the field $\hat \phi(\theta)$ are preserved.
One way to understand this is that the coefficient of the commutation relation of the $\phi$ field can be understood as the anomaly coefficient of the chiral boson. If we normalize this coefficient to one for $\phi$, then we find that the $\phi^I$ have anomaly equal to one, and the $\tilde \phi^J$ have anomaly minus one. The fact that there is one more $\phi^I$ than $\tilde \phi^J$ locally is the statement that the combined effective field theory of the $\phi^I, \tilde \phi^J$ has the same anomaly as the UV theory of $\phi$. In this sense, the effective field theory near our reference states has anomaly matching between the UV representation and the collective degrees of freedom.

Now, we will call $\delta \phi$ the quantum field relative to the background field. That is, we call
\begin{equation}
\delta \phi^I =\phi^I-\vev{\hat \phi^I}
\end{equation}
and similar for $\tilde \phi^J$ and $\hat\phi$. 

To proceed further, we can consider breaking our field into raising and lowering operator pieces for each of the indices $\delta\hat\phi=\delta\phi^+ +\delta\phi^-$ as was done in the multi-edge
regular  geometry by Fourier transforms. Our problem in this case is that we do not know the vacuum state. In the standard multi-strip geometries this is achieved by requiring that in the translationally invariant multi-strip geometry, the Fourier modes of the  operators for the  edge modes are raising/lowering operators depending on if the mode adds or subtracts energy relative to the reference state, with a provision that changes the assignment for anti-edges relative to edges. This is demonstrated for the strip geometries with the explicit construction based on the Young diagram representation of states.
Here we have to choose a vacuum, because the multi-fold geometries are
not translation invariant. We would still want the raising/lowering operators of the field $\phi$ to be locally in the same decomposition as for multi-edge states and the lowering operators of the $\hat \phi^I, \hat\phi^J$ to act trivially on the reference states. 

 Combing this with our edge/anti-edge decomposition, we make the approximation that
\begin{equation}
\delta\phi^+ (\theta)\simeq \sum_{I=0}^{n_{\rm{edges}}(\theta)} \delta\phi^{I+}(\theta) - \sum_{J=0}^{n_{\rm{anti-edges}}(\theta)}  \delta\tilde\phi^{J-}(\theta)
\end{equation}
where we know that the lowering operator on anti-edges adds energy, so should be grouped with the raising operator piece. We will call this the Local Multi-Edge approximation (LME approximation).
 Notice that this is a statement about raising and lowering operators relative to the state, so we have subtracted the classical value of the fields and only included the fluctuation piece. This is important because we want to compute the uncertainty in the measurement relative to the classical state (we did a similar step on the multi-edge geometries when we argued that the shift by the field vevs is an automorphism of the Weyl algebra).

Notice that the splitting into raising and lowering operators is not really local. However, it can be approximately local under some circumstances: as long as the region over which the splitting is done is much larger than the typical wavelength of the fields considered, the approximation makes sense. This means that the vacuum is very similar to the multi-edge setups locally on length scales that are small relative to where the folds begin and end. The finite size corrections should depend on the separation between the folds.

To proceed further, we start by convoluting the field operator $\hat{\phi}$ with a test function $f(\theta)$ to build a new operator
\begin{equation}
\hat\phi_{f}=\frac{1}{2\pi}\int_{0}^{2\pi}f\left(\theta\right)\hat\phi\left(\theta\right) d\theta
\end{equation}
Our goal is to use these operators to estimate uncertainties and correlators with particular choices of $f$. The role that $f$ plays is to select the correct wavelengths of the modes. If we are worried about the details where the folds are located, we can take $f$ to be nearly vanishing near these regions, but we will not do so here.

It will turn out that the operators given by $\hat\phi_f,\hat\phi_g$ with $f=e^{im\theta}$ and $g=e^{-ik\gamma}$ is just what we need.  Notice that, in a more familiar form, this is simply
\begin{equation}
\hat\phi_{e^{im\theta}}\hat\phi_{e^{-ik\gamma}}=a_m a^{\dagger}_k
\end{equation}
Now, let's try to compute the expectation value of this for a folded state, which we will call $\ket{\rm{fold}}$.
\begin{equation}
\bra{\rm{fold}} \hat\phi_{e^{im\theta}}\hat\phi_{e^{-ik\gamma}}\ket{\rm{fold}} =\frac{1}{(2\pi)^2} \int d\theta\, d\gamma\, e^{im\theta} e^{-ik\gamma} \bra{\rm{fold}} \hat\phi (\theta)\hat\phi (\gamma)\ket{\rm{fold}}
\end{equation}
We can break this up into its raising and lowering pieces
\begin{equation}
=\frac{1}{(2\pi)^2} \int d\theta\, d\gamma\, e^{im\theta} e^{-ik\gamma} \bra{\rm{fold}} \left(\phi^+ (\theta)+\phi^- (\theta)\right) \left(\phi^+ (\gamma)+\phi^- (\gamma)\right)\ket{\rm{fold}}
\end{equation}
Because of our choice of sign for the exponentials of $f$ and $g$, if we actually write out the raising and lowering pieces, we find that upon integration, $\phi^+ (\theta)$ and $\phi^- (\gamma)$ will vanish, leaving
\begin{equation}
=\frac{1}{(2\pi)^2} \int d\theta\, d\gamma\, e^{im\theta} e^{-ik\gamma} \bra{\rm{fold}} \phi^- (\theta) \phi^+ (\gamma)\ket{\rm{fold}}
\end{equation}
We can now expand this in terms of pieces that act on each edge and anti-edge
\begin{equation}
\begin{split}
=\frac{1}{(2\pi)^2} \int d\theta\, d\gamma\, e^{im\theta} e^{-ik\gamma} \bra{\rm{fold}} \left( \sum_{I=0}^{n_{\rm{edges}}(\theta)} \phi^{I-} (\theta) -\sum_{J=0}^{n_{\rm{anti-edges}}(\theta)}  \tilde\phi^{J+} (\theta) \right) \\
 \times \left( \sum_{I=0}^{n_{\rm{edges}}(\theta)} \phi^{I+} (\gamma) -\sum_{J=0}^{n_{\rm{anti-edges}}(\theta)}  \tilde\phi^{J-} (\gamma) \right)\ket{\rm{fold}}
\end{split}
\end{equation}
Here, if we zoom in on any small region in $\theta$, we imagine that the folded state will look like a striped state.  So, locally, we expect that these operators will act on the folded state, just as they would on a young tableaux state, giving
\begin{equation}
\simeq\frac{1}{(2\pi)^2} \int d\theta\, d\gamma\, e^{im\theta} e^{-ik\gamma} \bra{\rm{fold}} \sum_{I=0}^{n_{\rm{edges}}(\theta)} \left(  \phi^{I-} (\theta) \phi^{I+} (\gamma)\right) \ket{\rm{fold}}
\end{equation}
We notice now that we can write this as
\begin{equation}
=\frac{1}{(2\pi)^2} \int d\theta\, d\gamma\, e^{im\theta} e^{-ik\gamma} \bra{\rm{fold}} \sum_{I=0}^{n_{\rm{edges}}(\theta)} \left[  \phi^{I} (\theta), \phi^{I} (\gamma)\right] \ket{\rm{fold}}
\end{equation}
This, again, only works in this way because of the sign of the exponentials in $f$ and $g$, causing some of the terms to vanish upon integration. Others vanish because as you zoom in the state will look like a vacuum young tableaux state.  Finally, we can use the known commutation relation, giving
\begin{equation}
=\frac{1}{(2\pi)^2} \int d\theta\, d\gamma\, e^{im\theta} e^{-ik\gamma} \bra{\rm{fold}} \sum_{I=0}^{n_{\rm{edges}}(\theta)} \frac{i\left(2\pi\right)}{2}\left[\partial_{\theta}\delta\left(\gamma-\theta\right)-\partial_{\gamma}\delta\left(\gamma-\theta\right)\right]\delta^{II} \ket{\rm{fold}}
\end{equation}
Again, remember the decomposition is location dependent, with the index $I$ running over each edge at each location.  We see then that summing over $\delta^{II}$ simply gives the number of edges at each location.
\begin{equation}
=\frac{1}{(2\pi)^2} \int d\theta\, d\gamma\, e^{im\theta} e^{-ik\gamma} \bra{\rm{fold}} \frac{i\left(2\pi\right)}{2}\left[n_{\rm{edges}}(\gamma)\partial_{\theta}\delta\left(\gamma-\theta\right)-n_{\rm{edges}}(\theta)\partial_{\gamma}\delta\left(\gamma-\theta\right)\right] \ket{\rm{fold}}
\end{equation}
Integrating by parts off the delta functions onto the exponentials gives
\begin{equation}
=\frac{(m+k)}{2(2\pi)} \int d\theta\,  e^{i(m-k)\theta}  n_{\rm{edges}}(\theta) 
\end{equation}
Recalling that the original operator was equivalent to $a_m a_k^{\dagger}$, we find
\begin{equation}
\bra{\rm{fold}}a_m a_k^{\dagger}\ket{\rm{fold}}\simeq\frac{(m+k)}{2(2\pi)} \int_0^{2\pi} d\theta\,  e^{i(m-k)\theta}  n_{\rm{edges}}(\theta) 
\end{equation}
So, we see that these matrix elements give the Fourier transform of the number of edges.  Notice that if we had done the same computation starting with $f=e^{-im\theta}$ and $g=ik\gamma$, everything would have followed similarly, leading to
\begin{equation}
\bra{\rm{fold}}a_m^\dagger a_k\ket{\rm{fold}}=\frac{(m+k)}{2(2\pi)} \int_0^{2\pi} d\theta\,  e^{i(m-k)\theta}  n_{\rm{anti-edges}}(\theta) 
\end{equation}

The LME approximation also produces the following two results:
\begin{equation}
\vev{a_k^\dagger a_m^\dagger}_{fold}= \vev{a_k a_m}_{fold}=0
\end{equation}
which result from no mixing between raising and lowering operators of the same collective fields.

As a consistency check, we get that
\begin{equation}
\bra{\rm{fold}}a_m^\dagger a_k\ket{\rm{fold}}= (\bra{\rm{fold}}a_m a_k^\dagger\ket{\rm{fold}})^*
\end{equation}
and that 
\begin{equation}
\bra{\rm{fold}}a_m^\dagger a_k\ket{\rm{fold}}=\bra{\rm{fold}} a_k a_m^\dagger\ket{\rm{fold}}
\end{equation}
for $m\neq k$, which follows because $n_{anti-edges} = n_{edges}-1$, so they have the same Fourier coefficients for  the non-constant mode.

As a check of this proposal, we can evaluate it in the case of the multi-strip geometries. We would then expect this to agree with our previous result for young tableaux states. Here, we would find a constant number of edges, giving
\begin{equation}
\bra{\rm{YT}}a_m a_k^{\dagger}\ket{\rm{YT}}=\frac{(m+k)}{2(2\pi)} \int_0^{2\pi} d\theta\,  e^{i(m-k)\theta}  n_{\rm{edges}} =m \delta_{m,k} n_{\rm{edges}} 
\end{equation}
as computed previously.  And similarly for the anti-edge calculation.

The  upshot of the computation is that we can argue that $n(\theta)$ is encoded in the expectation values of $\bra{\rm{fold}}(a_m-\vev{a_m})( a_k^{\dagger}-\vev{a_k^\dagger})\ket{\rm{fold}}$. These are non-trivial correlations between the modes of the field, valid for modes $m$ that are larger than the inverse of the separation between the folds (measured with respect to $2\pi$). A pre-requisite for a state to have a non-trivial topology is that 
$n(\theta)\neq 1$ someplace. This means that at least one of these correlators (generalized uncertainties) is different than the ones in the vacuum. If $n(\theta)$ is non-constant, then some correlators 
with $m\neq k$ must be non-vanishing.

A test to determine if a state could in principle be geometric is that since the Fourier coefficients only depend on $m-k$, there has to be a large degree of consistency between the correlations of the different modes.
That is, even for semiclassical states with multi-edges, one can build a consensus measurement of $n(\theta)$: most of the modes that should be identical (with the proper normalization dependent on $m,k$) and determine the consensus measurement of the Fourier transform of $n(\theta)$, but again, it is not a single operator measurement on the Hilbert space of states that produces the result.

At this stage  we do not have a clear understanding on how to incorporate the finite size corrections. They should be small for short wavelengths. We should also expect that there are additional small corrections to the LME approximation related to how the transitions from one value of $n$ to another are handled. At this stage the LME approximation is not systematic and is used instead to set a benchmark for how such correlations should behave. This issue needs to be studied further and is beyond the scope of the present paper.

 Notice that eventually we also run into trouble at high frequencies, because we expect there to be a local bound on the momentum. In the Young tableaux case this occurs when the hooks at different corners start interfering with each other. A similar phenomenon should appear locally at each location of $\theta$ and should indicate a local stringy exclusion principle.

\section{Entanglement measurements of topology}

The particular model we have studied, as a quantum mechanical theory is a free field theory. The Hilbert space is a Fock space with a natural set of raising and lowering operators (a set of commuting algebras), and we can think of this Fock space as having a canonical product structure
\begin{equation}
{\cal H}_{\rm{tot}} \simeq {\cal H}_1 \otimes {\cal H}_2 \otimes \dots \otimes {\cal H}_\infty
\end{equation}
where the Hilbert space basis is defined by kets that have have all but finitely many oscillators in the ground state. The Hilbert space itself is the $L^2$ completion of states made from this basis. We use the monicker ${\cal H}_\infty$ to indicate the infinite product for all sufficiently large ${\cal H}_n$. Because we have this canonical factorization structure, it is possible to take a pure state in ${\cal H}$ and find a reduced density matrix for each of the modes, $\hat\rho_1$, $\hat\rho_2$ etc,  or for more modes grouped together  $\hat\rho_{ij}, \hat\rho_{ijk}$ etc. It is the structure under this class of factorizations   that we will try to understand for the states studied in the previous section.
The total Hamiltonian is 
\begin{equation}
\hat H = \sum _s  a_s^\dagger a_s = \sum _s  s \hat N_s
\end{equation}
We will aim to measure the entanglement entropies for each of these density matrices to characterize the state. The idea is to understand the entanglement structure of the state and to try to use that structure to measure the topology of the geometry etc. In a sense, this is putting to flesh the idea that entanglement can be the source of geometry a la Van Raamsdonk \cite{VanRaamsdonk:2010pw}. 
It is unclear how to implement this calculation as a calculation of entanglement entropy in gravity by a minimal surface \cite{Ryu:2006bv}.  This is mainly  studying the momentum space entanglement in 
the quantum field theory (in \cite{Balasubramanian:2011wt} this information is related to the Wilsonian effective action  ).

Consider a state that is an energy eigenstate of the full Hamiltonian (e.g. $\ket{\square}_{LM}$, which has energy $E= ML$).  We can decompose any such state, singling out the $j^{\rm{th}}$ oscillator, to find the reduced density matrix.  If we do this, we can write the state as follows
\begin{equation}
\ket \psi = \sum_n \xi_n \ket n_j \otimes \ket{w_n} 
\end{equation}
where $\ket{w_n}$ includes occupation information about all the other oscillators. We can now use
\begin{equation}
 \hat H \ket \psi= \sum_n \xi_n \hat H \ket n_k \otimes \ket{w_n} = \sum_n \xi_n (n  + \sum_{k\neq j}  k \hat N_k) \ket n_j \otimes \ket{w_n} =E_{\psi} \ket \psi
\end{equation}
We see the unevaluated part of the Hamiltonian  $\sum_{k\neq j}  k \hat N_k$ acts only on $\ket w_n$.  When we dot this with $_j\bra {\tilde n}$, we find that the states $\ket { w_n}$ are different eigenstates of a Hermitian operator.
\begin{equation}
 \sum_{k\neq j}  k \hat N_k \ket {w_{\tilde n}} =(E_{\psi}-\tilde n)\ket {w_{\tilde n}}
\end{equation}
As such, they are orthogonal and can be made orthonormal. The reduced density matrix for $\rho_j$ can be obtained from tracing with any orthonormal basis and we choose the $\ket w_n$ themselves.
We find then that the reduced density matrix is diagonal in the energy basis
\begin{equation}
\hat \rho_j = \sum_n |\xi_n|^2 \ket{n}_j \,_j\bra n= \sum_n p_n \ket{n}_j \,_j \bra n
\end{equation}
where the $p_n$ are the probabilities to find the state $\ket{\psi}$ has the $j^{\rm{th}}$ oscillator with occupation $n$ upon a measurement of $\hat N_j$.
Our goal is to compute the coefficients $1\geq p_n \geq 0$, and therefore the entropy
\begin{equation}
s_j = -\sum_n p_n \log p_n
\end{equation}
for each mode $j$.

We will start by considering the state $\ket \bigtriangleup_n$, for simplicity. The idea is that we can compute quantities like 
\begin{equation}
_n\vev {\Delta | a_1^{\dagger k} a_1^k | \Delta}_n = 1
\end{equation}
if $k \leq n$. Otherwise it vanishes. This also can equivalently be written as $a_1^{n+1}\ket \bigtriangleup_n=0$ because there are no states with negative energy. This in particular implies that $p_k=0$ for $k>n$ so we only have to determine finitely many of the $p_k$. 
This results trivially from applying the Murnaghan-Nakayama rule in operator form repeatedly. 

Because this number is only made of expectation values of operators that act on ${\cal H}_1$, the complete information is encoded in the density matrix $\hat \rho_1$. This equation can be written as
\begin{equation}
\hbox{tr} (\hat \rho_1 a_1^{\dagger k} a_1^k) =1 \label{eq:constraints}
\end{equation}
One can easily evaluate the matrix elements of the operator in the number basis
\begin{equation}
a_1^{\dagger k} a_1^k \ket j =  \frac {j !}{(j-k)!} \ket j 
\end{equation}
The equation \eqref{eq:constraints} reads
\begin{equation}
\sum_{j=0}^n p_j   \frac {j !}{(j-k)!} = 1
\end{equation}
 for all $0\leq k\leq n$ and together with $p_j=0$ for $j>n$ is a complete linear system so all of the $p_j$ can be determined.  The first of these equations is $\hbox{tr}(\hat\rho_1)=1$. The second one is 
 $\vev {\hat N_1}=1$. We can preform the same computation for the other modes, say the $j^{\rm{th}}$.  Computing $\hat \rho_j$ leads to
 \begin{equation}
\sum_{l=0}^{[n/k]} p^{(j )}_l   \frac {l !}{(l-k)!} = 1/j^k
\end{equation}
which can be solved to give us the probabilities, $p^{(j )}_l$ (the probability for the $j^{\rm{th}}$ oscillator to have occupation $l$).

Now, let us consider $\ket{\square_{LM}}$ states. Again, the simplest way to understand reduced density matrices starts by understanding that the density matrices $\hat \rho_j$ are diagonal in the occupation number basis. So, our job is to compute 
\begin{equation}
\vev {a_j^{\dagger k} a_j^k}_{LM} = \hbox{tr} (\hat \rho_j a_j^{\dagger k} a_j^k) 
\end{equation}
and to use these equations to compute the elements of $\hat \rho_j$. It will prove helpful to use our partial Bogoliubov transformation  \eqref{eq:Bog}. From
\begin{equation}
a_j= b^{(1)}_j+ b^{(2)}_j- c_j^{(1)\dagger}
\end{equation}
where, recall that each piece now acts on a particular edge or anti-edge.  The computations can be performed for small enough $k$ (below the stringy exclusion principle) by using Wick contractions of the $c$ oscillators only
\begin{equation}
\vev {a_j^{\dagger k} a_j^k}_{LM}\simeq \;_{b,c} \bra 0\left(c_j^{(1)} \right) ^k \left(c_j^{(1)\dagger}\right)^k \ket 0_{b,c}
\end{equation}
where we explicitly write the $LM$ state as the vacuum state for the $b_j^{(I)}$ and $c_j^{(J)}$ oscillators. This can be computed and gives
\begin{equation}
 _{b,c}\bra 0\left( c_j^{(1)} \right) ^k \left( c_j^{(1)\dagger}\right)^k \ket 0_{b,c} = k!j^k \label{eq:uppbound}
\end{equation}
for small enough $k$. At larger $k$ eventually we find that acting with too many $a_j$ kills the state, and that generically the equation \eqref{eq:uppbound} is an upper bound for the quantities we want.
For the case of more edges, we get a similar answer 
\begin{equation}
 _c\bra 0\left(\sum_{I=0}^{n_{\rm{edges}}} c_j^{(I)} \right) ^k \left(\sum_{J=0}^{n_{\rm{anti-edges}}}  c_j^{(J)\dagger}\right)^k \ket 0_c = k! j^k n^k_{\rm{anti-edges}}
\end{equation}
This is because now there are $ n_{\rm{anti-edges}}$ $c$ fields. We write this set of equations as
\begin{equation}
\sum_{J=0}^{n_{\rm{anti-edges}}}  p_j   \frac {j !}{(j-k)!} =  k!  n^k_{\rm{anti-edges}} \label{eq:denelemsum1}
\end{equation}
Notice, we have the same equation for each oscillator. This can be seen easily from writing it in terms of canonical normalization oscillators, rather than oscillators with the field theory normalization that was computed in equation \eqref{eq:oscnorm}, and this is exact below the stringy exclusion principle. After the stringy exclusion principle is crossed we do not understand sufficiently well how the
values on the right hand side taper off and eventually vanish. 

This suggests that a good approximation to the entropies can be had if we don't impose the stringy exclusion principle at all, assuming that for sufficiently high values of $k$ the elements of the density matrix are already sufficiently suppressed that their contribution to the entanglement entropy is negligible when we make them slightly smaller. 

To solve \eqref{eq:denelemsum1} when we don't have bounds, we consider a (thermal) partition function given by
\begin{equation}
Z[x] = \sum_{j=0}^\infty  x^j
\end{equation}
where we are saying $p_j\propto x^j$. Consider acting with $x^k \partial_ x^k$ on $Z[x]$. We get
\begin{equation}
x^j \partial_x^k Z[x] = \sum_{j=0}^\infty x^k j(j-1) \dots (j-k+1) x^{j-k} = \sum_{j=0}^\infty  \frac {j!}{(j-k)!} x^{j}
\end{equation}
which is of the form we want.
To normalize the answer, we should divide by the sum of the non-normalized $p$ and we get that the following should be true
\begin{equation}
\mathcal{N} \sum_{j=0} p_j   \frac {j !}{(j-k)!} = Z[x]^{-1}x^k \partial_x^k Z[x] =  k!  n^k_{\rm{anti-edges}}
\end{equation}
with $\mathcal{N}$ the normalization.  Now, the sums are straightforward to compute, and give
\begin{equation}
 Z[x]^{-1}x^k \partial_x^k Z[x] = (1-x) x^k  k! \frac{1}{(1-x)^{k+1}} = k!\left( \frac{x}{1-x}\right)^k
\end{equation}
so that the equations are solved if
\begin{equation}
n_{\rm{anti-edges}}= \frac x{1-x}
\end{equation}
or equivalently
\begin{equation}
x= \frac{n_{\rm{anti-edges}}}{1+n_{\rm{anti-edges}}}
\end{equation}
We should take this to mean that the moments of the density matrix $\hat \rho_j$ are identical to those of a thermal density matrix for all low enough moments. This means that the two density matrices should
be very similar. In the thermal density matrix the large $p_k$ are exponentially suppressed. Indeed, the thermal density matrix is the one that maximizes the entropy if we fix the number operator $\vev {\hat N}$, so at worst we get an upper bound for the entanglement entropy mode per mode.

Let us try to explain this. At first sight, this seems strange. The reason why having an approximately thermal density matrix is  strange is that the state we started with $\ket{\square_{LM}}$ has a supergravity dual that is free from horizons. However,  notice that the method of computing the moments based on \eqref{eq:Bog} starts from a partial Bogolubov transformation where three modes $b^{(1,2)}$ and $c^1$ are mixed. We can find the other two linear combination of modes that gives rise to a full Bogolubov transformation. Use for example
\begin{equation}
d^\dagger_s = \frac 1{\sqrt 2} ( b_s^{(1)\dagger}- b_s^{(2)\dagger}), \;\;\;\; e^\dagger_s = \sqrt 2 c^\dagger _s - \left(\frac {b_s^{(1)}+b_s^{(2)}}{\sqrt 2}  \right ) 
\end{equation}
We can compute the moments of the distribution in two different ways. In one, we use the oscillator basis $b, c$ and the vacuum $\ket 0_{bc}$ to get the answer. In the other way, we integrate out the fields $e, f$ in the vacuum $\ket 0_{bc}$ and compute a density matrix for the $a$ modes directly. This Bogolubov transformation  generically produces a squeezed state, and integrating the $d,e$ modes gives rise to a Gaussian density matrix  \footnote{ This is explained for example in  \cite{Asplund:2015osa} and references therein, where this statement follows from a simple generalization of eq. 40,  and see also appendix B}. This Gaussian density matrix is not pure, but thermal, as is typical in gravitational computations \cite{Hawking:1974sw}. 
Notice that this density matrix is not computed directly in the full Hilbert space ${\cal H}$, but is rather computed in the small (nearby) Hilbert space from equation \eqref{eq:nearby}
\begin{equation}
{\cal H}_{\rm{nearby}} \simeq {\cal H}_{\rm{blue}}\otimes {\cal H}_{\rm{green}} \otimes {\cal H}_{\rm{red}} 
\end{equation}
which is readily generalized to the other more general multi-edge solutions.

What is interesting is that the factorization of the full Hilbert space induces a factorization in the nearby Hilbert space. This is because the algebra of observables $a^\dagger, a$ acts simply on $H_{\rm{nearby}}$.
That is, the operators do not  take the states out of $H_{\rm{nearby}}$ when we use simple observables made of few such $a$ below the stringy exclusion principle. 

Indeed, this factorization structure and the corresponding quasi-thermal structure of the state persists even when we consider coherent states of the $b,c$ modes. These can be obtained by a shift in the
algebra of the $b,c$ fields. This is an automorphism of the algebra. Similarly, the answer is simple in terms of the shifted modes 
$
a_s -\vev {a_s} = \sum b_s^{(i)}- \vev{b_s^{(i)}} -(c_s^{(i)\dagger}
-\vev{c_s^{(i)\dagger}})
$ that appeared in equation \eqref{eq:shiftedmodes} and similar for the $d,e$ modes. Integrating out the $d, e $ modes or the shifted $d,e$ modes gives the same result. The density matrix for the 
shifted $a$  modes will still be Gaussian, and the entanglement entropy of $\hat \rho_1$ does not change: it is independent of the choice of basis in which we perform the computations. This entropy will only depend on the expectation value of the (shifted) occupation number.
Since this will be roughly the same for all modes below the stringy exclusion principle, we find that after a straightforward computation
\begin{equation}
 s_i= (n_{\rm{edges}}\log(n_{\rm{edges}}))- n_{\rm{anti-edges}} \log(n_{\rm{anti-edges}})
\end{equation}
and we can measure $n_{\rm{edges}}$ by consensus of the entanglement entropies of the different modes. This is not too different from the analysis in the previous section.

An additional interesting fact we observe is that in the small Hilbert space, the different $a$ modes arise from integrating out different $d, e$ modes. The density matrices $\hat \rho_{ij}$, $\hat \rho_{ijk}$ etc. are factorized! This means that there is no mutual information between the different modes $a_s$ for sufficiently small $s$. One expresses this by saying that 
there is no  entanglement between the  modes $a_s$. The entanglement occurs between these low-momentum modes and very high momentum modes (at or beyond the stringy exclusion principle).
We can say that these geometries arise from a  special kind of UV-IR entanglement, but that there is no IR-IR entanglement contributing to the geometry. 

Indeed, to the naive classical holographic observer that can only measure simple combinations of the low $a$ modes, the information of the $d,e$ modes is almost completely hidden (except for the total energy and that they act to purify the state).  This suggests that for these backgrounds the reconstruction procedure \cite{Hamilton:2005ju, Kabat:2011rz} will fail to construct excitations of the $d,e$ modes, which are clearly contributing to bulk fields
of supergravity modes. The precise way in which this could happen in these geometries is very interesting but  is also beyond the scope of the present paper. A partial answer has been discussed in 
\cite{Berenstein:2016mxt}.

Another interesting calculation to do is to understand how big an overlap between a state like $\ket{\square_{LM}}$ and a general coherent state of the free field theory can be. The best way to estimate this is to realize that the mode per mode entropy bounds how much overlap there is mode per mode. As coherent states are factorized between the modes, we get rather easily that 
\begin{equation}
|\braket{\square_{LM}}{{ \rm{Coh}}}|^2 << \exp(-\sum_{i=1}^L s_i) \sim\exp(-L s_1)
\end{equation}
which is exponentially suppressed in the dynamically generated cutoff. This means that when we think of the states $\ket{\square_{LM}}$ and other multi-edge geometries they are always an exponentially large superposition of coherent states around the trivial geometry. This is important in other setups with black holes \cite{Almheiri:2016blp}.

\subsection{Entanglement measurements of Geometries with folds}

We can do a similar analysis of the entanglement entropy, mode per mode, for the geometries with folds. On a first pass, because of the LME approximation, for a single mode we find that the uncertainty of each mode is given by
\begin{equation}
\vev{(a_k -\vev{a_k})(a^\dagger_k-\vev{a_k^\dagger})} = \frac k{2\pi} \int_0^{2\pi} n(\theta) d\theta =  k n_{av} 
\end{equation}
with the average number of edges denoted by $n_{av}$. Moreover we have that  $\vev{(a^\dagger_k-\vev{a_k^\dagger} )^2}=0$.

The state for mode $k$ results from a Bogolubov transformation of the collective modes with a shift. Integrating out the 'orthogonal' modes, the result is a regular thermal state for the shifted mode $k$. Such a thermal state is completely determined by the $n_{av}$. The entanglement entropy of such a mode is
\begin{equation}
s_k= (n_{av})\log(n_{av})- (n_{av}-1) \log(n_{av}-1)
\end{equation}
so again, mode per mode, the entanglement entropy of a single mode is constant and measures the average number of edges over $\theta$. This again makes it possible to measure
$n_{av}$ by a  consensus measurement on semiclassical states (those that differ from a folded geometry by a finite number of collective excitations).

Now, we also find that the LME approximation implies that there are non-trivial correlations between the different modes. Therefore the density matrix does not factorize anymore. This implies that there is mutual information between 
the different collections of modes that one can produce. This is determined uniquely by the Fourier transform of $n(\theta)$. Studying the detailed structure information of all of these correlations is beyond the scope of the present paper.

We should note that as is usual with entanglement entropy (mode per mode) and mutual information between modes, one can do unitary transformations on each of the subfactors without changing 
the answer. A particularly interesting unitary is $\exp(i \alpha_s  a_s^\dagger a_s/ s)$, which rotates the (shifted)  $s$ oscillator by a phase $\alpha _s$. These unitaries preserve the entanglement entropy mode per mode, but they modify the correlators as follows
\begin{eqnarray}
\vev{a_m a^\dagger_k} \to \vev{a_m a^\dagger_k} \exp(i (\alpha_k-\alpha_m))
\end{eqnarray}
etc. Now the phase of the correlation $\vev{a_m a^\dagger_k} $ should match the phase of $ \vev{a_{m+w} a^\dagger_{k+w}} $, but generically these unitaries do not do that, however, the mutual information of the factorization is not changed. This means that a simple unitary operator destroys the 'uncertainty measurement' of geometry without changing the entanglement entropy measurement. This indicates that the entanglement entropy measurement of topology is much weaker than the uncertainty measurement of topology: states that are (clearly) non  geometric would pass the entanglement entropy consensus measurement of topology, but not the uncertainty measurement tests. 
 
Notice also that if $n_{av}=0$, then the state is a coherent state: a minimum uncertainty packet in each of the Hilbert spaces and the entropy for each sub-Hilbert space for a mode $k$ or any collection of them is zero. In general this implies that the other correlators should vanish. This means that there are inequalities between the generalized correlators that need to be satisfied. Violations of these inequalities should in general lead to `negative probabilities': violations of unitarity. Some examples of such violations can be understood in generalized half-BPS solutions of type IIB supergravity that have closed time-like curves \cite{Caldarelli:2004mz,Milanesi:2005tp}. Studying these inequalities would also be very interesting, but again, this is beyond the scope of the present article.

\section{Discussion}

We have discussed topology changes in the set of LLM geometries and their dual realization. We focused on a particular simple limit where the full mini-superspace of half BPS geometries is quantum mechanically given by a free theory: the free chiral boson. 

We found that since the coherent states of the free chiral boson are overcomplete, any state in the quantum theory can be written as a superposition of this class of states. These coherent states all have the same `trivial topology' as the vacuum. It is curious that one can construct states with different topology (also known as bubbling solutions) just by superposing states with a  trivial topology, and the new topologically distinct states are macroscopically very different from any of the states that we are superposing. The overlaps between the new state and the elements of the overcomplete basis of  coherent states are all exponentially suppressed.    We state this by saying that topology changes can be triggered by superposition. This is a superposition of an exponentially large number of states, not a naive Sch\"odinger cat state that superposes just two distinct geometries.

A simple, yet deep, consequence  of this fact is that topology can not be measured by a single operator measurement. That is, the Hilbert space of states does not admit an orthogonal decomposition into different topological types. So if topology cannot be measured by an operator, it seems reasonable that finer geometric information might suffer the same fate. This puts into question how (a seemingly unitary) effective field theory of gravity can be compatible with this non-operator property.

To understand the physics of our example, it became  important to understand the physical states in more than one basis of vectors for the Hilbert space of states.
The set of wave functions can be written either in an oscillator basis for the chiral modes, or in terms of a Young Tableaux basis (the free fermion realization of matrix models). We carefully developed the 
dictionary between them, which is a generalized Fourier transform. An important result is that the set of raising and lowering operators of the free chiral boson act simply on the Young tableaux basis. We were able to show that this action encodes the Murnaghan-Nakayama rule for evaluating characters of the symmetric group, and the sign that is needed for this rule is supplied by Fermi statistics.

Armed with these tools we were able to show that topology changes are characterized by two important properties. First, a local field in the free field chiral boson becomes effectively multivalued so that in the simplest case 
\begin{equation}
\phi(\theta) = \phi^1(\theta)+\phi^2(\theta) -\tilde \phi^1(\theta)
\end{equation}
The total field, which in our case can be identified with  the charge current density,  can be written effectively as contributions from edges of the droplet distribution. This decomposition is valid only for nearby states to a reference state with a different topology from the vacuum. The important result for us is that the fields $\phi^{1,2}$ and $\tilde \phi^1$ on the subspace of nearby states states are each given by a chiral free boson, and they all commute with each other. The $\tilde \phi$ field has negative energy states rather than positive energy. When we mode expand, this decomposition is a partial Bogolubov transformation. This result puts into firmer footing observations that have been made in \cite{Koch:2008ah}, where we can also extend the ideas straightforwardly to fairly general coherent states.

On the face of it, this dynamical generation of new degrees of freedom is a violation of the Zamolodchikov c-theorem. A theory with central charge $c=1$ in the UV flows to a theory which is seemingly of central charge $c=3$ in the IR. 
It is better to write this  central charge as follows $c_{IR}=(2,1)$,  where we are indicating 
by the decomposition the fact that the first two act to increase the energy, and the other set of oscillators acts to decrease the energy. This is the signature of the energy as a quadratic form, similarly as is done with spacetime dimensions. It is the fact that we can lower the energy around the new vacuum that allows the violation of the c-theorem: the 
vacuum of the new state is not stable. This property essentially arises from trying to do effective field theory in a very special non-vacuum state. Notice that $c_{UV}= 2-1$, so something similar to an index is preserved in the flow, indeed we found that this is the chiral anomaly of the system. The positive energy bosons carry anomaly one, and the negative energy bosons carry anomaly $(-1)$. 

The second property of the solutions with new topology is that they are characterized by having low uncertainty mode per mode in the mode expansion of the field $\phi$. In this sense, the states can be said to be classical. We found that this uncertainty can be used as an order parameter to measure the topology. One can also similarly use the entanglement entropy of these modes to characterize the topology. The type of measurement that gets the topology is either an uncertainty measurement or an entropy measurement. Numerically, this results from measuring several quantum observables that don't commute with each other. We use this information to form an algebraic combination of the measurements that can be used to measure the topology. This ends up being a non-linear measurement on the wave functions.
To offset the possibility that one or a few of the new  modes is excited in a non-classical  
state, the non-linear measurement needs to be performed on a large number of modes. The majority rule decides the topology by what we have termed a consensus measurement.

It is clear that this can be generalized beyond the simple stripe geometries we consider for more intricate bubbles. The multi-valuedness should then be thought of on a local basis in the coordinate $\theta$, and the partial Bogolubov transformation should also be thought of in terms of a local expansion. We did a partial analysis of this setup with an approximation that describes the state as a locally multi-edge geometry. we observed that within this approximation one could find that generalized correlators encoded the fourier transform of the number of edges.

 Also, the partial Bogolubov transformation makes it clear that the set of nearby states is somewhat compatible with effective field theory. The effective quantum fields are the new collective modes $\phi^I, \tilde \phi^J$.
 They exist on a neighborhood of the reference state. These collective fields do not stretch all the way to the UV. They have an effective cutoff given by the stringy exclusion principle, which depends
 on the details of the reference state. Because these modes only exist relative to some reference background they should be thought of as being background dependent. This seems to get around the 
 problem of  geometry being quantized by operators  in a semiclassical approximation:  the operators that are needed to do so are state dependent in a way that depends weakly on the state, which is not too different from the background field method.

 \subsubsection*{To bubble or not to bubble}

Given that we can trigger changes in topology by superposition and that we can get all possible topologies this way, we can argue that thinking of a quantum gravity theory as a sum (or path integral) over all topologies is at best ambiguous. 

There are cases where this sum over topologies is absolutely correct. For example, in the topological string one can sum over topologies associated with crystal melting \cite{Iqbal:2003ds}. In that case,  each shape of the melted crystal is a different topology. The limit shape of the crystal is the geometry of the string at large distances. The partition function depends on a probe brane 
and there is a parameter $a$ that describes how far the probe brane is from the crystal. At large $a$, the molten crystal can be ignored. For $a\simeq 1$ one gets the quantum corrected stringy geometry, and for $a\simeq g_s$ one is in the quantum foam regime. 

We can form an analogy with this setup. The basis of Young tableaux states is a complete basis of states. Each of these is topologically distinct in the naive classical supergravity approximation. The reason for this is that even though one might have the same spacetime  topology for two configurations (same number of disks), one should also count the quantized flux through each cycle as topological data (this can not be deformed by small amounts, but only by integers, due to the Dirac quantization condition). In this basis, {\em any state is in principle a superposition of an infinite number of distinct topologies that can be measured}. This is not a complete set of all the possible topologies, only those that can be realized by rotationally invariant configurations.
One can expect that for sufficiently classical states, like those that are close to our reference states, one can define an average (coarse-grained) topology that only counts the big corners, but not the small indentations of corners (these are small semiclassical excitations around the reference state). The distinction between geometry and excitations about a geometry depends on energy (this replaces the parameter $a$ of the topological string). Our consensus measurement of topology necessitates a discussion of where we set the stringy exclusion principle. Unless we already know the state,  this is not  known a priori. Indeed, in our discussion in this paper, this is usually state dependent. When we try to go to lower energies for the probes, the geometry appears to bubble more, while when we go to much higher energies than the stringy exclusion principle dictates, the topology looks trivial. In our case, this is tied to how entangled the different long wavelength modes of the oscillators are to the UV degrees of freedom in the free chiral boson. We can not discuss this entanglement without first defining what we mean by long wavelength versus short wavelength.

On the other hand, we can also define the theory entirely in terms of classical coherent states of the trivial topology as we argued in this paper. This is a different partition of unity (a different choice of basis states for the Hilbert space). In this case, the definition of topology depends on the precise superposition of states  that we take. More precisely, it is contingent  on our ability to find a reasonable nearby space of excitations to a given reference state that can be associated to small deformations of a classical geometry. This is a background dependent formulation of the dynamics, similar to how one treats the background field method. What is curious is that the existence of the  new topologies implies  that there is more than one classical limit of the free chiral boson field theory. This is to be understood in a double scaling limit. 

In a certain sense, what we should be doing instead is to argue that the topology is not meaningful on its own: all versions of topology that we have discussed so far should be allowed at the same time, but most of them will not be useful descriptions of the system. This is very familiar when we think about dualities in field theory and string theory. Different duality frames are more or less useful depending on the size of particular cycles, or in the strength of certain coupling constants. 
The prescription that is more classical and permits us to get results with the least effort should be the preferred duality frame. In this sense, we should argue that  at least some aspects of topology in the study of bubbling solutions correspond to a choice of duality frame. The frame that most easily describes a configuration should be preferred. When we move away from simple configurations maybe none of the descriptions is useful on their own, but there we have a picture of a duality web where as we move between configurations, we get natural transitions in the topology of spacetime without any apparent singularity.

What is obvious is that if we want to have it both ways, we are double counting. This has implications for the fuzzball proposal \cite{Mathur:2012zp} (see also \cite{Bena:2013dka}) and the counting of 
states in those geometries. It might be the case that in these other setups all different geometries and topologies are superpositions of more basic coherent states with a  fixed  topology. Understanding this intriguing possibility is beyond the scope of the present work.

\subsubsection*{Decoding the hologram}

A natural question to ask is to what extent, given a state with a different topology than the vacuum of $AdS_5\times S^5$, is one able to recover the geometry from naive holographic data on the boundary. 
The techniques that usually permit one to do so are elaborations on the Fefferaman-Graham expansion of the metric, extended to other fields \cite{Witten:1998qj}. A hologram would give us the solutions for the vacuum expectation values of single trace operators in the boundary, and that data should be useable to decipher the geometry of the solution. 

As noticed in \cite{Skenderis:2007yb}, this data seems to be insufficient to understand the geometry of circularly symmetric solutions of supergravity, as there is some ambiguity in how to do that.
For us, this data is given by the expectation values of the modes of the chiral boson.
For standard coherent states around the vacuum topology, this data is sufficient to reconstruct the coherent state. 

For solutions around a geometric circularly symmetric solution with non-trivial topology, it was first noticed in \cite{Koch:2008ah} that the excitations of the mode expansion of the "trace" modes on the UV theory (the free chiral boson) decompose into linear combinations of modes at each edge. In this paper, we have proven this result and argued that the partial Boguliubov transformation can be completed to a full Boguliubov transformation. We have also seen  that the long wavelength modes of the mode expansion of the chiral boson of the UV theory act simply on these geometric states, and only a subset of the nearby Hilbert space is accessible by these actions. This is a general property of having a partial Boguliubov transformation. If we restore factors of $N$, and we have a solution with many annuli and energy of order $N^2$, the wavelength of the modes  of the chiral boson become dependent at energies of order $N$ (this is the stringy exclusion principle scale). More precisely, they fail to be planar at energies of order $\sqrt N$ (see for example \cite{Kristjansen:2002bb}), which are still larger than the naive Planck scale $N^{1/4}$.

A naive low energy observer would probe wavelengths up to the order of the Planck scale. When extended to the boundary, the modes would all have long wavelengths with  respect to the stringy exclusion principle. The other modes of the Boguliubov transformation become invisible and have to be treated as being traced over.  We cannot decode the hologram with the naive boundary data. This is true even if we have measured the approximate radii of the 
circular droplets. In essence, we only measure a linear combination of the geometric modes, and the other linear combinations are not accessible to the holographic observer at infinity. The modes that are visible are effectively in a generalized thermal state mode per mode and they are very entangled with the UV modes.

This suggests that the reconstruction of local fields in the bulk from the boundary, a la \cite{Hamilton:2005ju}, is generically suspect in a low energy approximation for states with non-trivial topologies.
This setup ignores the information of the transplanckian modes and the underlying UV theory, which at this scale is not really geometric in the classical sense any longer. If one acts with these types of mode operators of very high energy,  one disturbs the underlying geometry by either adding a D-brane or making excitations that make the droplets meet with each other. A fluctuation this large is non-local any longer.

Also, entanglement by itself is a very coarse description of the state and is not necessarily very useful.  Although we have been  able to realize horizon free geometries, where measuring the momentum space entanglement can be used as an order parameter to describe the topology, and we realize precisely some ideas in \cite{VanRaamsdonk:2010pw} in a different context, the precise set of states for which we get such a geometry are not uniquely determined by this information. It is the construction of the modes that describe the fluctuations to the nearby states to a reference state that actually represent 
the full details of the physics.

\subsubsection*{Final remarks}

One of the main conclusions of this paper is that even though we have a complete Hilbert space of states in which quantum mechanics is valid, the measurement of topology is not the result of an operator measurement. If topology is measured classically by integrating out a density made of  polynomials of the  curvature of the metric  over the manifold, as we expect for gauge invariant operators in gravity,
 the fact that the topology cannot be measured by an operator seems to indicate that the metric (even modulo gauge invariance issues) is also not described by  an operator. In our construction, the metric fluctuations around a sufficiently classical state exist relative to that state, but the construction of such operators does not extend to the full phase space of the theory. The generic state is non-geometric, but the semiclassical analysis is valid where it should be. This seems to be one of the properties that we need in order to claim that spacetime is emergent and not fundamental. 
 
The holographic modes at infinity always exists. In our case, they are the mode expansion of the free chiral boson.  These (boundary) modes give an approximation to something that looks geometric in a Fefferman-Graham expansion. However, the modes in the interior 
do not necessarily exist as operators. They might only be constructible around particular classical configurations. The proposal we have for this phenomenon is inherently non-linear: the modes that may exists in a superposition of states,  do not exist in any one of the states that we are superposing. This emergence of modes depends on the entanglement of the soft modes with the UV and with each other. A non-linear proposal for quantum mechanics defining the  physics inside of the horizon has been put forward by Papadodimas and Raju  \cite{Papadodimas:2012aq}. This proposal depends crucially on this entanglement, and essentially only on this entanglement (the state is pure but typical, so the details of the state are fairly random). 
 For us, the entanglement of the modes is clearly not enough. The modes that we build are all 
outside the horizon and their existence depends on the state being just right.

\acknowledgments

D.B. would like to thank D. Gross, G. Horowitz,  A. Puhm, S. Ramgoolam, M. Romo and C. Vafa for various discussions, S. Raju for correspondence,  and especially D. Marolf, who has contributed enormously to our insight in this paper. 
Work  supported in part by the department of Energy under grant {DE-SC} 0011702. 
\appendix

\section{The Murnaghan-Nakayama rule}\label{app:MN}

A standard problem in the group theory of the symmetric group $S_n$ is the computation of the characters of conjugacy classes $[\sigma]$ in a given representation $R$. That is, we want to compute 
\begin{equation}
\chi_{R}([\sigma])
\end{equation}
and as is explicitly presented in section \ref{sec:gt}, we need these characters to implement the Fourier transform relating the ``string basis"  and the ``D-brane basis" of our Hilbert space. 

The representations $R$ of $S_n$ will be labeled by Young diagrams with $n$ boxes, while the conjugacy classes will be presented in a cycle form $[\sigma] = t_1^{w_1} \dots t_k ^{w_k}$ with $\sum k w_k = n$. The Murnaghan-Nakayama rule gives a recursive way to evaluate $\chi_{R}([\sigma])$ in terms of $\chi_{\tilde R}([\tilde \sigma])$, where we have that $[\sigma] = [\tilde \sigma] t_s$ for some $t_s$, and the $\tilde R$ is a set of representations of $S_{n-s}$ related to $R$ and $s$ in a particular way. 

The rule is easiest to explain with an example first.

Consider for example  the tableaux with $10$ boxes given by
\begin{equation}
\ydiagram{4,2,2,1,1}
\end{equation}
which is a partition of 10. Assume that we want to compute the characters for splitting into two traces (only group elements with two different cycles) of the diagram. There are $5$ such possibilities: $5+5, 6+4, 7+3, 8+2,9+1$. Let us compute the splitting into $5+5$. The first step is to find the hooks of the diagram, and to decorate the diagram with the hook lengths, in the standard way
\begin{equation}
\begin{ytableau}
8&5&2&1\\
5&2\\
4&1\\
2\\
1
\end{ytableau}
\end{equation}

This will be useful, as hooks are in one to one correspondence with skew hooks, which are of interest to us, and the corresponding pairs have the same length.  We see that there are two (skew) hooks of length $5$.  We first remove one of the two length $5$ skew hooks, as shown in the figure
\begin{equation}
\ydiagram[*(white) \bullet]{1+3,1+1,1+1,0,0}*[*(white)]{4,2,2,1,1}
\end{equation}
the other option is
\begin{equation}
\ydiagram[*(white) \bullet]{4+0,1+1,0+2,0+1,0+1}*[*(white)]{4,2,2,1,1}
\end{equation}
As you can see, a skew-hook is a regular hook that has been moved to the edge of the tableaux. Like standard hooks, they do not contain a $2\times 2$ squares. As previously stated, the set of skew-hooks and regular hooks are in one to one correspondence, so one reads of the allowed skew-hook lengths by reading the lengths of the ordinary hooks.

A sign is  assigned to each such skew hook. The sign  is $(-1)$ if the number of rows covered by the skew hook is even, or $(+1)$ if this is odd.  After removing the skew-hook we are left with a proper partition, and we are also left with a cycle decomposition of the remnant of the group element (or conjugacy class). The Murnaghan-Nakayama rule states that to obtain the character of a diagram, sum over the characters of the remnant of the group element on the remnant tableaux with the sign of the skew-hook accounted for. Let us do so for the example above.

To simplify, we will notate a young diagram by the length of its rows. 
Thus, the full diagram is $(4,2,2,1,1)$, and we will notate the conjugacy class by the lengths of the cycles. That is, $[5,5]$. 

For the example above, one skew hook has an odd number of  rows and the other has an even number of rows, so the sign is $(+1)$ and $(-1)$ respectively. The rule then gives 
\begin{equation}
\chi_{4,2,2,1,1} ([5,5]) = \chi_{1,1,1,1,1} ([5]) - \chi_{4,1}([5])= 1+1=2
\end{equation}
where the character of the remnant character is also $\pm 1$ and is determined by the number of rows it contains.  Since there are no hooks of length $6, 7, 9$ (as made explicit by our diagram with the hook lengths notated), we also find immediately that
\begin{equation}
\chi_{4,2,2,1,1} ([6,4])=\chi_{4,2,2,1,1} ([7,3])=\chi_{4,2,2,1,1} ([9,1])=0
\end{equation}
while for the last one, we find 
\begin{equation}
\chi_{4,2,2,1,1}([8,2]) =\chi_{1,1}([2]) = -1
\end{equation}

A convenient way to think about the Murnaghan-Nakayama rule is that it gives the action of the lowering operators $s\partial_{t_s}$ on the Young diagram basis, and so it does not just compute the 
characters, but the action of the lowering operator on the Hilbert space of states. Let us discuss this with a few  examples. Consider first the state

\begin{equation}
|t_{1} t_{2}^{2} \rangle = 
 \ytableausetup{boxsize=1em, aligntableaux = center}
   \begin{ytableau}
   *(white) & *(white) & *(white) & *(white) & *(white)
  \end{ytableau}
+
 \ytableausetup{boxsize=1em}
   \begin{ytableau}
   *(white) & *(white) & *(white) \\
   *(white) & *(white)
  \end{ytableau}
-2\;
 \ytableausetup{boxsize=1em}
   \begin{ytableau}
   *(white) & *(white) & *(white) \\
   *(white)  \\
   *(white)
  \end{ytableau}
+
 \ytableausetup{boxsize=1em}
   \begin{ytableau}
   *(white) & *(white) \\
   *(white) & *(white) \\
   *(white)
  \end{ytableau}
+
 \ytableausetup{boxsize=1em}
   \begin{ytableau}
   *(white) \\
   *(white) \\
   *(white) \\
   *(white) \\
   *(white)
  \end{ytableau}
\end{equation}
We could act on this state with the lowering operator $a_{1}$.  The conjugacy class states are just Fock space states, so we know

\begin{equation}
a_{1} |t_{1} t_{2}^{2} \rangle = |t_{2}^{2} \rangle
\end{equation}

We could further compute the necessary characters using the MN rule and expand the new state as 

\begin{equation}
|t_{2}^{2} \rangle = 
 \ytableausetup{boxsize=1em}
   \begin{ytableau}
   *(white) & *(white) & *(white) & *(white)
  \end{ytableau}
-
 \ytableausetup{boxsize=1em}
   \begin{ytableau}
   *(white) & *(white) & *(white) \\
   *(white)
  \end{ytableau}
+2\;
 \ytableausetup{boxsize=1em}
   \begin{ytableau}
   *(white) & *(white) \\
   *(white) & *(white)
  \end{ytableau}
-
 \ytableausetup{boxsize=1em}
   \begin{ytableau}
   *(white) & *(white) \\
   *(white) \\
   *(white)
  \end{ytableau}
+
 \ytableausetup{boxsize=1em}
   \begin{ytableau}
   *(white) \\
   *(white) \\
   *(white) \\
   *(white)
  \end{ytableau}
\end{equation}
Alternatively, we could have considered applying the lowering operator directly to the diagrams.  Our lowering operator is $a_{1}$, so it removes skew hooks of length 1 as shown 

\begin{equation}
a_{1}\;
 \ytableausetup{boxsize=1em}
   \begin{ytableau}
   *(white) & *(white) & *(white) \\
   *(white)
  \end{ytableau}
  =
   \ytableausetup{boxsize=1em}
   \begin{ytableau}
   *(white) & *(white) & *(white)
  \end{ytableau}
  + 
  \ytableausetup{boxsize=1em}
   \begin{ytableau}
   *(white) & *(white) \\
   *(white)
  \end{ytableau}
\end{equation}

There were two possible ways to remove skew hooks of length 1, so we end with a sum of the two possibilities.  On the other hand, if we apply this to the trivial representation, there is only one way to remove a hook, so we have

\begin{equation}
a_{1}\;
 \ytableausetup{boxsize=1em}
   \begin{ytableau}
   *(white) & *(white) & *(white) & *(white)
  \end{ytableau}
  =
 \ytableausetup{boxsize=1em}
   \begin{ytableau}
   *(white) & *(white) & *(white)
  \end{ytableau}
\end{equation}

Applying the lowering operator to the full state, we find

\begin{equation}
a_{1} |t_{1} t_{2}^{2} \rangle = a_{1} \left(
 \ytableausetup{boxsize=1em, aligntableaux = center}
   \begin{ytableau}
   *(white) & *(white) & *(white) & *(white) & *(white)
  \end{ytableau}
+
 \ytableausetup{boxsize=1em}
   \begin{ytableau}
   *(white) & *(white) & *(white) \\
   *(white) & *(white)
  \end{ytableau}
-2\;
 \ytableausetup{boxsize=1em}
   \begin{ytableau}
   *(white) & *(white) & *(white) \\
   *(white)  \\
   *(white)
  \end{ytableau}
+
 \ytableausetup{boxsize=1em}
   \begin{ytableau}
   *(white) & *(white) \\
   *(white) & *(white) \\
   *(white)
  \end{ytableau}
+
 \ytableausetup{boxsize=1em}
   \begin{ytableau}
   *(white) \\
   *(white) \\
   *(white) \\
   *(white) \\
   *(white)
  \end{ytableau} \; \right)
\end{equation}

\begin{equation}
=
 \ytableausetup{boxsize=1em}
   \begin{ytableau}
   *(white) & *(white) & *(white) & *(white)
  \end{ytableau}
+
 \ytableausetup{boxsize=1em}
   \begin{ytableau}
   *(white) & *(white) \\
   *(white) & *(white)
  \end{ytableau}
+
 \ytableausetup{boxsize=1em}
   \begin{ytableau}
   *(white) & *(white) & *(white) \\
   *(white)
  \end{ytableau}
-2\;
 \ytableausetup{boxsize=1em}
   \begin{ytableau}
   *(white) & *(white) & *(white) \\
   *(white)
  \end{ytableau}
-2\;
 \ytableausetup{boxsize=1em}
   \begin{ytableau}
   *(white) & *(white) \\
   *(white)  \\
   *(white)
  \end{ytableau}
+
 \ytableausetup{boxsize=1em}
   \begin{ytableau}
   *(white) & *(white) \\
   *(white) \\
   *(white)
  \end{ytableau}
+
 \ytableausetup{boxsize=1em}
   \begin{ytableau}
   *(white) & *(white) \\
   *(white) & *(white)
  \end{ytableau}
+
 \ytableausetup{boxsize=1em}
   \begin{ytableau}
   *(white) \\
   *(white) \\
   *(white) \\
   *(white)
  \end{ytableau}
\end{equation}
We see that if we simplify this, we get the same expression for the state $| t_{2}^{2}\rangle$ shown above.

We would expect that if we apply any $a_j$ to our original state $|t_{1} t_{2}^{2} \rangle $ with $j>2$ that this should kill the state.  Let's see how this works diagrammatically.  Of course, there are no skew hooks of length greater than 5, as each diagram has only five boxes, so any $a_{j}$ with $j>5$ will kill the state.  As a less trivial example, let's consider acting with $a_3$.  This gives

\begin{equation}
a_{3} |t_{1} t_{2}^{2} \rangle = a_{3} \left(
 \ytableausetup{boxsize=1em}
   \begin{ytableau}
   *(white) & *(white) & *(white) & *(white) & *(white)
  \end{ytableau}
+
 \ytableausetup{boxsize=1em}
   \begin{ytableau}
   *(white) & *(white) & *(white) \\
   *(white) & *(white)
  \end{ytableau}
-2\;
 \ytableausetup{boxsize=1em}
   \begin{ytableau}
   *(white) & *(white) & *(white) \\
   *(white)  \\
   *(white)
  \end{ytableau}
+
 \ytableausetup{boxsize=1em}
   \begin{ytableau}
   *(white) & *(white) \\
   *(white) & *(white) \\
   *(white)
  \end{ytableau}
+
 \ytableausetup{boxsize=1em}
   \begin{ytableau}
   *(white) \\
   *(white) \\
   *(white) \\
   *(white) \\
   *(white)
  \end{ytableau}\; \right)
\end{equation}

\begin{equation}
=
 \ytableausetup{boxsize=1em}
   \begin{ytableau}
   *(white) & *(white)
  \end{ytableau}
-
 \ytableausetup{boxsize=1em}
   \begin{ytableau}
   *(white)  \\
   *(white)
  \end{ytableau}
-
 \ytableausetup{boxsize=1em}
   \begin{ytableau}
   *(white) & *(white) 
  \end{ytableau}
+
 \ytableausetup{boxsize=1em}
   \begin{ytableau}
   *(white) \\
   *(white)
  \end{ytableau} =0
\end{equation}
where we remember that if the height of the hook is even, then the diagram changes sign.  Notice also that there were no allowed hooks of length 3 in the third diagram, so that piece vanishes.

Finally, we would expect that if we apply any $a_{j}$ on the state $|t_{j}\rangle$, then we should get $a_{j}|t_{j}\rangle=j|0\rangle$, where the factor of $j$ comes from the commutation rules for our raising and lowering operators: $[a_{i},a_{j}^{\dagger}]=\delta_{ij}j$.  We will represent the vacuum diagrammatically as $\bullet$.  As an example of this, we have

\begin{equation}
a_{4}|t_{4} \rangle = a_{4}\left(
 \ytableausetup{boxsize=1em}
   \begin{ytableau}
   *(white) & *(white) & *(white) & *(white)
  \end{ytableau}
-
 \ytableausetup{boxsize=1em}
   \begin{ytableau}
   *(white) & *(white) & *(white) \\
   *(white)
  \end{ytableau}
+
 \ytableausetup{boxsize=1em}
   \begin{ytableau}
   *(white) & *(white) \\
   *(white) \\
   *(white)
  \end{ytableau}
-
 \ytableausetup{boxsize=1em}
   \begin{ytableau}
   *(white) \\
   *(white) \\
   *(white) \\
   *(white)
  \end{ytableau}\;\right)
\end{equation}

\begin{equation}
= \bullet + \bullet +\bullet +\bullet = 4 \; \bullet= 4 \ket 0
\end{equation}
where each piece was exactly a skew hook of length 4 and so became the vacuum state when hit with $a_{4}$.  Also, note the sign changes come from the height of the hooks.

One can see that this is indicative of  the general case. This is handled by using the fact that 
\begin{equation}
\ket R= \sum \frac {\chi_R[\sigma]}{ \prod k^w_k w_k !} \prod t_k^{w_k}
\end{equation}
Now, let us try to remove one $t_s$ from the above equation, by acting with $s\partial_{t_s}$ on a monomial $\prod t_k^{w_k}$. We get   $s w_s t_s^{w_{s}-1}$. The extra factor of $s$ and $w_s$ cancel terms in the denominator, so that we get regular denominators as would correspond to $[\sigma]/t_s= [\tilde \sigma]$ for any sigma that has a $t_s$ in it.

Now we use that $\chi_R[\sigma]= \sum (-1)^* \chi_{\tilde R}([\tilde \sigma])$ where $(-1)^* $ is the sign assigned by the Murnaghan-Nakayama rule. That is, we find that 
\begin{equation}
s\partial_{t_s} \ket R= \sum \frac {(-1)^*\chi_{\tilde R}[\tilde \sigma]}{ \prod k^w_k w_k !} ( s w_s) [\sigma]/t_s = \sum_{\hbox{hooks of length } s} (-1)^* \ket {\tilde R}
\end{equation}
and because we recognize that the sum is over $[\tilde \sigma]$ unrestricted, we find that on the right hand side we sum over the states $\tilde R$ with the Murnaghan-Nakayama rule sign and nothing else.

\section{Assigning Young Tableaux to Fermions states}\label{app:fermions}

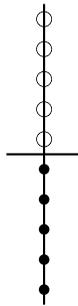
\begin{figure}[h]\begin{center}
\begin{pspicture}(2,4)
\psdots[dotstyle=o, dotsize=6pt](1,2.2)(1,2.6)(1,3)(1,3.4)(1,3.8)
\psdots[dotsize=4pt](1,1.8)(1,1.4)(1,1)(1,0.6)(1,0.2)
\psline(1,0)(1,4)
\psline(0.5,2)(1.5,2)
\end{pspicture}
\caption{Vacuum State of Fermi Sea}\label{fig:vacfer}
\end{center}
\end{figure}

The ground state of the multiple particle system $\bullet$ is defined by the Slater determinant
\begin{equation}
\bullet \simeq \lim_{N\to \infty} \ket{-1/2} \ket{-3/2} \dots \ket{-N/2}_{antisymmetrized} 
\end{equation}
where, as in the text, we shift our allowed energies so that particles sit at half integer levels.  And, as usual, the ground state has the full infinite tower of negative states occupied.  This can be represented pictorially as follows, as in figure \ref{fig:vacfer},
where dots represent filled states and circles represent holes.

A complete basis of states is given by
\begin{equation}
\ket{\{n \}}\simeq \lim_{N\to \infty} \ket {n_1} \ket{n_2} \dots \ket{n_{N}}_{\rm{antisymm}} 
\end{equation}
with $n_1>n_2>n_3 >\dots>n_N$, half integers, and for all sufficiently large $j$ we require that $n_j=-\frac{2j-1}{2}$.  This state is represented by filling in each $n_j$ energy level with a dot. If a particular value of $j$ is missing, it is left empty (circles in the drawings).

To each state, we will assign a Young diagram whose $j$-th row has $r_j=n_{j}-(\frac 12 - j)$ boxes (and if $r_j=0$ we leave those rows empty and without boxes).  By inverting this expression, we can go the other way, assigning a state of the Fermions to a Young diagram. We then relate these diagrams to representations of the symmetric group.

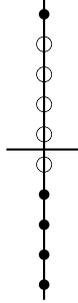
\begin{figure}[h]\begin{center}
\begin{pspicture}(2,4)
\psdots[dotstyle=o, dotsize=6pt](1,2.2)(1,2.6)(1,3)(1,3.4)(1,1.8)
\psdots[dotsize=4pt](1,1.4)(1,1)(1,0.6)(1,0.2)(1,3.8)
\psline(1,0)(1,4)
\psline(0.5,2)(1.5,2)
\end{pspicture}
\caption{Excitation with 5 units of energy on the highest Fermion}\label{fig:excit5row}
\end{center}
\end{figure}

A trivial representation (totally symmetric if considered as a representation of $U(N)$ instead of $S_n$) corresponds to an excitation out of the sea, with an energy equal to the number of boxes.  That is, the state given by

\begin{equation}
 \ytableausetup{boxsize=1em, aligntableaux = center}
\begin{ytableau}
*(white) & *(white) & *(white) & *(white) & *(white) 
\end{ytableau}
\end{equation}
corresponds to the figure \ref{fig:excit5row}

\begin{figure}[h]\begin{center}
\begin{pspicture}(2,4)
\psdots[dotstyle=o, dotsize=6pt](1,2.6)(1,3)(1,3.4)(1,3.8)(1,0.6)
\psdots[dotsize=4pt](1,1.8)(1,1.4)(1,1)(1,0.2)(1,2.2)
\psline(1,0)(1,4)
\psline(0.5,2)(1.5,2)
\end{pspicture}
\caption{Excitation with 1 unit of energy on the highest five fermions, or equivalent, a hole with 5 units of energy has been excited}\label{fig:excit5column}
\end{center}
\end{figure}
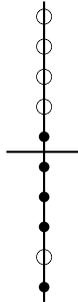

Further, the totally anti-symmetric state corresponds to a hole, where an $n$-box representation corresponds to a hole $n$ spaces below zero.  That is,

\begin{equation}
 \ytableausetup{boxsize=1em, aligntableaux = center}
\begin{ytableau}
*(white) \\
*(white) \\
*(white) \\
*(white) 
\end{ytableau}
\end{equation}
corresponds to the figure \ref{fig:excit5column}.

Basically, a tableaux is assigned by taking the highest fermion energy available and subtracting the energy of the highest occupied fermion in the ground state and assigning that many boxes to the first row
of the tableaux. We then do the same with the second highest energy fermion, and so on until all the subsequent fermions in the excited state have the same energy as the corresponding fermions in the vacuum, where we stop assigning boxes.

\end{document}